\documentclass[a4paper,fleqn,usenatbib]{mnras}
\usepackage{newtxtext,newtxmath}
\usepackage[T1]{fontenc}
\usepackage{ae,aecompl}

\usepackage{graphicx}	% Including figure files
\usepackage{amsmath}	% Advanced maths commands
\usepackage{amssymb}	% Extra maths symbols
\usepackage{xspace}
\usepackage{multirow}
\usepackage{anyfontsize}
\graphicspath{{graphics/}}
\newcommand{\kms}{kms$^{-1}$\xspace}	% km/s
\newcommand{\pa}{Par~1379\xspace} % Par 1379
\newcommand{\pb}{Par~2244\xspace} % Par 2244
\newcommand{\prot}{$P_{\text{rot}}$\xspace}
\newcommand{\sv}{Stokes~\emph{V}\xspace}
\newcommand{\si}{Stokes~\emph{I}\xspace}
\newcommand{\teff}{$T_{\text{eff}}$\xspace}
\newcommand{\rsun}{\text{R}\textsubscript{\sun}\xspace}
\newcommand{\rstar}{\ensuremath{R_{\star}}\xspace}
\newcommand{\msun}{\text{M}\textsubscript{\sun}\xspace}
\newcommand{\lsun}{\text{L}\textsubscript{\sun}\xspace}
\newcommand{\mjup}{\text{M}\textsubscript{Jup}\xspace}
\newcommand{\vrad}{$v_{\text{rad}}$\xspace}
\newcommand{\vsini}{$v\sin{i}$\xspace}
\newcommand{\radd}{$\text{rad~d}^{-1}$\xspace}

\newcommand{\msunyr}{\text{M}\textsubscript{\sun}\text{yr}$^{-1}$\xspace}

\title[The wTTSs Par~1379 and Par~2244]{Magnetic activity and radial velocity filtering of young Suns: \\The weak-line T Tauri stars Par~1379 and Par~2244}
\author[C. A. Hill et al.]{
C. A. Hill$^{1}$\thanks{E-mail: chill@irap.omp.eu}, A. Carmona$^{1}$, J.-F. Donati$^{1}$, G. A. J. Hussain$^{2,1}$, S. G. Gregory$^{3}$ \newauthor S. H. P. Alencar$^{4}$, J. Bouvier$^{5}$ and the MaTYSSE collaboration \\
$^{1}$IRAP, Universit\'{e} de Toulouse, CNRS, UPS, CNES, 14 Avenue Edouard Belin, Toulouse, F-31400, France \\
$^{2}$ESO, Karl-Schwarzschild-Str. 2, D-85748 Garching, Germany \\
$^{3}$SUPA, School of Physics and Astronomy, Univ. of St Andrews, St Andrews, Scotland KY16 9SS, UK \\
$^{4}$Departamento de F\`{i}sica - ICEx-UFMG, Av. Ant\^{o}nio Carlos, 6627, 30270--901 Belo Horizonte, MG, Brazil \\
$^{5}$Universit\'{e} Grenoble Alpes, CNRS, IPAG, F-38000 Grenoble, France \\}

\date{Accepted XXX. Received YYY; in original form ZZZ}

\pubyear{2017}

\begin{document}
\label{firstpage}
\pagerange{\pageref{firstpage}--\pageref{lastpage}}
\maketitle

\begin{abstract}
We report the results of our spectropolarimetric monitoring of the weak-line T-Tauri stars (wTTSs) \pa and \pb, within the MaTYSSE (Magnetic Topologies of Young Stars and the Survival of close-in giant Exoplanets) programme. Both stars are of a similar mass (1.6 and 1.8~\msun) and age (1.8 and 1.1~Myr), with \pa hosting an evolved low-mass dusty circumstellar disc, and with \pb showing evidence of a young debris disc. We detect profile distortions and Zeeman signatures in the unpolarized and circularly polarized lines for each star, and have modelled their rotational modulation using tomographic imaging, yielding brightness and magnetic maps. We find that \pa harbours a weak (250~G), mostly poloidal field tilted $65\degr$ from the rotation axis. In contrast, \pb hosts a stronger field (860~G) split 3:2 between poloidal and toroidal components, with most of the energy in higher order modes, and with the poloidal component tilted $45\degr$ from the rotation axis. Compared to the lower mass wTTSs, V819 Tau and V830 Tau, \pb has a similar field strength, but is much more complex, whereas the much less complex field of \pa is also much weaker than any other mapped wTTS. We find moderate surface differential rotation of $1.4\times$ and $1.8\times$ smaller than Solar, for \pa and \pb, respectively. Using our tomographic maps to predict the activity related radial velocity (RV) jitter, and filter it from the RV curves, we find RV residuals with dispersions of 0.017~\kms and 0.086~\kms for \pa and \pb, respectively. We find no evidence for close-in giant planets around either star, with $3\sigma$ upper limits of 0.56 and 3.54~\mjup (at an orbital distance of 0.1~au).
\end{abstract}

%%%%%%%%%%%%%%%%%%%%%%%%%%%%%%%%%%%%%%%%%%%%%%%%%%

%%%%%%%%%%%%%%%%% BODY OF PAPER %%%%%%%%%%%%%%%%%%
\section{Introduction}
Stellar magnetic fields have their largest impact during the early evolution of low-mass stars. T-Tauri stars (TTS) are late-type pre-main-sequence (PMS) stars that are gravitationally contracting towards the MS, and are initially surrounded by a protoplanetary disc \citep[e.g.][]{andre2009,donati2009}. The radial contraction increases the density of the stellar interior, and eventually leads to the development of a radiative core (for stellar masses $\gtrsim0.35$~\msun, \citealt{chabrier1997}), where the core boundary is known as the tachocline. This change in stellar structure is thought to significantly alter the stellar dynamo mechanism, from one that is distributed throughout the convective zone, to a solar-like dynamo concentrated at the tachocline \citep[e.g.][]{durney1993}. Such changes to the underlying dynamo mechanism may exhibit themselves as variations in the resulting magnetic field topologies, and so observations of magnetic fields on young stars can lead to a better understanding of the behaviour and evolution of the underlying stellar dynamo mechanisms as the star evolves towards the zero-age main-sequence (ZAMS).

At an age of 0.5--10~Myr, TTS may be classified into two groups (based on their accretion status), known as classical T-Tauri stars (cTTSs) and weak-line T-Tauri stars (wTTSs). cTTSs are defined as TTS that are accreting material from a massive (presumably planet forming) disc, whereas wTTSs have exhausted the gas in their inner discs and are no longer accreting (or are accreting at a low level). 

Magnetic fields of cTTSs play a vital role in controlling accretion process and triggering outflows, and largely dictate their angular momentum evolution \citep[e.g.][]{bouvier2007,frank2014}. In particular, large-scale fields of cTTSs can evacuate the central regions of accretion discs, funnel the material on to the star, and enforce co-rotation between the star and the inner disc Keplerian flows, causing cTTSs to rotate more slowly than expected from the contraction and accretion of the disc material \citep[e.g.][]{davies2014}. Furthermore, magnetic fields of cTTS and their discs can affect the formation and migration of planets \citep[e.g.][]{baruteau2014}. Moreover, fields of both cTTSs and wTTSs are known to trigger thermally driven winds through heating by accretion shocks and/or Alfv\'{e}n waves \citep[e.g.][]{cranmer2009,cranmer2011}, resulting in flares, coronal-mass ejections, and angular momentum loss \citep[e.g.][]{aarnio2012,matt2012}. It is clear then that characterizing magnetic fields in cTTSs and wTTSs is critical for testing and developing theoretical models (providing more physical realism), and trialling scenarios currently invoked to explain low-mass star and planet formation. 

Magnetic fields were first detected on cTTSs nearly 20 yr ago \citep[e.g.][]{johnskrull1999}, and through the MaPP (Magnetic Protostars and Planets) Large Observing Programme allocated on ESPaDOnS on the 3.6-m Canada-France-Hawaii Telescope (CFHT), the large-scale topologies of 11 cTTSs have been revealed \citep[e.g.][]{donati2007,hussain2009,donati2010a,donati2013}. Here, the large-scale stellar magnetic fields were mapped by performing an inversion of a time series of spectropolarimetric data, reconstructing the observed field with a spherical harmonic expansion, adopting a maximum-entropy approach. This technique (called Zeeman Doppler Imaging, ZDI) allows one to determine both the field strength and topology. This first survey revealed that the large-scale fields of cTTSs remain relatively simple and mainly poloidal when the host star is still fully or largely convective, but become much more complex when the host star turns mostly radiative \citep{gregory2012,donati2013}. This survey also showed that these fields are likely of dynamo origin, varying over time-scales of a few years \citep{donati2011, donati2012, donati2013}, and resembling those of mature stars with comparable internal structure \citep{morin2008b}.

In the case of wTTSs, a total of five stars have been magnetically mapped to date (using ZDI), namely V410 Tau, LkCa~4, V819~Tau, V830~Tau and TAP~26 \citep{skelly2010,carroll2012, donati2014, donati2015,yu2017}. These few stars show a much wider range of field topologies compared to cTTSs and MS dwarfs with similar internal structures; while V819~Tau and V830~Tau display a mostly poloidal field topology, V410~Tau and LkCa~4 show significant toroidal components despite being fully convective, in surprising contrast to fully convective cTTSs and mature M dwarfs that harbour relatively simple poloidal fields \citep{morin2008b,donati2013}. This toroidal component may develop by the stellar dynamo being influenced by the rapid spin-up that TTS experience, due to their gravitational contraction and angular momentum conservation, once their discs have dispersed \citep[see discussion in][]{donati2014}.

Given that wTTSs are the transitional phase between cTTSs and ZAMS low-mass stars, studies of their magnetic topologies and associated winds are of great interest, as these are the initial conditions in which disc-less PMS stars initiate their unleashed spin-up as they contract towards the ZAMS. This is the goal of the MaTYSSE (Magnetic Topologies of Young Stars and the Survival of close-in giant Exoplanets) Large Programme, allocated at CFHT over semesters 2013a--2016b (510~h) with complementary observations with the NARVAL spectropolarimeter on the 2-m Telescope Bernard Lyot at Pic du Midi in France (420~h) and with the HARPS spectropolarimeter at the 3.6-m ESO Telescope at La Silla in Chile (70~h). As well as determining the magnetic topologies of $\sim30$ wTTSs and monitoring the long-term topology variability of $\sim5$~cTTSs, MaTYSSE will allow us to broadly study magnetic winds of wTTSs and corresponding spin-down rates \citep[e.g.][]{vidotto2014}. Furthermore, we are able to filter-out most of the activity-related jitter from the radial velocity (RV) curves of wTTSs (using spectropolarimetry and doppler imaging to model surface activity) for potentially detecting hot Jupiters (hJs; see \citealt{donati2016a}), and thus verifying whether core accretion and migration is the most likely mechanism for forming close-in giant planets \citep[e.g.,][]{alibert2005}.

Here, we present full analyses of the phase-resolved spectropolarimetric observations of \pa and \pb, continuing our study of wTTSs in the framework of the MaTYSSE programme. For our study, we first document our observations in Section~\ref{sec:observations}, then review the stellar and disc properties of both stars in Section~\ref{sec:prop}. In Section~\ref{sec:tomography} we describe the results obtained after applying our tomographic modelling technique to the data, and in Section~\ref{sec:rv} we use these results to filter out the activity jitter from the RV curves, and look for the potential presence of hJs around both stars. In Section~\ref{sec:discussion} we discuss our results and their implications for understanding low-mass star and planet formation. Finally, in Section~\ref{sec:summary} we provide a summary of our results.

\section{Observations}
\label{sec:observations}
Spectropolarimetric observations of \pa were taken in January 2014, with observations of \pb taken in December 2014 and January 2015, both using ESPaDOnS at the 3.6-m~CFHT. Spectra from ESPaDOnS span the entire optical domain (from 370 to timesp1000~nm) at a resolution of 65,000 (i.e., a resolved velocity element of 4.6~\kms) over the full wavelength range, in both circular or linear polarization \citep{donati2003}. 

A total of 10 circularly-polarized (\sv) and unpolarized (\si) spectra were collected for \pa over a time-span of 14 nights, corresponding to around 2.3 rotation cycles. Time sampling was regular apart from a five night gap due to bad weather near the middle of the run. For \pb, 14 spectra were collected over a 25 night time-span, corresponding to around 9.2 rotation cycles. Time sampling for these spectra was irregular, with a six night gap after the first four nights, then an eight night gap after the next 3~nights, with regular spacing after that. These observational gaps, corresponding to two and three rotation cycles, respectively, did not significantly affect our ability to model surface features and magnetic fields (see Section~\ref{sec:tomography}), and only became significant when applying our RV filtering technique (as discussed in Section~\ref{sec:rv}).

All polarization spectra consist of four individual sub-exposures (each lasting 946.2~s for \pa, and 835~s for \pb), taken in different polarimeter configurations to allow the removal of all spurious polarization signatures at first order. All raw frames were processed using the \textsc{esprit} software package, which performs bias subtraction, flat fielding, wavelength calibration, and optimal extraction of (un)polarized \'{e}chelle spectra, as described in the previous papers of the series (\citealt{donati1997}, also see \citealt{donati2010b, donati2011, donati2014}), to which the reader is referred for more information. The peak signal-to-noise ratios (S/N, per 2.6~\kms velocity bin) achieved on the collected spectra range between 99 to 139 (median 126) for \pa, and between 96 and 164 (median 145) for \pb, depending mostly on weather/seeing conditions. All spectra are automatically corrected for spectral shifts resulting from instrumental effects (e.g., mechanical flexures, temperature or pressure variations) using atmospheric telluric lines as a reference. This procedure provides spectra with a relative RV precision of better than 0.030~\kms \citep[e.g.][]{moutou2007,donati2008a}. A journal of all observations is presented in Table~\ref{tab:obs} for both stars.

\begin{table}
\centering
\caption{Journal of ESPaDOnS observations of \pa (first 10 lines) and \pb (last 14 lines). Each observation consists of a sequence of 4 subexposures, each lasting 946.2~s for \pa and 835~s for \pb, respectively. Columns 1--4 list (i) the UT date of the observation, (ii) the corresponding UT time at mid exposure, (iii) the Barycentric Julian Date (BJD), and (iv) the peak S/N (per 2.6~\kms velocity bin) of each observation. Columns 5 and 6 respectively list the S/N in \si least-squares deconvolution (LSD) profiles (per 1.8~\kms velocity bin), and the rms noise level (relative to the unpolarized continuum level $I_{\text{c}}$ and per 1.8~\kms velocity bin) in the \sv LSD profiles. Column~7 indicates the rotational cycle associated with each exposure, using the ephemerides given in Equation~\ref{eq:ephemeris}.}
\label{tab:obs}
\setlength\tabcolsep{3pt} % default value: 6pt
\begin{tabular}{lcccccc}
\hline													
\multicolumn{1}{c}{Date} 	&	 UT 	&	BJD	&	S/N	&	S/N$_{\textsc{lsd}}$	&	 $\sigma_{\textsc{lsd}}$ 	&	Cycle	\\
	&	 (hh:mm:ss) 	&	(2,456,660+)	&	 	&		&	 (0.01\%) 	&		\\
\hline													
2014 Jan 07	&	10:29:51	&	5.94192	&	119	&	1484	&	5.6	&	0.011	\\
2014 Jan 08	&	10:36:33	&	6.94653	&	129	&	1518	&	5.1	&	0.191	\\
2014 Jan 09	&	08:44:42	&	7.86883	&	122	&	1521	&	5.5	&	0.356	\\
2014 Jan 10	&	09:56:19	&	8.91852	&	132	&	1559	&	4.9	&	0.544	\\
2014 Jan 15	&	10:25:58	&	13.93888	&	99	&	1459	&	7.1	&	1.443	\\
2014 Jan 16	&	08:49:59	&	14.87218	&	122	&	1521	&	5.4	&	1.610	\\
2014 Jan 17	&	08:01:30	&	15.83846	&	130	&	1518	&	5.0	&	1.783	\\
2014 Jan 18	&	09:36:06	&	16.90410	&	132	&	1511	&	5.0	&	1.974	\\
2014 Jan 19	&	07:37:37	&	17.82177	&	113	&	1466	&	6.1	&	2.139	\\
2014 Jan 20	&	08:14:06	&	18.84704	&	139	&	1563	&	4.5	&	2.322	\\
\hline													
2014 Dec 18	&	13:57:57	&	351.06756	&	125	&	1651	&	6.0	&	0.031	\\
2014 Dec 19	&	11:52:09	&	351.98019	&	126	&	1638	&	5.6	&	0.355	\\
2014 Dec 20	&	11:13:19	&	352.95321	&	154	&	1723	&	4.3	&	0.701	\\
2014 Dec 21	&	12:06:11	&	353.98993	&	146	&	1682	&	4.7	&	1.069	\\
2014 Dec 27	&	13:45:41	&	360.05892	&	96	&	1260	&	8.3	&	3.225	\\
2014 Dec 28	&	11:33:31	&	360.96712	&	164	&	1755	&	4.0	&	3.547	\\
2014 Dec 29	&	10:10:49	&	361.90967	&	160	&	1736	&	4.2	&	3.882	\\
2015 Jan 06	&	11:15:31	&	369.95436	&	141	&	1685	&	4.9	&	6.739	\\
2015 Jan 07	&	10:16:28	&	370.91331	&	144	&	1660	&	4.8	&	7.080	\\
2015 Jan 08	&	10:10:47	&	371.90933	&	157	&	1739	&	4.1	&	7.434	\\
2015 Jan 09	&	10:12:45	&	372.91066	&	139	&	1682	&	4.9	&	7.790	\\
2015 Jan 10	&	11:56:48	&	373.98288	&	151	&	1597	&	4.6	&	8.170	\\
2015 Jan 11	&	10:07:39	&	374.90704	&	159	&	1770	&	4.1	&	8.499	\\
2015 Jan 13	&	12:22:31	&	377.00060	&	138	&	1622	&	5.3	&	9.242	\\
\hline					
\end{tabular}
\end{table}

\section{Stellar and disc properties}
\label{sec:prop}
Both stars, belonging to the Orion Nebula Cluster (ONC) flanking fields \citep{furesz2008, tobin2009}, were classified as wTTSs by \cite{rebull2000} due to their young age (based on evolutionary tracks), combined with a lack of UV excess (suggesting low levels of accretion). Indeed, our spectra show strong, non-variable \ion{Li}{i}~6708~\AA\ absorption, confirming the young ages of these targets, with the lack of significant veiling supporting their status as wTTSs (see Section~\ref{sec:accretion} for further discussion). Moreover, \cite{rebull2001} report very regular periodic light curves for both \pa and \pb that do not appear like that of a cTTS, further supporting their non-accreting status. However, \cite{megeath2012} place \pa near the limit between wTTSs and cTTSs (based on mid-infrared photometry).

Photometric rotation periods of $5.620\pm0.009$~d and $2.820\pm0.002$~d were determined by \cite{rebull2001} for \pa and \pb, respectively, with \cite{carpenter2001} finding a similar period of 2.83--2.84~d for \pb. For the remainder of this work, however, we adopt the rotation periods that are determined from our tomographic modelling in Section~\ref{sec:tomography}. Here, the equatorial rotational cycles of \pa and \pb (denoted $E_{1}$ and $E_{2}$ in Equation~\ref{eq:ephemeris}) are computed from Barycentric Julian Dates (BJDs) according to the (arbitrary) ephemerides

\begin{align}
\text{BJD (d)} &= 2456665.9 + 5.585E_{1} & \text{(for Par 1379)} \nonumber \\ 
\text{BJD (d)} &= 2457011.0 + 2.8153E_{2} & \text{(for Par 2244)}
\label{eq:ephemeris}
\end{align}

\noindent where our rotational periods are in excellent agreement (within 1 and 2$\sigma$ of our values) with those determined previously by \cite{rebull2001} and \cite{carpenter2001}.

%mentioned photometric period previously published in Rebull et al 2001 
%par 1379 = star #1032 in R01 : Prot = 5.62\pm0.009d
%par 2244 = star #2350 in R01 : Prot = 2.82\pm0.002d 

\subsection{Stellar properties} 
\label{sec:evolution}
We applied our automatic spectral classification tool (discussed in \citealt{donati2012}) to several of the highest S/N spectra for both stars. This tool is similar to that of \cite{valenti2005}, and fits the observed spectrum using multiple windows in the wavelength ranges of 515--520~nm and 600--620~nm, using Kurucz model atmospheres \citep{kurucz1993}. This process yields estimates of \teff and $\log{g}$, where the optimum parameters are those that minimize $\chi^{2}$, with errors bars determined from the curvature of the $\chi^{2}$ landscape at the derived minimum. For \pa, we find that $T_{\text{eff}} = 4600\pm50$~K and $\log{g} = 3.9\pm0.2$ (with \emph{g} in cgs units). Likewise, for \pb we find that $T_\text{eff} = 4650\pm50$~K and $\log{g} = 4.1\pm0.2$. 

Our derived \teff is a good estimate in the approximation of a homogeneous star, however, there are likely large ranges in temperature across the photospheres of both stars, given the high level of spot coverage (see Section~\ref{sec:tomography}). A two-temperature model would provide a better \teff estimate of the immaculate photosphere, as well as the level of spot coverage \citep[see][]{gullysantiago2017}, however, a homogeneous model is sufficient for our purposes.

Given the measured $V$ magnitude of 12.8 for both stars, reported by \cite{rebull2000}, and taking into account a spot coverage of the visible stellar hemisphere of $\sim20$~per~cent (as is typical for active stars, see Section~\ref{sec:tomography}), we derive unspotted $V$ magnitudes of $12.6\pm0.2$ for both stars. We note that assuming a different spot coverage (such as 0 or 30~per~cent) places our derived parameters within our quoted error bars. 

Combining our spectroscopic \teff with the intrinsic colour and \teff sequence for young stars found by \cite{pecaut2013}, we find $(V-I_\textsc{c})_{0}$ to equal 1.09 and 1.07 for \pa and \pb, respectively. Using the standard $V$ band extinction of ${A_{\textsc{v}} = 3.1E(B-V)}$, combined with the conversion of $E(V-I_\textsc{c}) = 1.25E(B-V)$ from \cite{bessell1988}, and the measured colours of $V-I_{\textsc{c}}$ (equal to 1.28 and 1.52), we determine $A_{\textsc{v}}$ to be $0.5\pm0.2$ and $1.1\pm0.2$, for \pa and \pb, respectively. Given that we directly measure the spectroscopic temperature, our extinction values are more accurate than those found using colours alone by \cite{rebull2000} (equal to 0.2 and 0.23).

Using the visual bolometric correction $BC_{\text{v}}$ expected at these temperatures (equal to $-0.57\pm0.04$ and $-0.53\pm0.03$, \citealt{pecaut2013}), combined with $A_{\textsc{v}}$, and the trigonometric parallax distance to the ONC of $388\pm5$~pc (corresponding to a distance modulus of $7.94\pm0.03$, \citealt{kounkel2017}), we obtain absolute bolometric magnitudes of $3.6\pm0.3$ and $3.0\pm0.3$, or equivalently, logarithmic luminosities relative to the Sun of $0.45\pm0.11$ and $0.72\pm0.11$, for \pa and \pb, respectively. When combined with the photospheric temperature obtained previously, we obtain radii of $2.7\pm0.2$ and $3.5\pm0.2$~\rsun. 

Coupling the rotation periods of 5.585~d and 2.8153~d for \pa and \pb, respectively (see Equation~\ref{eq:ephemeris}), with the measured line-of-sight projected equatorial rotation velocity \vsini of $13.7\pm0.1$ and $57.2\pm0.1$~\kms (as determined from our tomographic modelling in Section~\ref{sec:tomography}), we can infer that $R_{\star}\sin{i}$ is equal to $1.52\pm0.04$~\rsun and $3.17\pm0.02$~\rsun, where $R_{\star}$ and $i$ denote the stellar radius and the inclination of its rotation axis to the line of sight. By comparing the radii derived from the luminosities, to that from the stellar rotation, we derive that $i$ is equal to $35\degr$ and $64\degr$ (to an accuracy of $\simeq10\degr$) for \pa and \pb, respectively.

\begin{figure}
\includegraphics[width=\columnwidth]{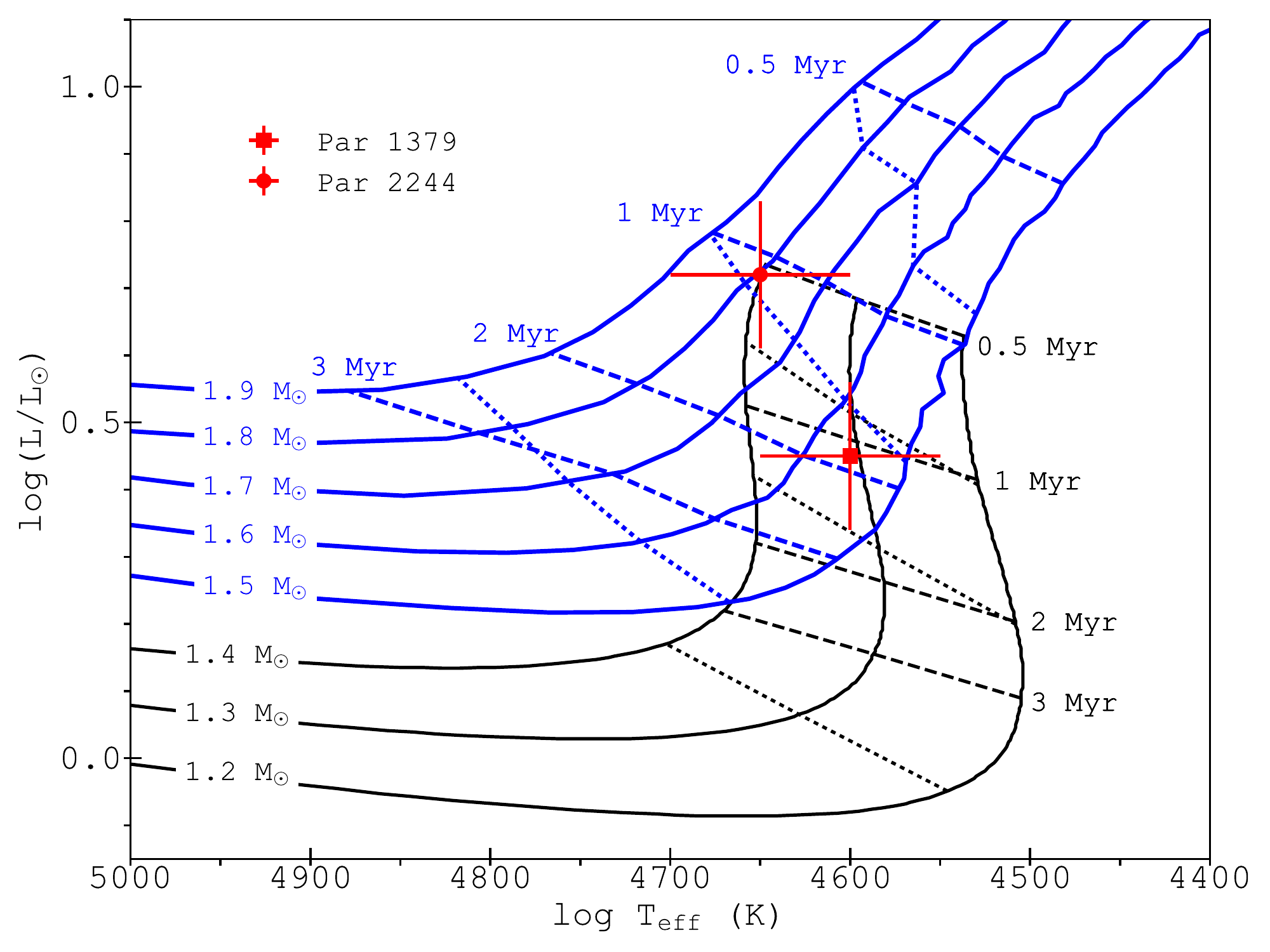}
\caption{H-R diagram showing the stellar evolutionary tracks provided by \citet[blue solid lines]{siess2000} for masses in the range 1.5--1.9~\msun, and those provided by \citet[black solid lines]{baraffe2015} for masses in the range 1.2--1.4~\msun. Dashed lines show the corresponding isochrones for ages 0.5, 1, 2 and 3~Myr (blue for \citealt{siess2000}, and black for \citealt{baraffe2015}). Dotted lines mark the 0, 30 and 50~per~cent fractional radius for the size of the radiative core (black, \citealt{baraffe2015}) and the bottom of the convective envelope (blue, \citealt{siess2000}.)}
\label{fig:hr}
\end{figure}

Using the evolutionary models of \cite{siess2000} (assuming solar metallicity and including convective overshooting), we find that the stellar masses are $1.6\pm0.1$~\msun and $1.8\pm0.1$~\msun, with ages of $1.8\pm0.6$~Myr and $1.1\pm0.3$~Myr, for \pa and \pb, respectively (see the H--R diagram in Figure~\ref{fig:hr}, with evolutionary tracks and corresponding isochrones). However, using the most recent evolutionary models of \cite{baraffe2015}, we obtain substantially different stellar masses of $1.3\pm0.1$~\msun and $1.4\pm0.1$~\msun, with ages of $1.0\pm0.5$~Myr and $0.5\pm0.5$~Myr, for \pa and \pb, respectively. Given the inherent limitations of these evolutionary models, we do not consider the formal error bars on the derived masses and ages to be representative of the true uncertainties, and thus both parameters may be considered to be similar for each star. However, we note that for internal consistency with previous MaPP and MaTYSSE results, the values from the \cite{siess2000} models should be referenced. We also note that the ages implied by both the \cite{siess2000} and \cite{baraffe2015} evolutionary models are somewhat lower than the mean age of the ONC of $2.2\pm2$~Myr, but are still within the spread of ages \citep{reggiani2011}.

As well as placing the stars at younger ages, with lower masses, the internal structure is also somewhat different between these two evolutionary models, as can be seen in Figure~\ref{fig:hr}. Here, the size of the convective envelope (by radius) for the \cite{siess2000} models is around 70~per~cent for both stars, whereas the \cite{baraffe2015} models suggest both stars are fully convective. The most critical change in large-scale magnetic field topology is expected to occur when the star evolves from one that is largely convective, to one that is largely radiative ($>50$~per~cent by radius, depending on the stellar evolution model; e.g. \citealt{gregory2012}). Given that the observed field topologies for both stars show non-axisymmetric, mid-to-weak strength poloidal fields (see Section~\ref{sec:magfield}), the models of \cite{siess2000} place the stars in better agreement with expectations. Furthermore, the weaker field strength and higher differential shear (see Section~\ref{sec:diffrot}) of \pa suggests that it is likely more structurally evolved than \pb, possibly to the point of being mostly radiative (despite the model predictions), whereas \pb is still in a largely convective state. 

A more complete study of the change in magnetic topology with internal structure will be performed for the wTTSs in the MATYSSE sample in a future publication, where we will also discuss the implications for different evolutionary models. 

\newpage

\subsection{SED fitting}
\label{sec:SED}
\subsubsection{Model and grid description}
Spectral Energy Distributions (SEDs) of \pa and \pb were constructed using photometric data collected by \cite{rebull2000}, \cite{megeath2012} and \cite{zacharias2015} as well as from the WISE, Spitzer and Gaia catalogues \citep{wise, spitzer, gaia2016}. Deep, sensitive sub-mm and mm photometry are not currently available for the sources. The SEDs reveal that \pa has an infrared excess starting at 2~$\mu$m, and that \pb has an infrared excess starting at 22~$\mu$m, indicating that both objects have dusty circumstellar discs (see Figure~\ref{fig:SED}).

To constrain the basic physical properties of the discs, we fit the SEDs using a passively irradiated dusty disc model. For the modelling, we used the stellar properties derived earlier (in Section~\ref{sec:evolution}), a parametric disc geometry, and to calculate the disc continuum emission, the Monte-Carlo dust-radiative transfer code \textsc{mcfost} \citep{pinte2006,pinte2009}. The parametric disc model has a flared geometry with a Gaussian density profile in the vertical direction $\rho(r,z)=\rho(r)\, {\rm exp}(-z^2/2h(r)^2)$. The dust surface density distribution and the scale height as a function of the radius are described with power-laws $\Sigma(r)=\Sigma_0(r/r_0)^\alpha$ and $h(r)=h_0(r/r_0)^\beta$ respectively, where $h_0$ is the scale height at the reference radius $r_0$. The value of $\Sigma_0$ is deduced from the assumed disc dust mass $M_{\rm dust}$. The free parameters for computing a model are therefore the dust size distribution and composition, the disc inner radius $R_{\rm in}$, the disc outer radius $R_{\rm out}$, the dust disc mass $M_{\rm dust}$, the aspect ratio $h_0/r_0$, the flaring angle $\beta$, and the
the surface density exponent $\alpha$.

We assumed astronomical silicate opacities for the dust \citep{draine1984} and tested models with two dust size distributions: a {\it small dust} distribution with 0.01--10~$\mu$m sized grains, and a {\it large dust} distribution with 0.01--2000~$\mu$m grains, assuming a size distribution slope of 3.5 for both distributions. We used a reference radius $r_0$ of 1~au and a flaring angle $\beta$ of 1.1. Furthermore, we assumed disc inclinations of $36\degr$ and $59\degr$ for \pa and \pb, respectively (from our tomographic modelling in Section~\ref{sec:tomography}), and a distance of 388~pc for both objects.

For each of the sources, we ran a uniform grid of SED models, varying the free parameters as described in Table~\ref{tab:grid}. The grid includes {\it disc-like} models extending tens to hundreds of au and narrow {\it ring-like} models with a width of a few au. In total we tested 3936 models for \pa, and 2430 models for \pb. For each SED model the $\chi^2$ statistic was computing using the logarithm of the flux in each photometric band. Each $\chi^2$ lead to a model probability defined as $P=\exp{\left(-\chi^{2} / 2\right)}$. The ensemble of probabilities $P$ of the grid were normalized such that the sum of $P$ in the entire grid is equal to 1. Armed with the probability for each model in the grid, we constructed histograms for each free parameter, by summing $P$ over all models in the grid that had the same value of the free parameter. In the case of $R_{\rm in}$ and $R_{\rm out}$, for calculating the histograms, the sum of $P$ in each histogram bin was divided by the number models in that bin, and the histograms were re-normalized such that their sum is equal to 1. This was necessary because models with $R_{\rm out}$ < $R_{\rm in}$ are not possible, and the histograms need to be corrected for the different number of models per bin. Additionally, in the case of $R_{\rm in}$ and $R_{\rm out}$, and $M_{\rm dust}$ and $R_{\rm out}$, we constructed 2-D $\chi^2$ plots to reveal cross correlations between these parameters.

\subsubsection{SED fitting results}
\begin{figure}
\includegraphics[width=0.24\textwidth]{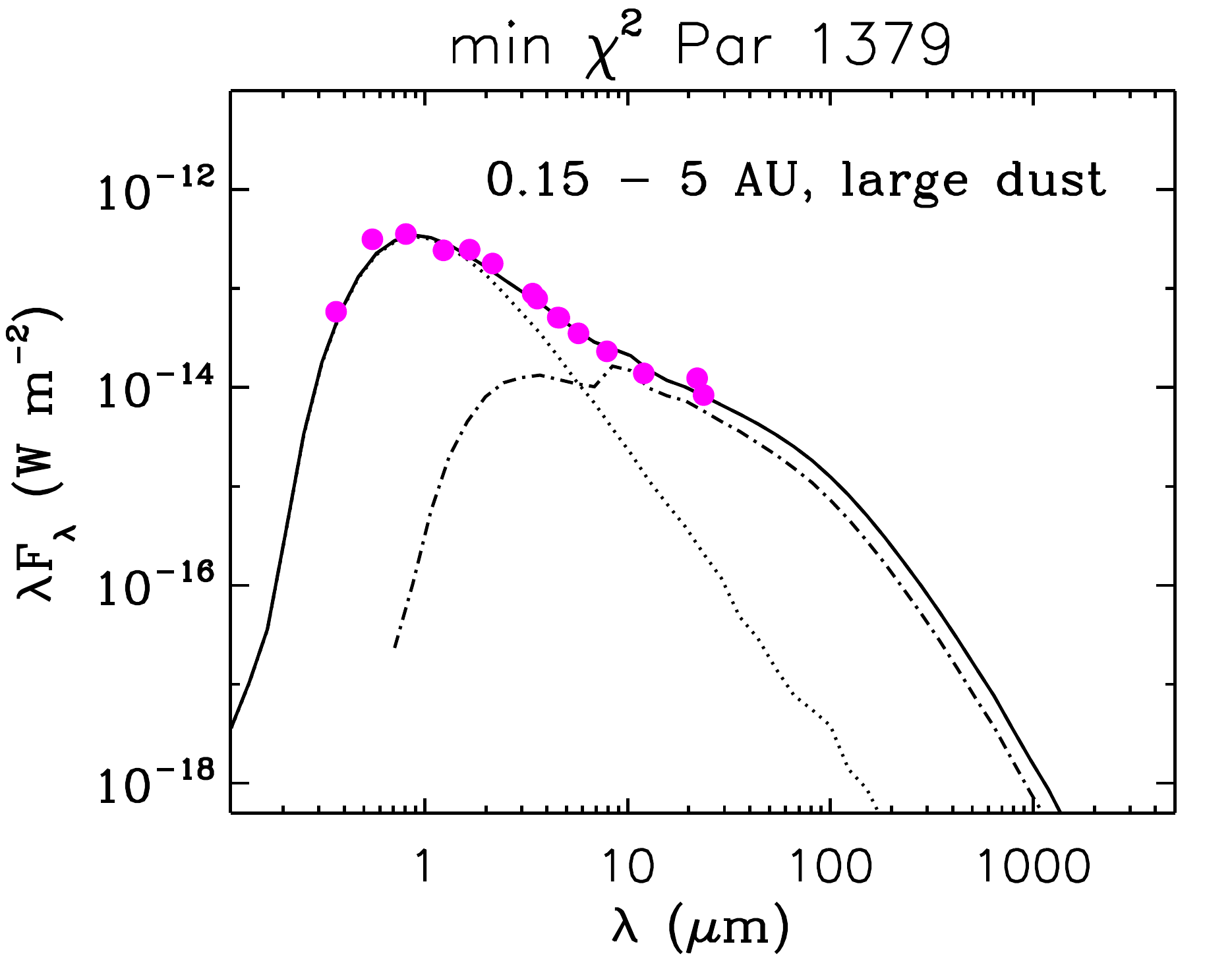}
\includegraphics[width=0.24\textwidth]{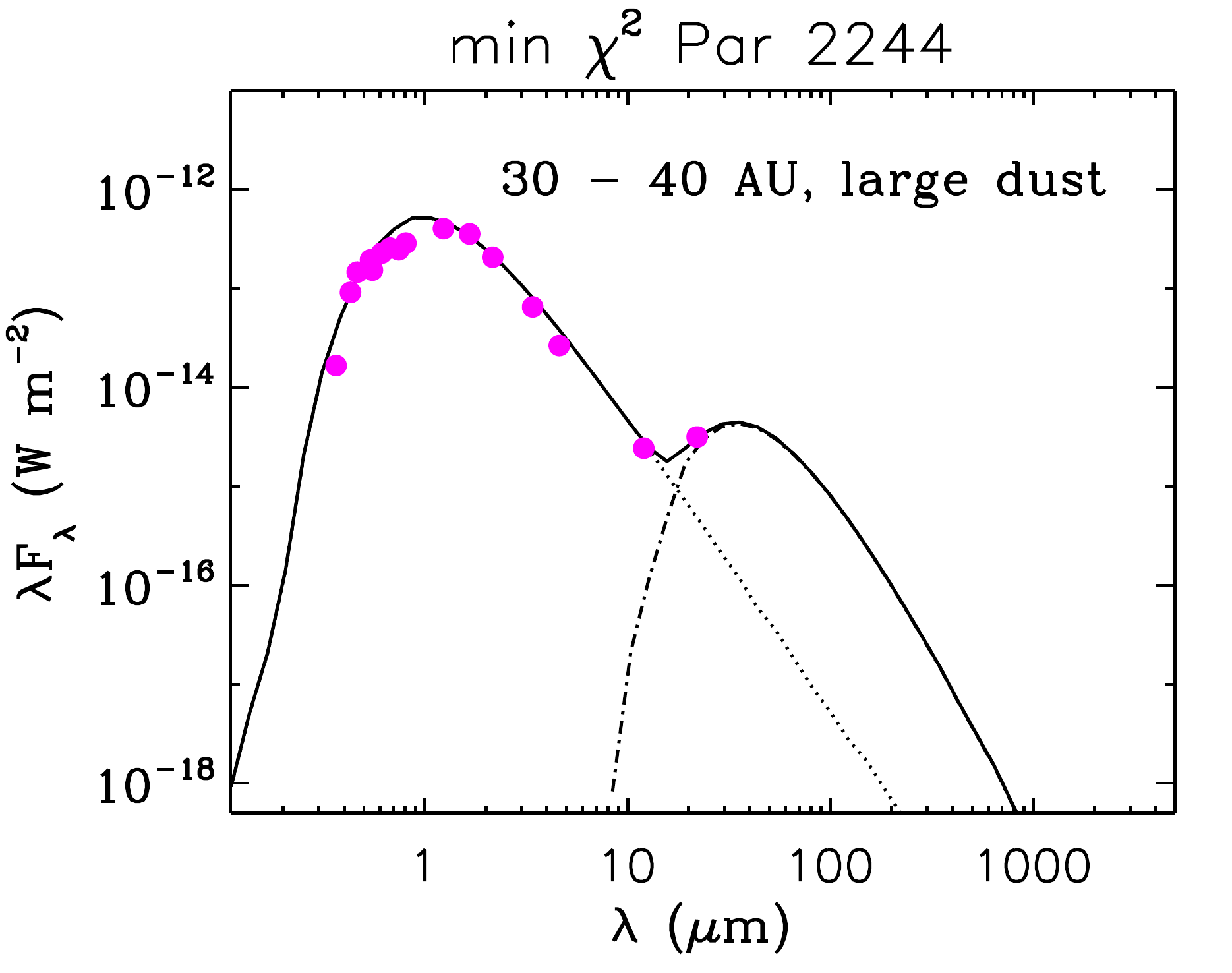} \\
\includegraphics[width=0.24\textwidth]{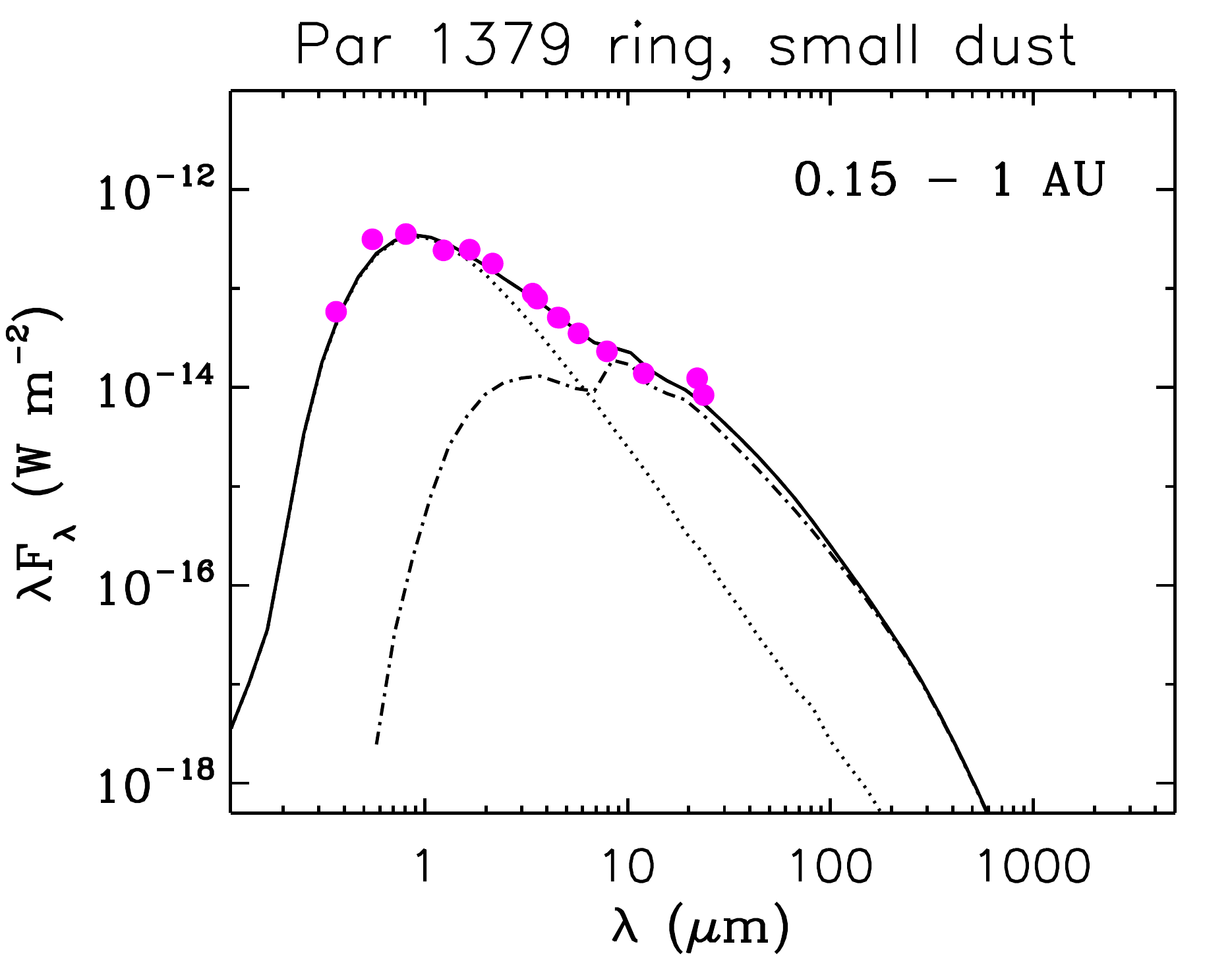}
\includegraphics[width=0.24\textwidth]{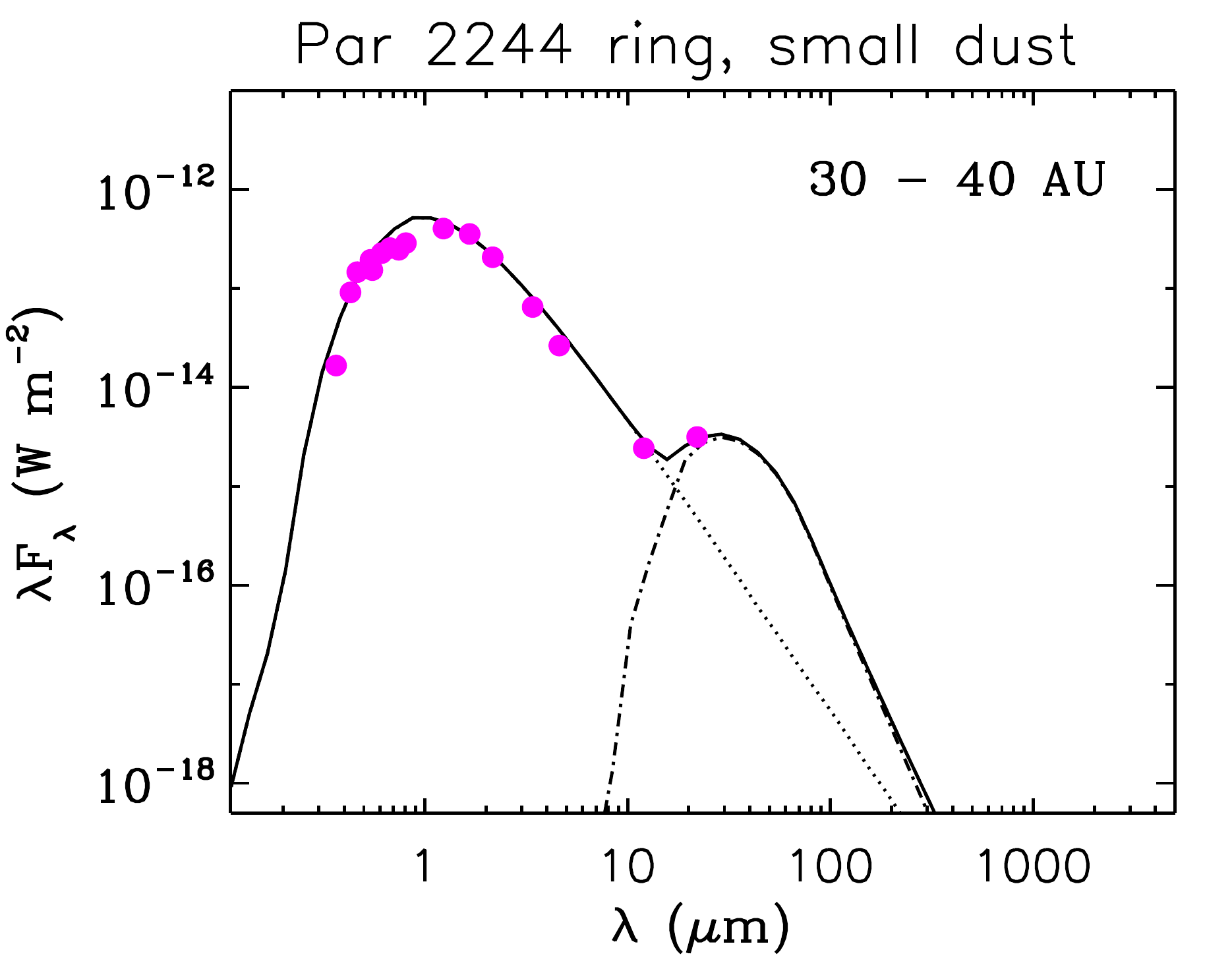} \\
\includegraphics[width=0.24\textwidth]{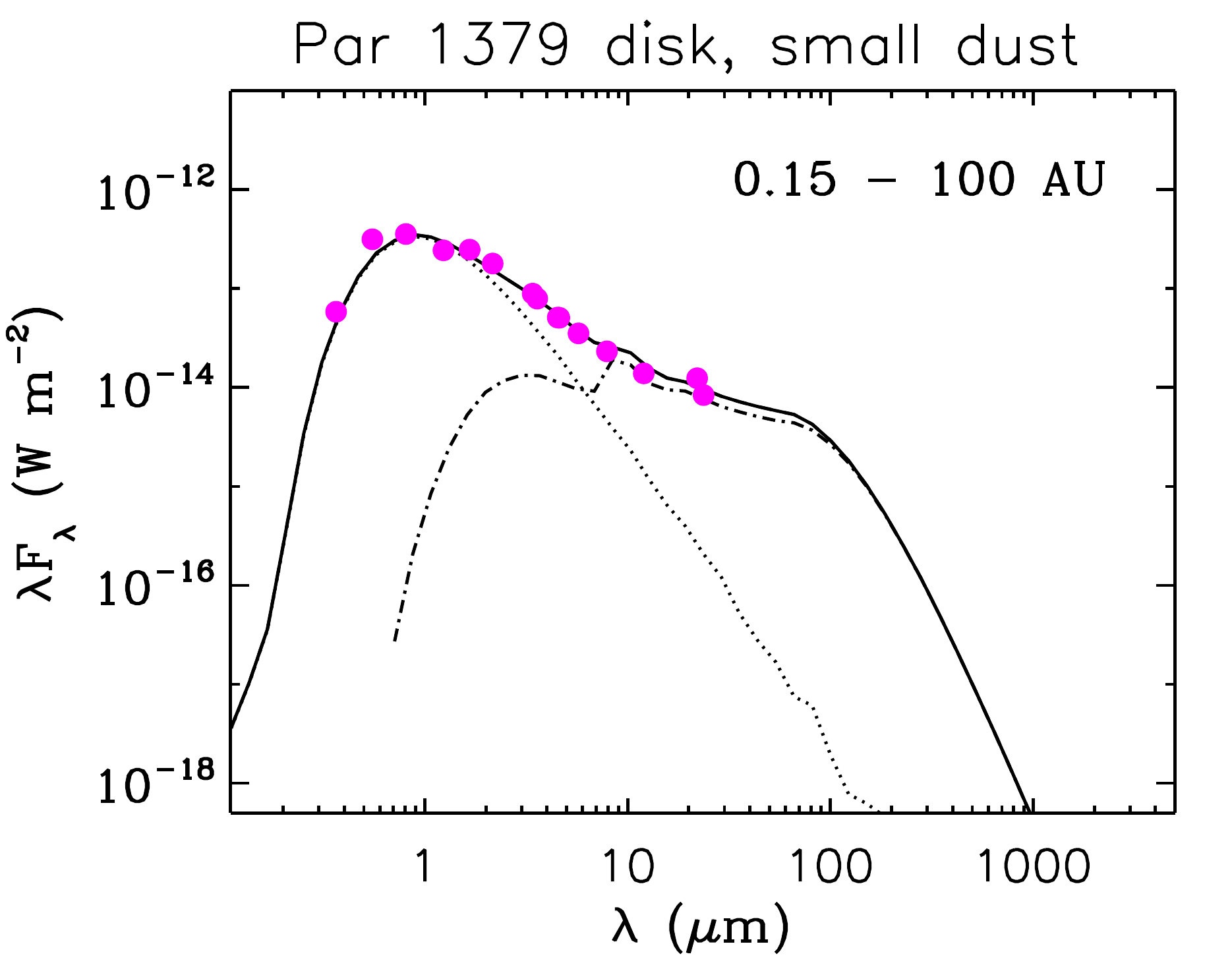}
\includegraphics[width=0.24\textwidth]{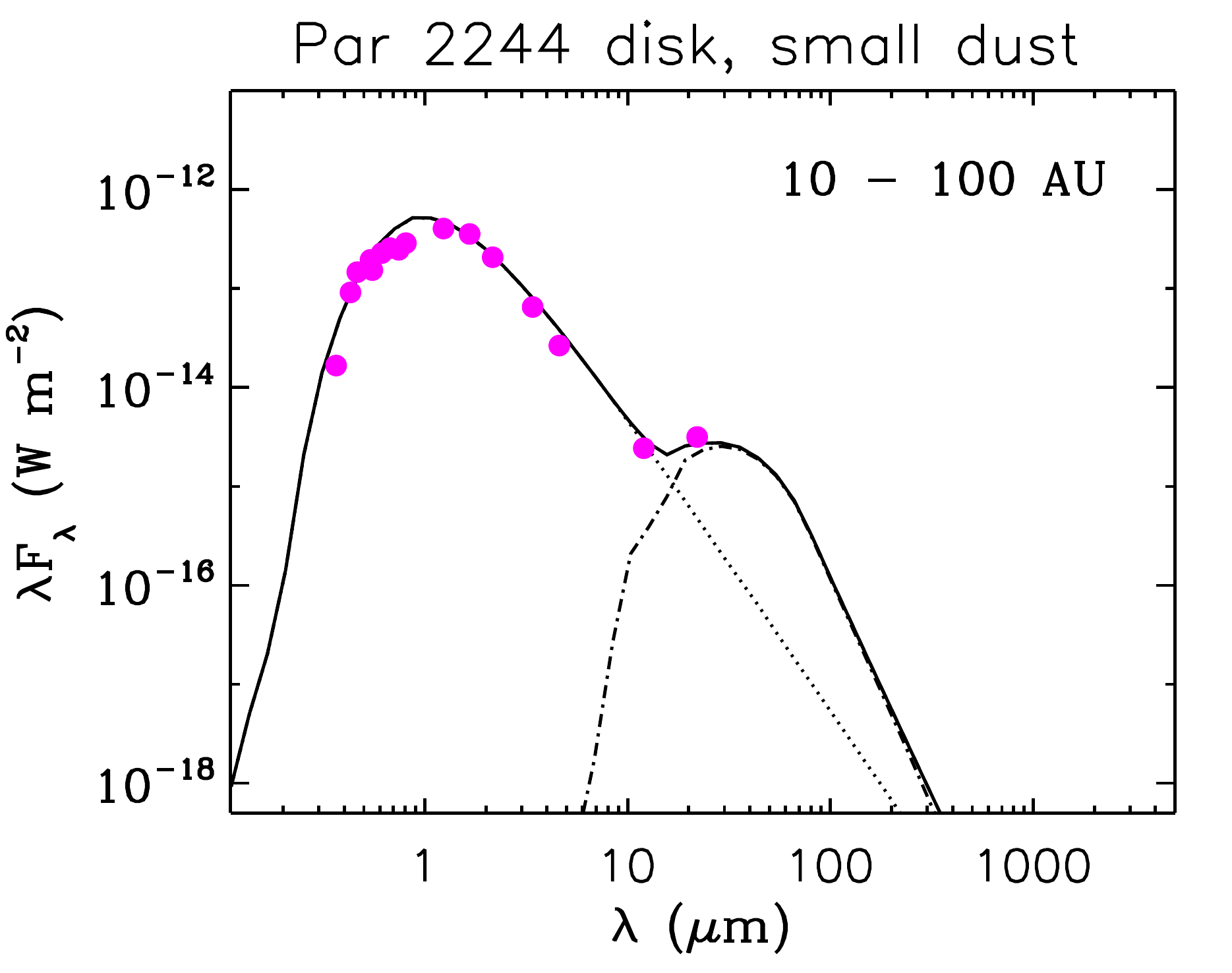} \\
\includegraphics[width=0.24\textwidth]{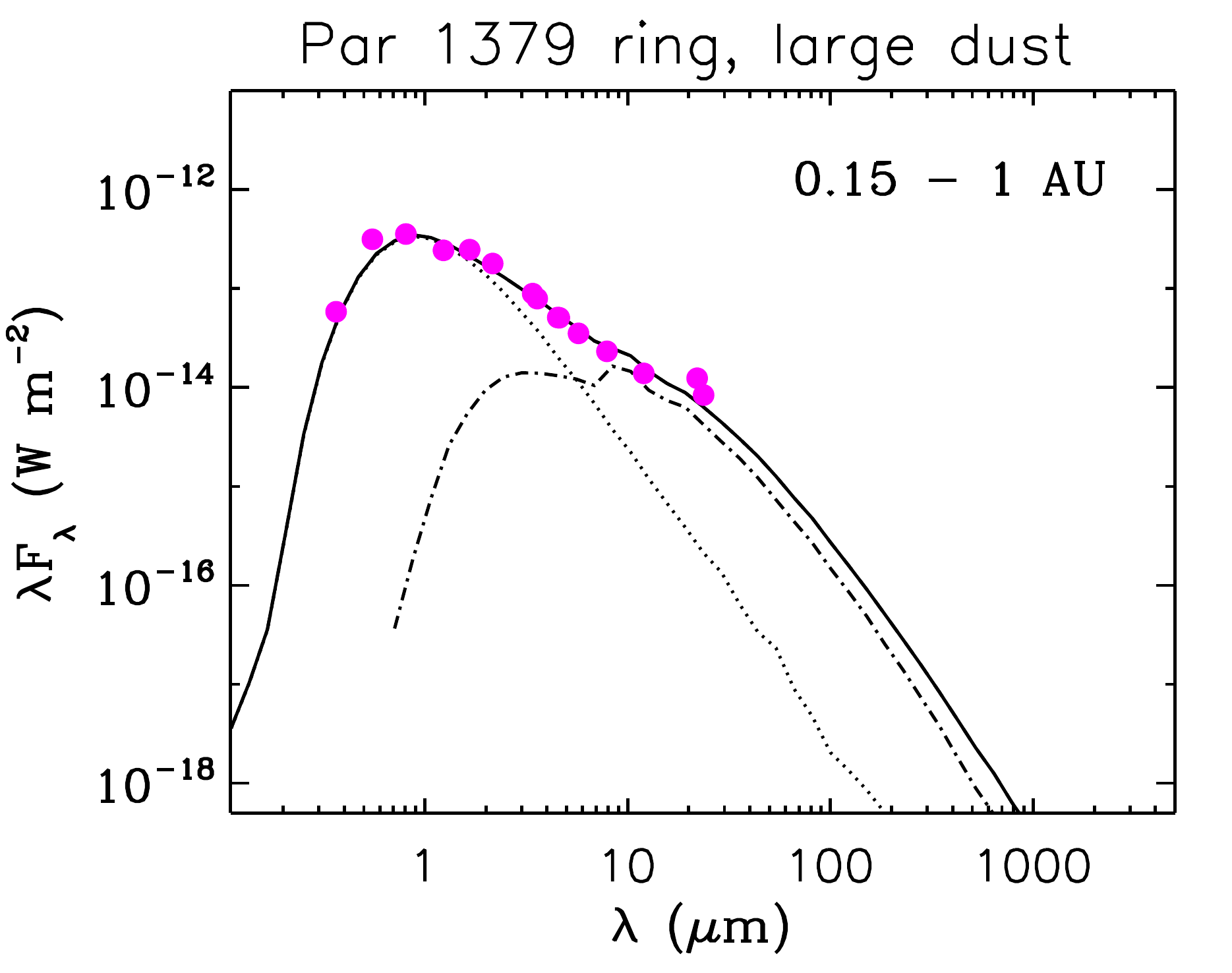}
\includegraphics[width=0.24\textwidth]{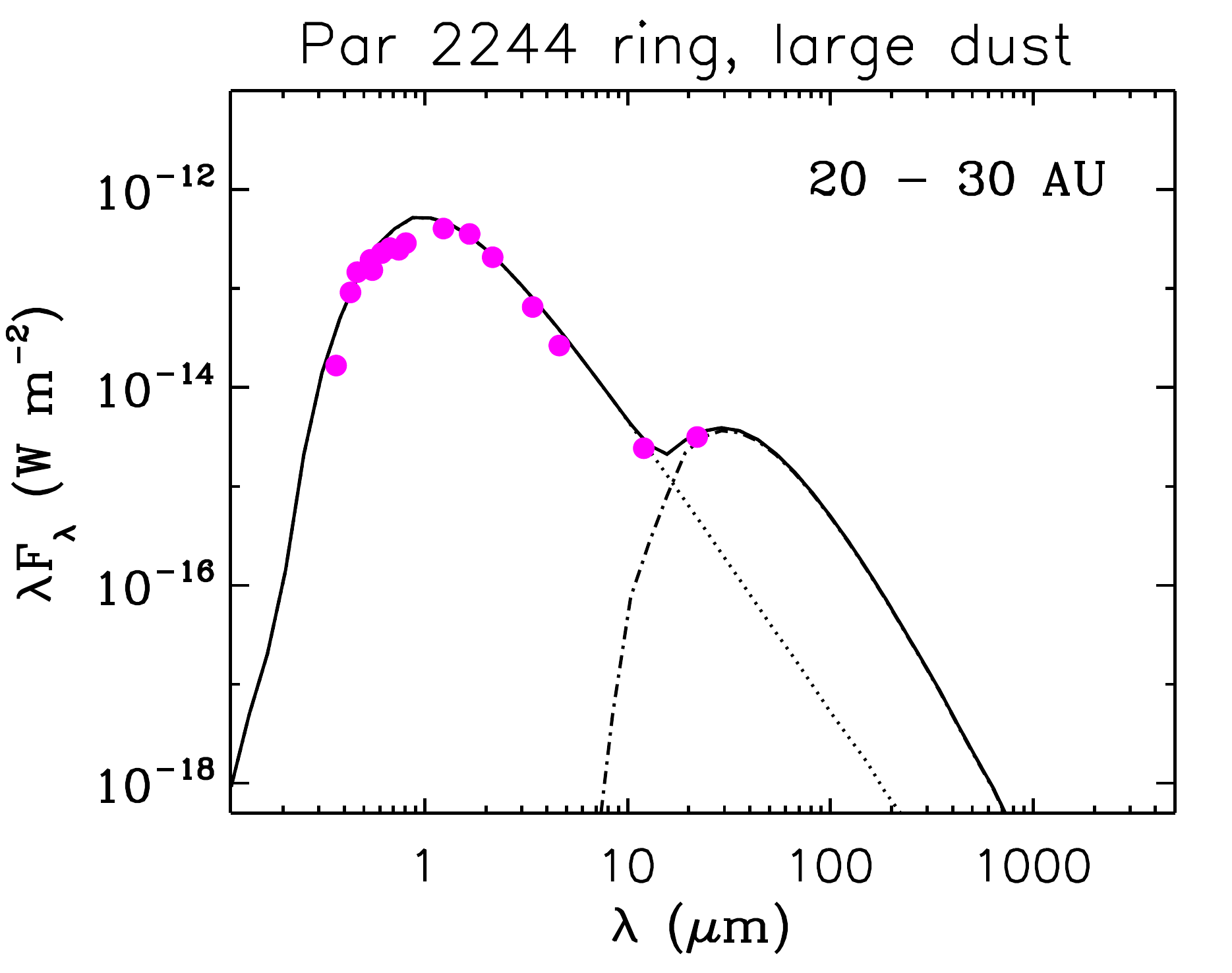} \\
\includegraphics[width=0.24\textwidth]{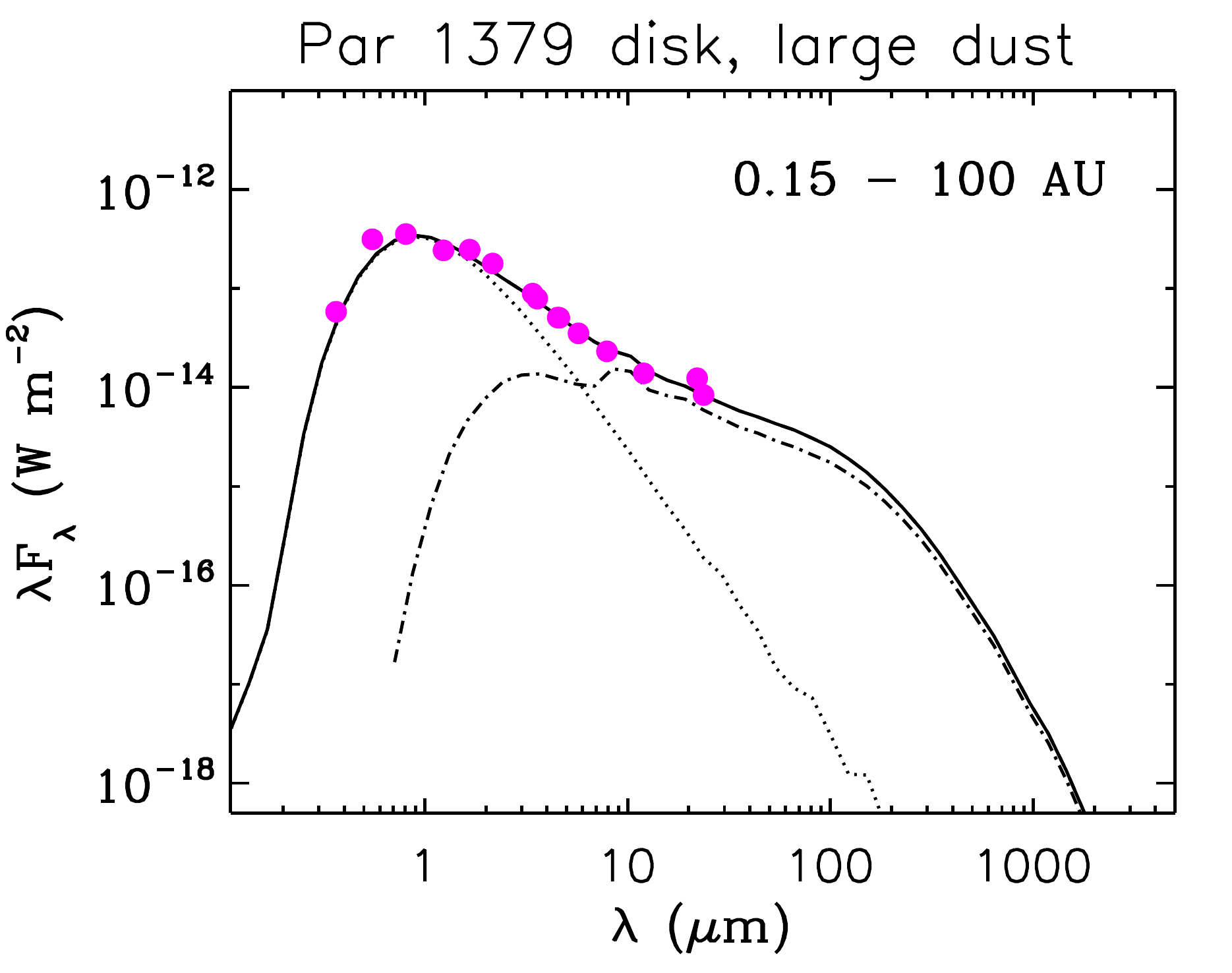}
\includegraphics[width=0.24\textwidth]{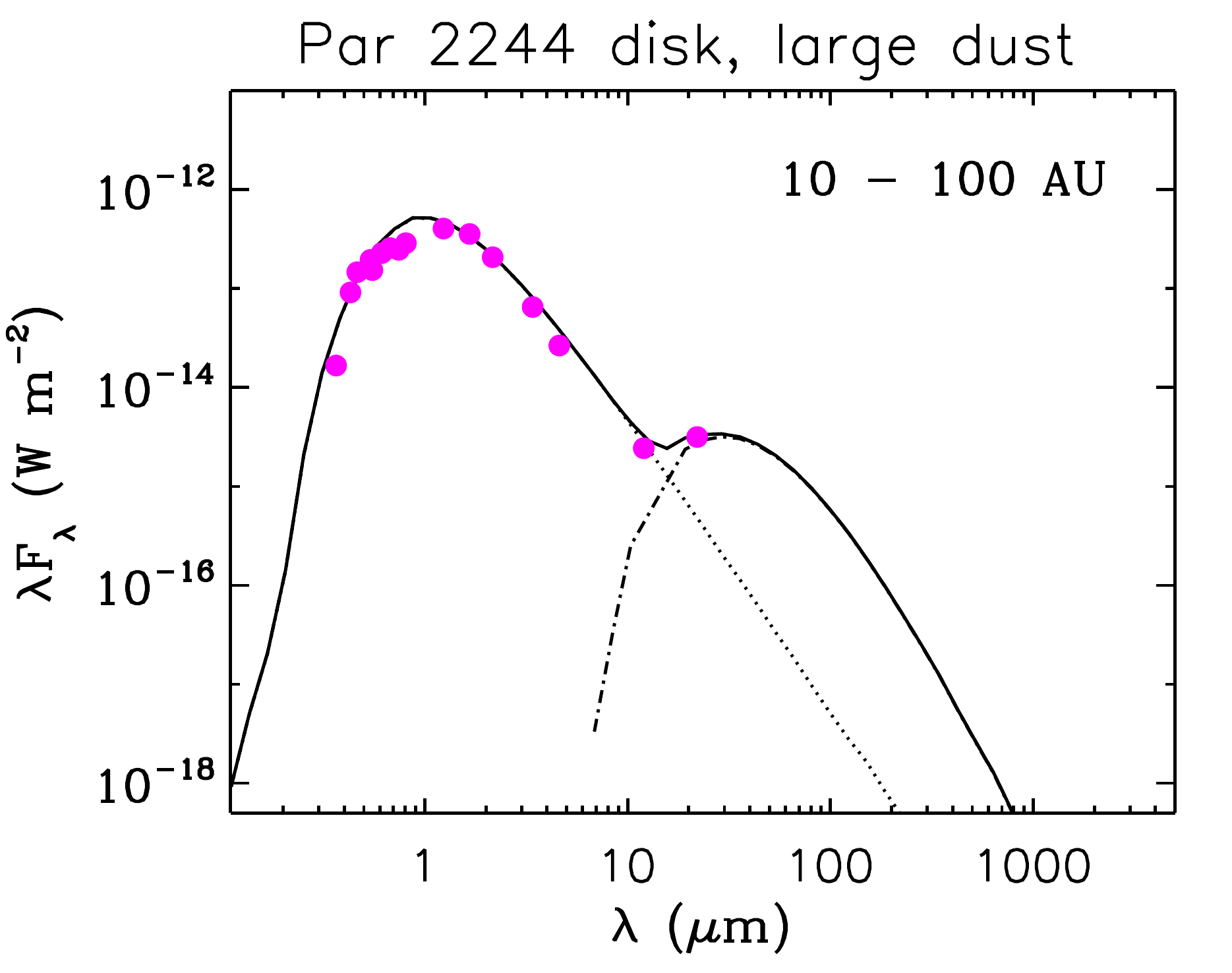} \\
\caption{Spectral Energy Distributions of \pa (left panels) and \pb (right panels). Photometric points are shown as dots, with the emission from the star marked as a dotted line, the thermal emission from the dust grains inside the disc marked as a dashed-dotted line, and the total marked with a solid line. The models of best fit are shown in the top row, with examples of data-compatible ring models (second and fourth rows) and disc models (third and fifth rows) for both \textit{small dust} and \textit{large dust}.}
\label{fig:SED}
\end{figure}

Our results are summarized in Figures~\ref{fig:SED} and \ref{fig:Histograms}, and Table~\ref{tab:SED}. Ring and disc solutions describe the SED equally well, highlighting that the solutions are degenerate. Indeed, to better distinguish between the models, one would require sensitive sub-mm and mm photometry as well as spatially resolved imaging. However, despite the degeneracies, we may draw general conclusions given that the near-IR SED enable us to constrain the inner radius of the IR-excess emission, as well as the disc's scale height (if the emission is optically thick). Furthermore, the SED provides us with an order-of-magnitude estimate of the dust mass, given a disc's geometry and extension.

In the case of \pa, the near-IR excess requires the presence of warm dust at R$<$ 1 au. The probability histogram (Figure~\ref{fig:Histograms}) clearly displays a higher probability for models with inner radius 0.15 and 0.2 au. The outer disc is less well constrained, but there is a preference for models with an outer disc radius smaller than 10 au. The solution with the minimum $\chi^2$ has $R_{\rm in} =  0.15$~au, and $R_{\rm out}= 5$~au. However, one should bare in mind that the preference for discs with an outer radius smaller than 10 au in the grid could be a bias due to the absence of sensitive sub-mm photometry. The SED could in fact be explained either by a narrow, low-mass ($10^{-7}$~\msun) ring of warm dust between 0.15--1~au, a disc with mass $10^{-6}$~\msun extending from 0.15 to 10 au, or a large 100~au disc with a dust mass of $10^{-5}$~\msun (note that in Figure~\ref{fig:Histograms}, for $R_{\rm in} = 0.15$~au, the probability of models with $R_{\rm out}>2$~au is similar). Large and dust grains distributions appear to have similar probabilities, and the dust surface density exponent is not well constrained, as all the values in the grid display a similar probability. However, the model with the minimum $\chi^2$ has a dust distribution with large grains, and a dust mass of $10^{-6}$~\msun with $\alpha = -1.0$.

Compared with the dust mass measured in the discs of cTTSs in Taurus \citep[$10^{-6}$--$10^{-4}$~\msun,][]{Andrews2005}, \pa has a lower dust mass than a typical cTTS, if the disc of \pa is smaller than 10 au. However, if the disc of \pa is large, then its dust mass would be comparable to those of standard cTTS discs. The scale height $h/r$ required to fit the SED of \pa is between 0.02--0.03 at 1 au - a value similar to the scale height of 0.03 of an isothermal disc in hydrostatic equilibrium around a 1.6~\msun star at 1 au, assuming a mid-plane temperature of 350~K$^($\footnote{$h/r=c_s/v_K= \sqrt{\frac{k_B T}{\mu m_H} \frac{r}{G M_\star}}$, where $c_s$ is the sound speed, $v_K$ the Keplerian velocity, $k_B$ the Boltzmann constant, $\mu$ the mean molecular weight (2.3), $m_H$ the proton mass, $r$ the radius, and $G$ the gravitational constant.}$^)$. The scale height suggested by the SED fit indicates that some gas might be still present in the disc. In summary, the SED of \pa suggests that it is surrounded by a dusty disc extending down to the silicate sublimation radius, and that this disc is likely primordial as its scale height is consistent with a disc in hydrostatic equilibrium.

In the case of \pb, the SED fit reveals that there is an inner dust cavity of at least 10 au in radius, with the models that best describe the data possessing an inner radii of 20 au. Dusty rings around 20--30 au with a width of 10--20 au have higher probabilities in the grid, however, a dust disc extending 10--100 au can also satisfactorily describe the data. The model with the lowest $\chi^2$ is a dusty ring extending from 30--40 au, with a dust mass of $10^{-7}$ ~\msun and large dust grains. However, a similar model with a small dust distribution and a mass of $5\times10^{-9}$ ~\msun describes the data equally well. The lack of sensitive sub-mm photometry limits our ability to distinguish between these two models, however, in the whole ensemble of the grid there is the statistical trend that the models that best reproduce the SED have dust masses of $10^{-9}$--$10^{-8}$~\msun, which are very low masses, when compared to the mass of cTTS in Taurus. The low dust-mass suggested by the SED fit (solutions are optically thin) together with the lack of strong accretion signatures (see Section~\ref{sec:accretion}), hints that \pb has either recently dissipated its gaseous disc and we only observe the remaining primordial dust, or alternatively, that \pb harbours a young debris disc. We note, however, that the grid slightly favours large grain solutions, lending weight to the hypothesis of remaining primordial dust. As the solutions are optically thin, stringent constraints on the scale height and surface density exponent can not be derived. In summary, the SED suggests that \pb is surrounded by a low mass $10^{-9}$--$10^{-8}$~\msun dusty disc, extending at least to 50 AU with a 10 -- 20 au cavity. \pb could be considered as a non-accreting transition disc, or a young-debris disc.

\begin{table}
\begin{center}
\caption{Free parameters explored in the grid of SEDs calculated for \pa (top) and \pb (bottom).}
\label{tab:grid}
\setlength\tabcolsep{8pt} % default value: 6pt
\begin{tabular}{ll}
\multicolumn{2}{c}{\textbf{\pa}} \\
\hline
dust size ($\mu$m)& 0.01$-$10 (small), 0.01$-$2000 (large) \\
$R_{\rm in}$ (au)	& 0.15,	0.2, 	0.5,	1,	5,	10,	20\\
$R_{\rm out}$ (au) 	& 0.5,	1,	5,	10,	20,	30,	50,	100\\
$M_{\rm dust}$ 	&  $10^{-7}, 	10^{-6},	5\times10^{-6},10^{-5}$\\
$h_0/r_0$ 		& 0.01,	0.02,	0.03,	0.05\\
$\alpha$			& -1.2,	-1.0,	-0.6\\
$\beta$			& 1.1 \\
{\bf models}		& {\bf 3936} \\
\hline
\\
\end{tabular}
\\
\begin{tabular}{ll}
\multicolumn{2}{c}{\textbf{\pb}} \\
\hline
dust size ($\mu$m)& 0.01$-$10 (small), 0.01$-$2000 (large) \\
$R_{\rm in}$ (au)	& 1,	5,	10,	20,	30,	40\\
$R_{\rm out}$ (au) 	& 5,		10,	20,	30,	40,	50,	100\\
$M_{\rm dust}$ & $10^{-9}, 5\times10^{-9}, 10^{-8}, 5\times10^{-8}, 10^{-7}$\\
$h_0/r_0$ 	& 0.01,	0.03,	0.05\\
$\alpha$ &  -1.2,	-0.6,	0.0\\
$\beta$ & 1.1 \\
{\bf models}		& {\bf 2430} \\
\hline
\\
\end{tabular} \\

\end{center}
\textbf{Note:} $\alpha$ corresponds to the exponent of the surface density power law, and $\beta$ to the exponent of the disc flaring, and $r_{0} = 1$~au.
\end{table}

\begin{table}
\begin{center}
\caption{Examples of models describing the SEDs of \pa (top) and \pb (bottom).}
\label{tab:SED}
\setlength\tabcolsep{8pt} % default value: 6pt
\begin{tabular}{lcccccc}
\multicolumn{6}{c}{\textbf{\pa}} \\
\hline											
Model 					& {\it min $\chi^2$} & {\it Ring} & {\it Disc} & {\it Ring} &  {\it Disc}\\[1mm]
Dust size 					& large & \multicolumn{2}{c}{small} & \multicolumn{2}{c}{large} \\	
\hline											
$R_{\rm in}$ (au)			&	0.15	& 0.15	&	0.15	&	0.15 &	0.15	\\
$R_{\rm out}$ (au) 			&	5	& 1		&	100	&	1	&	100	\\
$h_0/r_0$ 				&	0.03	& 0.03	&	0.03	&	0.03	&	0.03	\\
$M_{\rm dust}$ ($10^{-6}$~\msun)&1 	& 0.1		&	5	&	1.0	&	10	\\
$\alpha$					&	-1.0	& -0.6	&	-1.2	&	-1.2	&	-1.2			\\
$\beta$					&	1.1	& 1.1		&	1.1	&	1.1	&	1.1	\\
\hline
\\[1mm]										
\end{tabular}
\begin{tabular}{lccccccc}
\multicolumn{6}{c}{\textbf{\pb}} \\
\hline											
Model 					& {\it min $\chi^2$} & {\it Ring} & {\it Disc} & {\it Ring} &  {\it Disc}\\[1mm]
Dust size 					&  	large			& \multicolumn{2}{c}{small} & \multicolumn{2}{c}{large} \\
\hline											
$R_{\rm in}$ (au)			&	30	&	30	&	10	&	20	&	10	\\
$R_{\rm out}$ (au) 			&	40	&	40	&	100	&	30	&	100	\\
$h_0/r_0$ 				&	0.03	&	0.03	&	0.01	&	0.01	&	0.03	\\
$M_{\rm dust}$ ($10^{-8}$~\msun)&	10	&	0.5	&	1	&	5	&	10	\\
$\alpha$					&	-1.2	&	-0.6	&	0.0	&	-1.2	&	-0.6	\\
$\beta$					&	1.1	&	1.1	&	1.1	&	1.1	&	1.1	\\
\hline											
\end{tabular} \\
\end{center}
\textbf{Note:} $\alpha$ corresponds to the exponent of the surface density power law, and $\beta$ to the exponent of the disc flaring, and $r_{0} = 1$~au.
\end{table}

Given that we detect discs around both our target stars, we may estimate the age at which disc locking ceased by assuming a disc-locking period of 6--12~d \citep[for stars with $M > 0.3$~\msun, see][]{landin2016} with constant angular momentum evolution since then. Given these assumptions, we find ages around 0.7--1.50~Myr and 0.3--0.5~Myr for \pa and \pb, respectively (using \citealt{siess2000} evolutionary tracks). These ages are consistent with typical disc dissipation timescales of $\simeq1$~Myr for the ONC (ranging between 0.2--3~Myr, see \citealt{landin2016}), with \pb in particular dissipating its disc at a relatively young age. For the \cite{baraffe2015} evolutionary models, we are only able to place upper limits of 0.5~Myr on the age at which disc locking ceased, for both stars.

\subsection{Accretion status}
\label{sec:accretion}
The presence of dusty discs in \pa and \pb raises the question of whether the discs still have gas, and whether the stars are still accreting. Given that we have obtained multiple high-quality spectra of both targets, we may determine their accretion status using several metrics.

We find that \pa and \pb both show core \ion{Ca}{ii} infrared triplet (IRT) emission, with mean equivalent widths (EWs) of around 0.44~\r{A} (15.5~\kms) and 0.20~\r{A} (7.3~\kms), similar to what is expected from chromospheric emission for such PMS stars, and somewhat lower than that for accreting cTTSs \citep[e.g.][]{donati2007}. We also find that both \pa and \pb show H$\alpha$, H$\beta$ and \ion{He}{i}~$D_{3}$ emission, with \pa in particular displaying a time-varying absorption component in both H$\alpha$, H$\beta$.

One may estimate the level of surface accretion in TTSs by adopting the relations between line luminosity $L_{\text{line}}$ and the accretion luminosity $L_{\mathrm{acc}}$ of \cite{alcala2017}. For this purpose we determined $L_{\text{line}}$ by assuming blackbody scaling using the stellar radius $R_{\star}$ and \teff given in Table~\ref{tab:systempars}. Then, following \cite{gullbring1998}, the mass accretion rate ${\dot{M}_{\mathrm{acc}}}$ was calculated using the relationship
\begin{equation}
\dot{M}_{\mathrm{acc}} = \frac{L_{\mathrm{acc}} R_{\star}}{GM_{\star}(1-\frac{R_{\star}}{R_{\text{in}}})}
\end{equation}

\noindent where $R_{\text{in}}$ denotes the truncation radius of the disc, and is taken to be $5R_{\star}$ \citep{gullbring1998}. 

For \pa, we detect weak \ion{He}{i}~$D_{3}$ emission, with an EW of around 0.012~\r{A} (0.6~\kms), corresponding to ${\log{\dot{M}_{\mathrm{acc}}} \simeq -10.7}$~\msunyr. For H$\beta$, we detect absorption components in several line profiles between phases 0--0.4 (see Figure~\ref{fig:halpha}), and we find the EW of the H$\beta$ emission to range between 0.01--0.34~\r{A} (average of 0.16~\r{A}, equivalent to 9.8~\kms, corresponding to ${\log{\dot{M}_{\mathrm{acc}}} \simeq -10.6}$~\msunyr). Moreover, we detect a redshifted absorption component (in addition to the emission) in several H$\alpha$ line profiles between phases 0.0--0.4 (see Figure~\ref{fig:halpha}). We find the EW of the emission ranges between 1.4--3.6~\r{A} (average of 2.4~\r{A}, equivalent to 108~\kms, corresponding to ${\log{\dot{M}_{\mathrm{acc}}} \simeq  -9.9}$~\msunyr), and that of the absorption ranges between 0.1--1.4~\r{A} (average of 0.5~\r{A}, equivalent to 24~\kms). In the case of \pb, we detect very weak \ion{He}{i}~$D_{3}$ emission, with an EW of around 0.005~\r{A} (0.25~\kms), corresponding to $\log{\dot{M}_{\mathrm{acc}}}$ of around -11.1~\msunyr. Furthermore, we find weak H$\beta$ emission, with an average EW of 0.06\r{A} (3.7~\kms, corresponding to ${\log{\dot{M}_{\mathrm{acc}}} \simeq -10.9}$~\msunyr). Moreover, we do not find any absorption in the H$\alpha$ profiles of \pb, with the mean emission EW of 1.3\r{A} (60~\kms, corresponding to ${\log{\dot{M}_{\mathrm{acc}}} \simeq -10.1}$~\msunyr, see Figure~\ref{fig:halpha}).

At such low EWs and accretion rates, chromospheric activity becomes a significant influence on the strength and width of emission lines \citep{ingleby2011}. For TTSs in particular, their large convective turnover times \citep{gilliland1986} combined with their rapid rotation means they possess a low Rossby number, placing them well within the saturated activity regime (e.g., \citealt{reiners2014}). For young stars, the H$\alpha$ line luminosity is observed to saturate at around $\log[L(H\alpha)/L_{\text{bol}}] = -3.3$ or lower (around $-3.8$ for K4 spectral types, \citealt{barrado2003}; around $-4.0$ for later spectral types, see \citealt{newton2017}). As our target stars both show line luminosities below these levels, with $\log[L(H\alpha)/L_{\text{bol}}]$ equal to $-4.5$ for \pa and $-5.0$ for \pb, any measurements of accretion rates based on the line luminosities (especially H$\alpha$, and to a lesser extent H$\beta$) must be considered to be significantly influenced or even dominated by chromospheric activity.

The limit at which one can distinguish between accretion and chromospheric activity has been explored by several authors. \cite{barrado2003} propose a spectral-type dependant relationship between the EW(H$\alpha$) and the accretion rate. Their empirical criterion is based on the saturation limit of chromospheric activity (taken to be $\log[L(H\alpha)/L_{\text{bol}}] = -3.3$) and is used to distinguish between accreting and non-accreting stars and brown dwarfs. Adopting the spectral type of K4 for both \pa and \pb, as determined from our spectroscopic temperature in Section~\ref{sec:evolution}, the criterion of \cite{barrado2003} define a cTTS threshold of EW(H$\alpha) = 4.4$~\AA. Given that the EW(H$\alpha$) of \pa and \pb are both below these limits (with maxima of 3.6 and 2.1~\AA\ for \pa and \pb, respectively), both stars fall into the non-accreting regime (where line broadening is dominated by chromospheric activity).

Elsewhere, the impact of chromospheric activity on accretion rate measurements was explored by \cite{manara2013}, where the authors determined the point at which the line emission may be dominated by the contribution of chromospheric activity (termed chromospheric accretion `noise'). Using the empirical relationships from \cite{manara2013}, we have determined that the level of chromospheric noise for \pa and \pb is equal to $\log{(L_{\text{acc,noise}}/L_{\text{star}})} = -1.9\pm0.1$. For \pa, the measured line luminosities $\log{(L_{\text{acc}}/L_{\text{star}}})$ for H$\alpha$, H$\beta$ and \ion{He}{i}~$D_{3}$ are respectively equal to $-2.3\pm0.3$, $-2.9\pm0.3$ and $-3.1\pm0.3$. For \pb, we find that $\log{(L_{\text{acc}}/L_{\text{star}})}$ for H$\alpha$, H$\beta$ and \ion{He}{i}~$D_{3}$ are respectively equal to $-1.9\pm0.3$, $-2.0\pm0.3$, and $-3.3\pm0.3$. Comparing these values, we see that the luminosity of all three emission lines (for both stars) are significantly below the threshold where the line is dominated by chromospheric emission (apart from H$\alpha$ and H$\beta$ for \pb, which are at a similar level to the chromospheric noise). Thus, the accretion rates determined above for \pa and \pb must be taken to be upper limits, given that chromospheric emission is likely the dominant broadening mechanism. Indeed, \cite{manara2013} derive a limit for the detectable accretion rate of $\dot{M}_{\text{acc}} \sim 3\times10^{-10}$~\msunyr for a 1.1~\msun mass star at 3~Myr old (the closest mass and age available for our targets). Moreover, as higher mass stars have a higher detection limit, the fact that our derived $\dot{M}_{\text{acc}}$ are lower than this limit indicates that our target stars are likely not accreting, or are accreting at a low (undetectable) level. Thus, as the accretion rates are likely very low, we refer to our target stars as wTTSs.

Given that the H$\alpha$ emission in our targets is likely dominated by chromospheric emission, it is somewhat unreliable as a classification and accretion diagnostic. Nevertheless, \cite{white2003} have proposed using the width of H$\alpha$ emission at 10~per~cent intensity to distinguish between cTTSs and wTTSs, with stars possessing a width $>270$~\kms classed as cTTSs (also see e.g., \citealt{fang2009}). In the case of \pa, we find that the H$\alpha$ width at 10~per~cent intensity ranges between 163--313~\kms, with an average of 226~\kms, thus placing it well within the wTTS regime (H$\alpha$ line profiles are shown in Figure~\ref{fig:halpha}). Indeed, only three of the ten spectra show a width above 270~\kms (at cycles 1.610, 1.783 and 1.974 in Figure~\ref{fig:halpha}). Compared to the average spectrum, these three line profiles are additionally broadened by a blueshifted emission component. This broadening may be related to accretion, however the blueshifted emission may also be explained by prominences rotating into view (see discussion in Section~\ref{sec:discussion}), given the reconstructed magnetic field topology (see Section~\ref{sec:magfield}). For \pb, we find that the H$\alpha$ width at 10~per~cent intensity ranges between 187--430~\kms, with an average of 318~\kms (see Figure~\ref{fig:halpha}), thus placing it within the cTTS regime. However, its high \vsini of 57.2~\kms (see Section~\ref{sec:tomography}) additionally broadens the H$\alpha$ line, and if this and the significant chromospheric emission is accounted for, \pb likely falls below the threshold for accretion.

As well as classing our targets as cTTSs or wTTSs, one may also estimate the mass accretion rate using the H$\alpha$ width at 10~per~cent intensity and the relationship found by \cite{natta2004}. Here, the accretion rate $\log{\dot{M}_{\text{acc}}}$ for \pa ranges between -9.9 and -11.3~\msunyr, with an average of -10.7~\msunyr. For \pb, the accretion rate ranges between -8.7 and -11.1~\msunyr, with an average of -9.8~\msunyr. However, the authors note that, due to the large dispersion, accretion rates derived from this relationship are necessarily inaccurate for individual objects, and so should be used with care. Indeed, given that H$\alpha$ is likely dominated by chromospheric emission (see discussion above), this method for deriving accretion rates is unreliable.

\section{Tomographic modelling}
\label{sec:tomography}
Having characterized the atmospheric properties, and the evolutionary and accretion status of \pa and \pb, we now apply our dedicated stellar-surface tomographic-imaging package to the spectropolarimetric data set described in Section~\ref{sec:observations}. In using this tool, we assume that the observed variability in the data is dominated by rotational modulation (and optionally differential rotation). Then, the imaging code simultaneously inverts a time series of \si and \sv profiles into brightness maps (featuring both cool spots and warm plages, known to contribute to the activity of very active stars) and magnetic maps (with poloidal and toroidal components, using a spherical harmonic decomposition). For brightness imaging, a copy of a local line profile is assigned to each pixel on a spherical grid, and the total line profile is found by summing over all visible pixels (at a given phase), where the pixel intensities are scaled iteratively to fit the observed data. For magnetic imaging, the Zeeman signatures are fit using a spherical-harmonic decomposition of potential and toroidal field components, where the weighting of the harmonics are scaled iteratively \citep{donati2001}. The data are fit to an aim $\chi^{2}$, with the optimal fit determined using the maximum-entropy routine of \cite{gull1991}, and where the chosen map is that which contains least information (maximum entropy) required to fit the data. For further details about the specific application of our code to wTTSs, we refer the reader to previous papers in the series \citep[e.g.,][]{donati2010b,donati2014,donati2015}.

Given that typical Zeeman signatures have relative amplitudes of $\sim0.1$~per~cent, with relative noise levels of around $10^{-3}$ in a typical spectrum (for a single line), we require some means to improve the S/N to a sufficient level for reliable mapping of the stellar magnetic fields. To achieve this, Least-Squares Deconvolution \citep[LSD,][]{donati1997} was applied to all spectra. This technique involves cross-correlating the observed spectrum with a stellar line-list, and results in a single `mean' line profile with a dramatically improved S/N, with accurate errorbars for the Zeeman signatures \citep{donati1997}. The stellar line list used for LSD was sourced from the Vienna Atomic Line Database \citep[VALD,][]{vald}, and was computed for $T_{\text{eff}} = 4500$~K  and $\log{g} = 4.0$ (in cgs units), appropriate for both \pa and \pb (see Section~\ref{sec:evolution}). Only moderate to strong atomic spectral lines were included (with line-to-continuum core depressions larger than 40~per~cent prior to all non-thermal broadening). Furthermore, spectral regions containing strong lines mostly formed outside the photosphere (e.g. Balmer, He, \ion{Ca}{ii}~H~\&~K and infra red triplet (IRT) lines) and regions heavily crowded with telluric lines were discarded (see e.g. \citealt{donati2010b} for more details), leaving 6671 spectral lines for use in LSD . Expressed in units of the unpolarized continuum level $I_{\text{c}}$, the average noise level of the resulting \sv signatures range from 4--8.3$\times10^{-4}$ per 1.8~\kms velocity bin, with a median value of $4.9\times10^{-4}$ for both stars.

\begin{figure}
\includegraphics[width=\columnwidth]{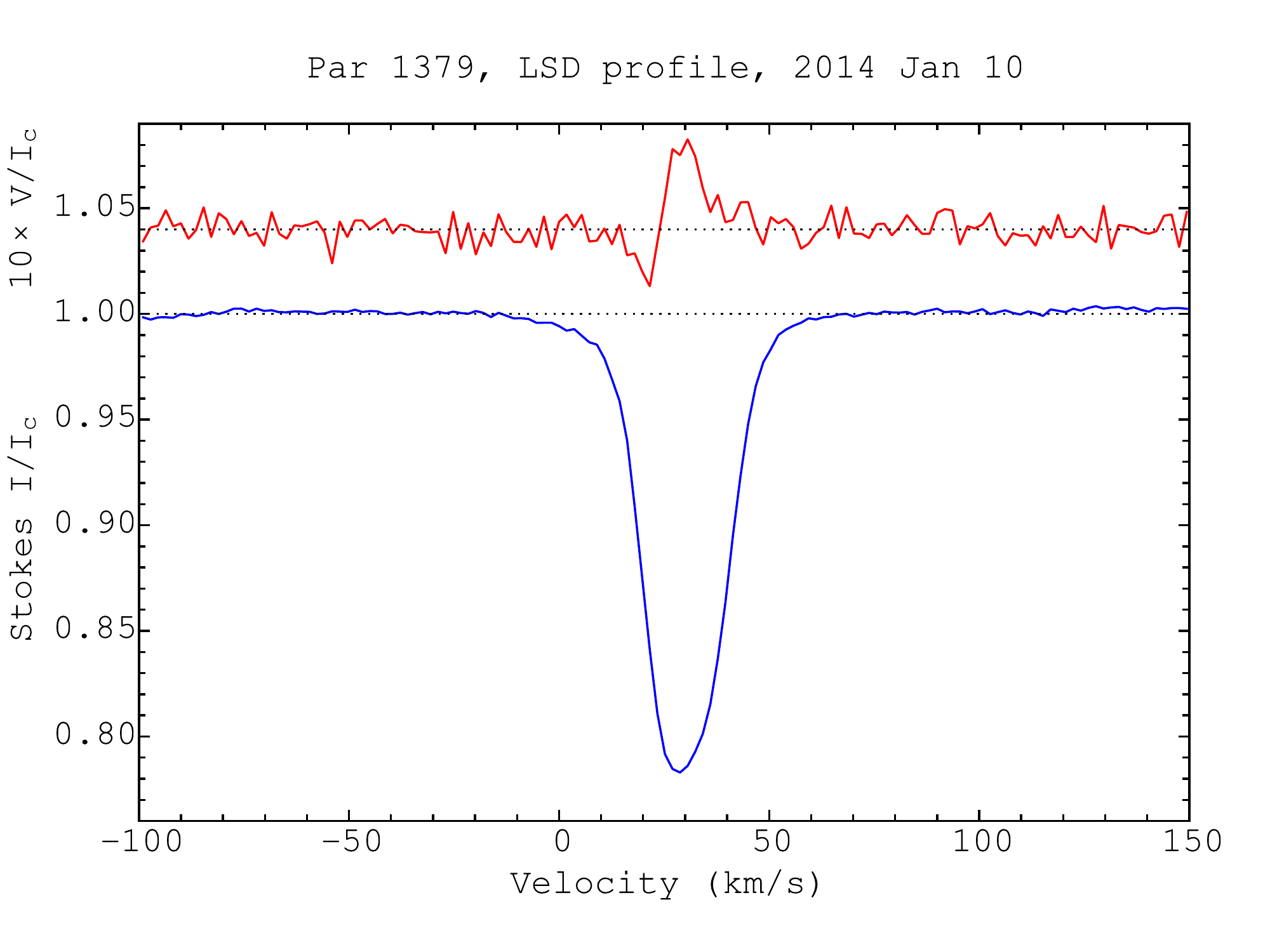}
\includegraphics[width=\columnwidth]{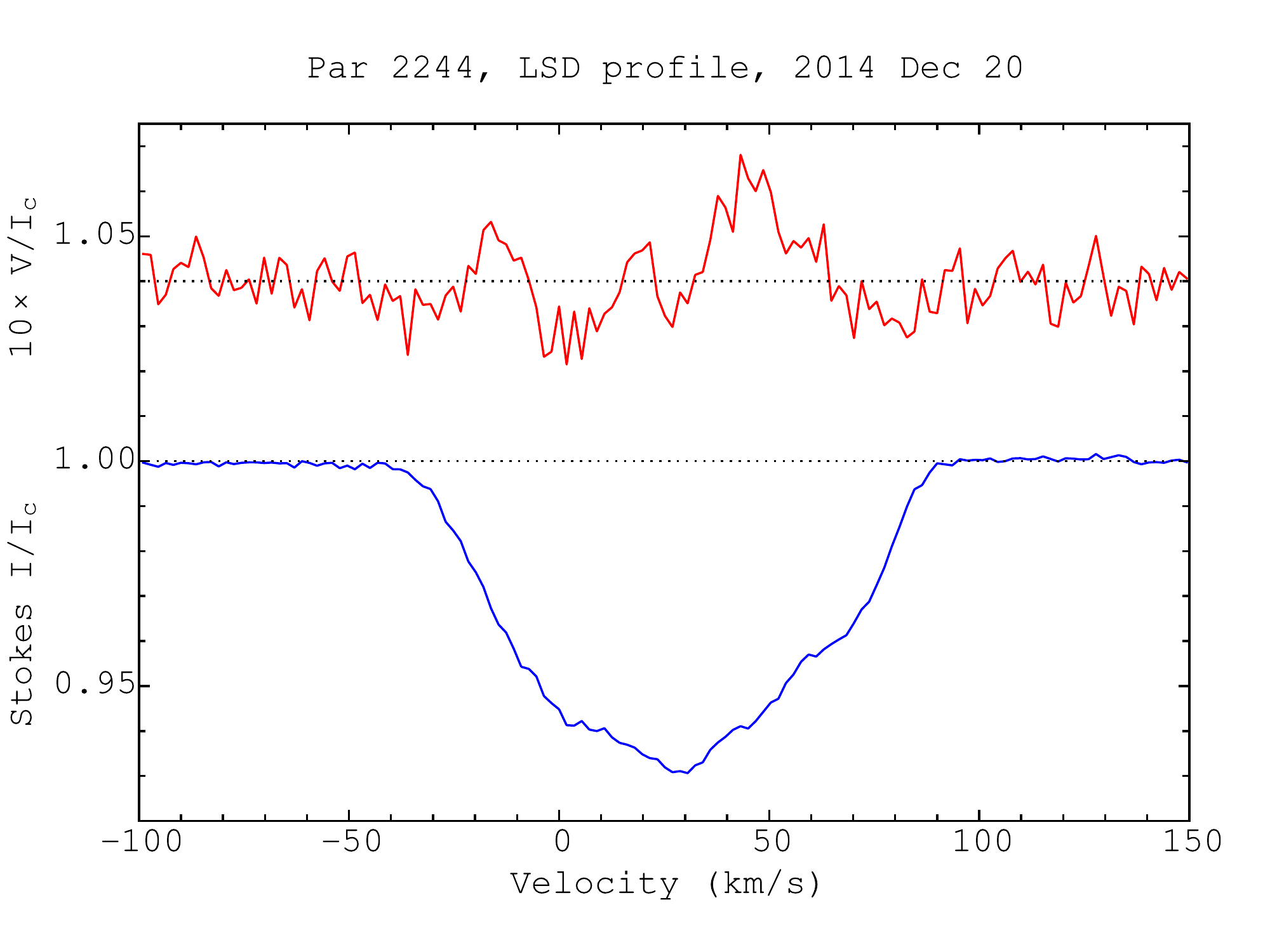}
\caption{LSD circularly-polarized (\sv, top/red curve) and unpolarized (\si, bottom/blue curve) profiles of \pa (top panel) and \pb (bottom panel) collected on 2014 Jan. 10 (cycle 0.54) and 2014 Dec. 20 (cycle 0.7). Clear Zeeman signatures are detected in the LSD \sv profile of both stars in conjunction with the unpolarized line profiles. The mean polarization profiles are expanded by a factor of 10 and shifted upwards by 1.04 for display purposes.}
\label{fig:lsd}
\end{figure}

Zeeman signatures are detected at all times in \sv LSD profiles (see Figure~\ref{fig:lsd} for an example), featuring typical amplitudes of 0.2--0.3~per~cent for both \pa and \pb, with the latter showing more complex field structures. Distortions are also visible in \si LSD profiles for both stars, suggesting the presence of brightness inhomogeneities on the surface of both stars, with the larger distortions for \pb suggesting a higher number of surface structures.

The disc-integrated average photospheric LSD profiles are computed by first synthesizing the local Stokes $I$ and $V$ profiles using the Unno-Rachkovsky analytical solution to the polarized radiative transfer equations in a Milne-Eddington model atmosphere, taking into account the local brightness and magnetic field. Then, these local line profiles are integrated over the visible hemisphere (including linear limb darkening) to produce synthetic  profiles for comparison with observations. This method provides a reliable description of how line profiles are distorted due to magnetic fields (including magneto-optical effects, e.g., \citealt{landi2004}). The main parameters of the local line profiles are similar to those used in our previous studies; the wavelength, Doppler width, equivalent width and Land\'{e} factor being set to 670~nm, 1.8~\kms, 3.9~\kms and 1.2, respectively.

As part of the imaging process we obtain accurate estimates for \vrad (the RV the star would have if unspotted), equal to ${30.40\pm0.07}$~\kms and ${27.00\pm0.05}$~\kms, the \vsini, equal to ${13.7\pm0.1}$~\kms and ${57.2\pm0.1}$~\kms, and the inclination $i$ of the rotation axis to the line of sight, equal to ${36\degr\pm10\degr}$ and ${59\degr\pm10\degr}$ (in excellent agreement with the values derived in Section~\ref{sec:evolution}) for \pa and \pb, respectively (see Table~\ref{tab:systempars}).

\begin{table}
\centering
\caption{Main parameters of \pa and \pb as derived from our study (plus their distance), with \vrad noting the RV that the star would have if unspotted (as inferred from the modeling of Section~\ref{sec:tomography}). Note, the stellar masses and ages are those determined from \protect \cite{siess2000} models, with values from \protect \cite{baraffe2015} given in parenthesis.}
\label{tab:systempars}
\begin{tabular}{lll}
\hline					
	&	Par 1379	&	Par 2244	\\
\hline					
$M_{\star}$~(\msun)	&	$1.6\pm0.1$ ($1.3\pm0.1$)	&	$1.8\pm0.1$ ($1.4\pm0.1$)	\\
$R_{\star}$~(\rsun)	&	$2.7\pm0.2$	&	$3.5\pm0.2$	\\
age (Myr)	&	$1.8\pm0.6$ ($1.0\pm0.5$)	&	$1.1\pm0.3$ ($0.5\pm0.5$)	\\
$\log{g}$ (cgs units)	&	$3.9\pm0.2$	&	$4.1\pm0.2$	\\
\teff~(K)	&	$4600\pm50$	&	$4650\pm50$	\\
log($L_{\star}$/\lsun)	&	$0.45\pm0.11$	&	$0.72\pm0.11$	\\
\prot~(d)	&	$5.585\pm0.035$	&	$2.8153\pm0.0023$	\\
\vsini~(\kms)	&	${13.7\pm0.1}$	&	${57.2\pm0.1}$	\\
\vrad~(\kms)	&	$30.40\pm0.07$	&	$27.00\pm0.05$	\\
$i$~(\degr)	&	${36\degr\pm10\degr}$	&	${59\degr\pm10\degr}$	\\
distance (pc)	&	$388\pm5$	&	$388\pm5$	\\
\hline					
\end{tabular}
\end{table}

\subsection{Brightness and magnetic imaging}
\label{sec:imaging}
\begin{figure*}
\includegraphics[width=0.23\textwidth]{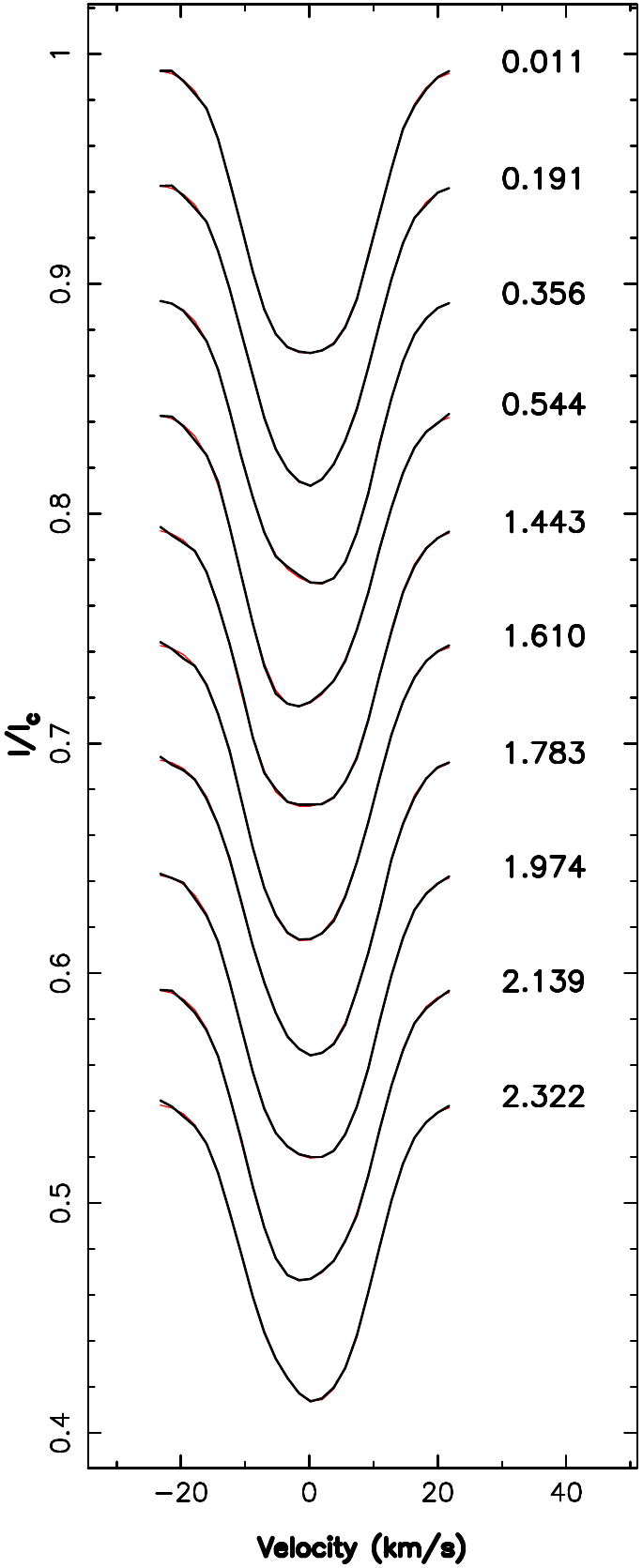}
\includegraphics[width=0.23\textwidth]{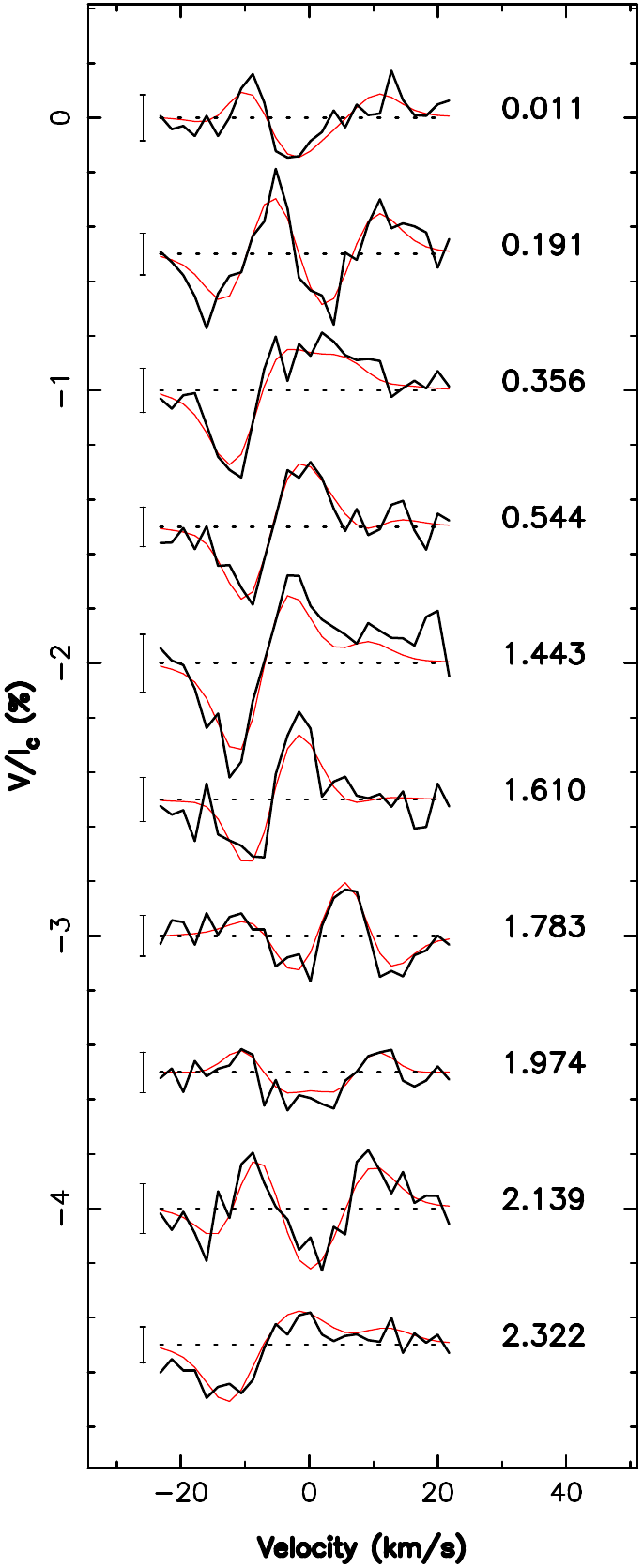}
\hspace{0.04\textwidth}
\includegraphics[width=0.23\textwidth]{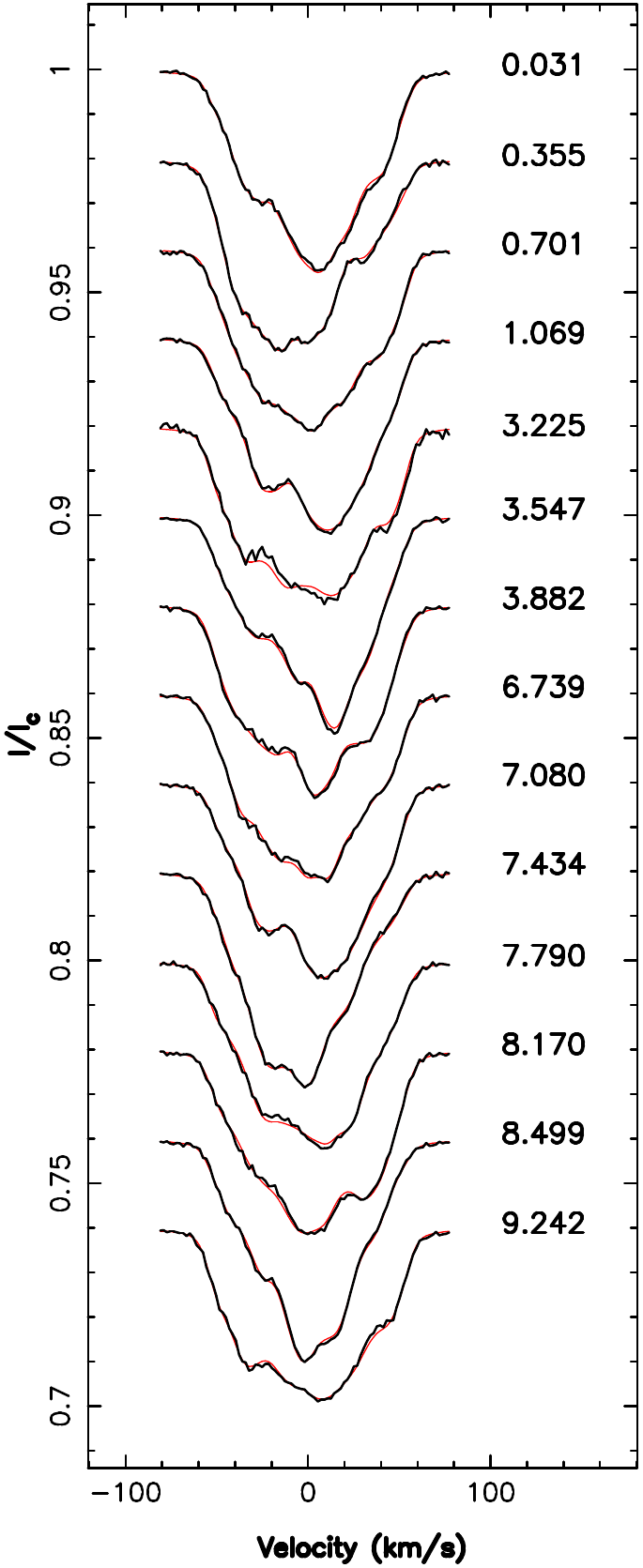}
\includegraphics[width=0.23\textwidth]{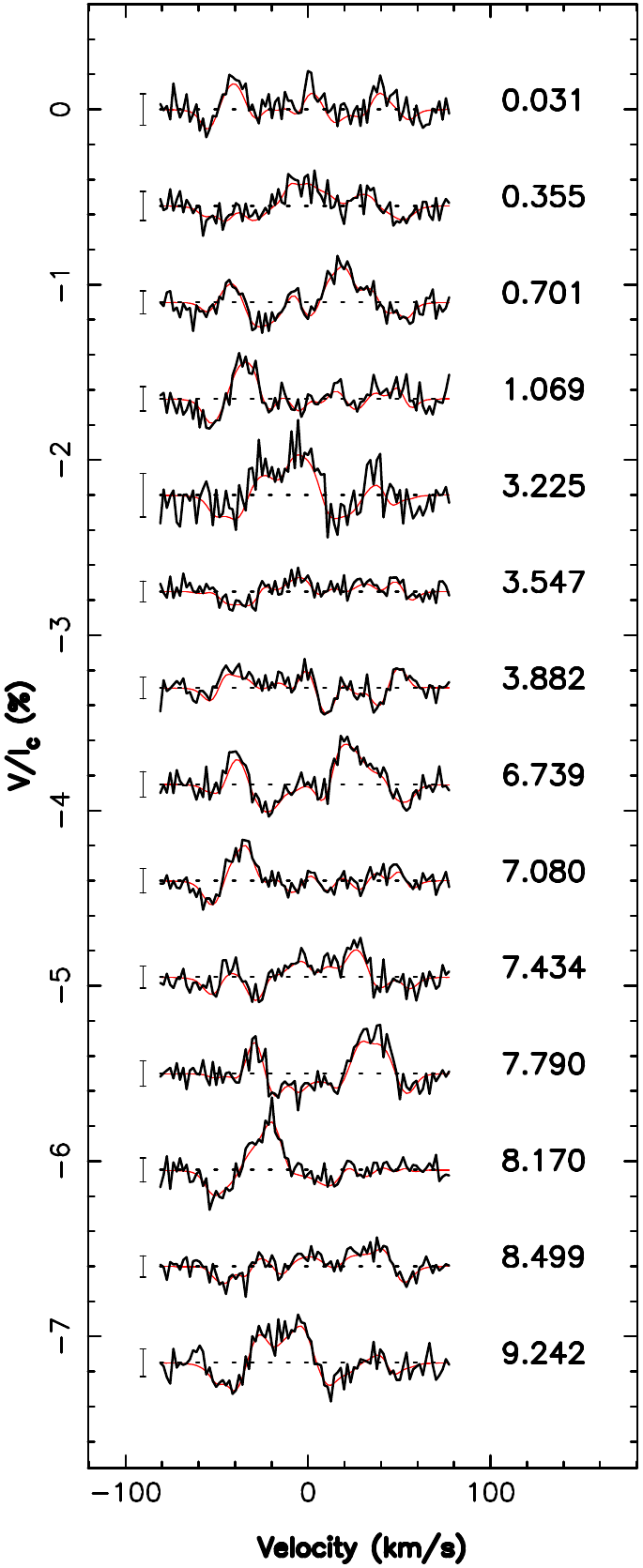}
\caption{Maximum-entropy fit (thin red line) to the observed (thick black line) \si (first and third panels) and \sv (second and fourth panels) LSD photospheric profiles of \pa (first two panels) and \pb (last two panels). Rotational cycles are shown next to each profile. This figure is best viewed in colour.}
\label{fig:lsdprofiles}
\end{figure*}

Figure~\ref{fig:lsdprofiles} shows the Stokes $I$ and $V$ LSD profiles of \pa and \pb, as well as our fits to the data. All our fits correspond to a reduced chi-squared $\chi_{\text{r}}^{2}$ equal to 1 (i.e., where $\chi^{2}$ equals the number of fitted data points, equal to 260 for \pa, and 1246 for \pb), emphasising the high quality of our data set and our modelling technique at reproducing the observed modulation of the LSD profiles. While the phase coverage for our two stars is less dense than that for LkCa~4, V819~Tau or V830~Tau \citep[see][]{donati2014,donati2015}, the small rms of the RV residuals for both stars (see Section~\ref{sec:rv}) and the large \vsini for \pb (providing $\sim4\times$ the resolution compared to \pa), means we can safely claim that our maps include no major imaging artefact nor bias. 

\begin{figure*}
\includegraphics[width=0.33\textwidth]{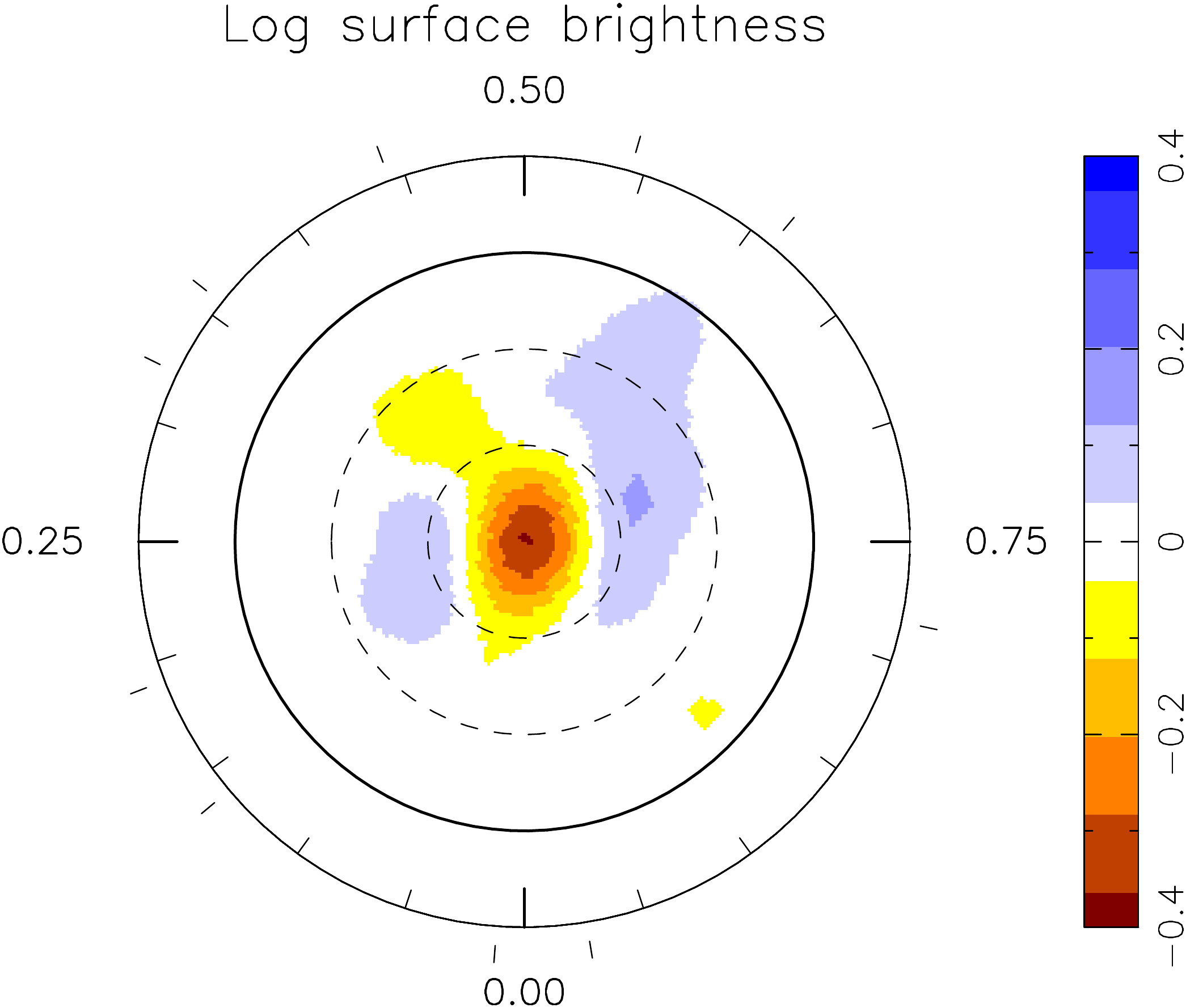}
\hspace{2cm}
\includegraphics[width=0.33\textwidth]{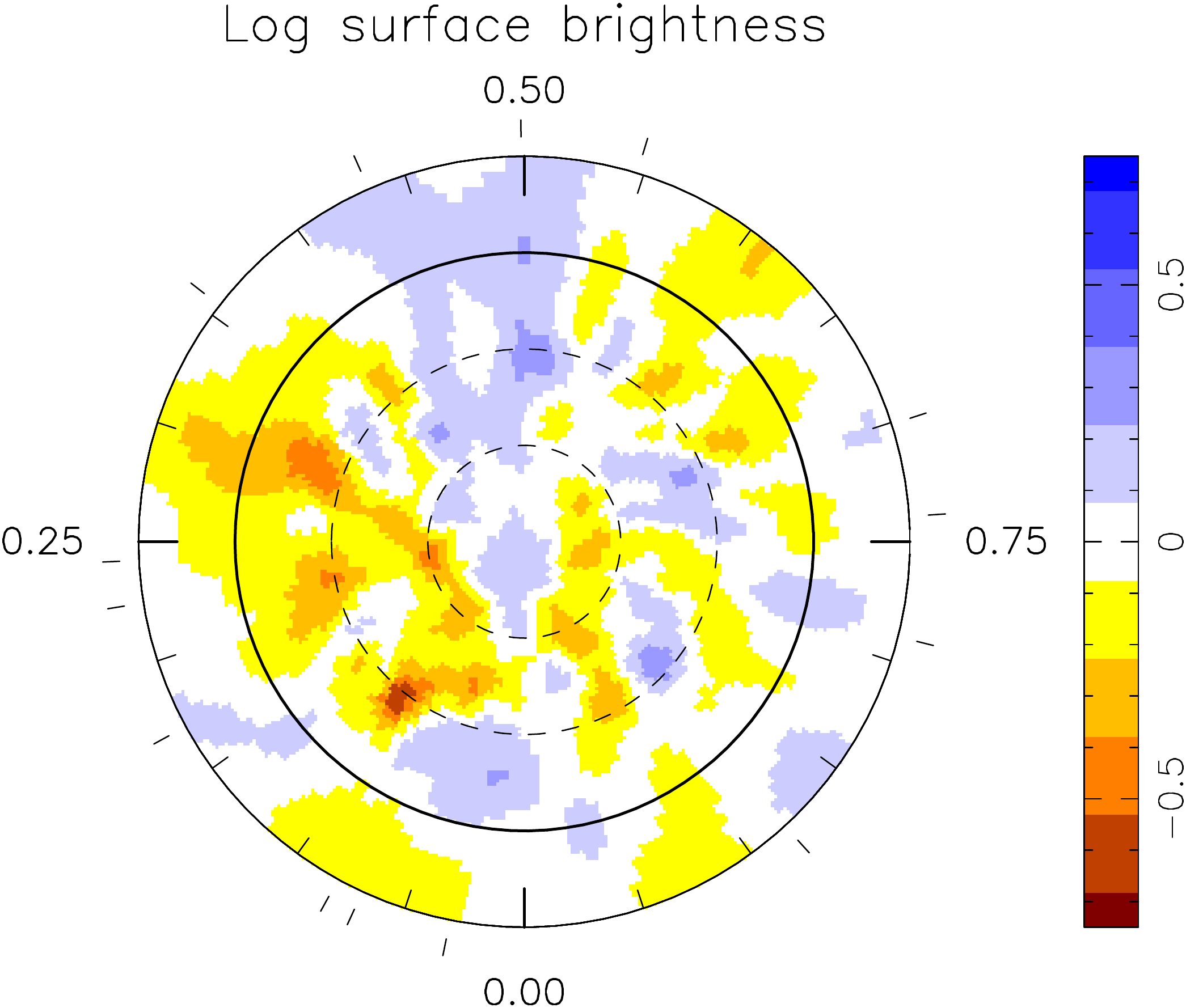}\\
\caption{Maps of the logarithmic brightness (relative to the quiet photosphere), at the surfaces of \pa (left) and \pb (right). The stars are shown in flattened polar projection down to latitudes of $-30\degr$, with the equator depicted as a bold circle, and 30\degr and 60\degr parallels as dashed circles. Radial ticks around each plot indicate phases of observations. This figure is best viewed in colour.}
\label{fig:brightnessmaps}
\end{figure*}

The brightness maps of \pa and \pb include both cool spots and warm plages (see Figure~\ref{fig:brightnessmaps}), with \pb showing significantly more contrast between these features. The brightness map of \pa features a single dark, circular spot over the polar region, with a region of bright plage extending from around $70\degr$ to the equator (centred around phase 0.65). These relatively simple features are required to reproduce the observed \si profile distortions for \pa (concentrating in the line cores rather than in the wings, see Figure~\ref{fig:lsdprofiles}), and in particular, the small RV variations (of maximum amplitude 0.28~\kms, see Section~\ref{sec:rv}). We find an overall spot and plage coverage of ${\simeq5}$~per~cent (${\simeq2.5}$~per~cent each for spots and plages), much lower than that for LkCa~4 \citep{donati2014}, and around half that for V819~Tau and V830~Tau \citep{donati2015}.

The brightness map of \pb (right panel of Figure~\ref{fig:brightnessmaps}) features spots and plages distributed across the entire surface, with no polar spot. Spotted regions extend from around $70\degr$ to below the equator around phase 0.2--0.35, with a region of bright plages extending from the pole to below the equator around phase 0.4--0.5. We find an overall spot and plage coverage of $\simeq19$~per~cent ($\simeq11$ and $\simeq8$~per~cent for spots and plages, respectively), where the numerous and complex surface features are required to fit the large RV fluctuations (of maximum amplitude 5.1~\kms, see Section~\ref{sec:rv}) of the observed line profiles. We note that the relatively poor fit at cycle 3.225 (providing the largest RV residual, see Figure~\ref{fig:rvcurves}) coincides with strong, relatively narrow H$\alpha$ emission (see Figure~\ref{fig:halpha}).

Note that the estimates of spot and plage coverage should be considered as lower limits only, as Doppler imaging is mostly insensitive to small-scale structures that are evenly distributed over the stellar surface (hence the larger minimal spot coverage assumed in Section~\ref{sec:evolution} to derive the location of the stars in the H-R diagram).

\subsection{Magnetic field imaging}
\label{sec:magfield}
The reconstructed magnetic fields are described as a sum of a poloidal and toroidal fields, each expressed a spherical-harmonic (SH) expansion, with $\ell$ and $m$ denoting the mode and order of the SH \citep{donati2006b}. For a given set of complex coefficients $\alpha_{\ell, m},~\beta_{\ell, m}$ and $\gamma_{\ell,m}$ (where $\alpha_{\ell, m}$ characterizes the radial field component, $\beta_{\ell, m}$ the azimuthal and meridional components of the poloidal field term, and $\gamma_{\ell, m}$ the azimuthal and meridional components of the toroidal field term), one can construct an associated magnetic image at the surface of the star, and thus derive the corresponding \sv data set. The inverse is carried out here, where we reconstruct the set of coefficients that fit the observed data.

\begin{figure*}
\includegraphics[width=\textwidth]{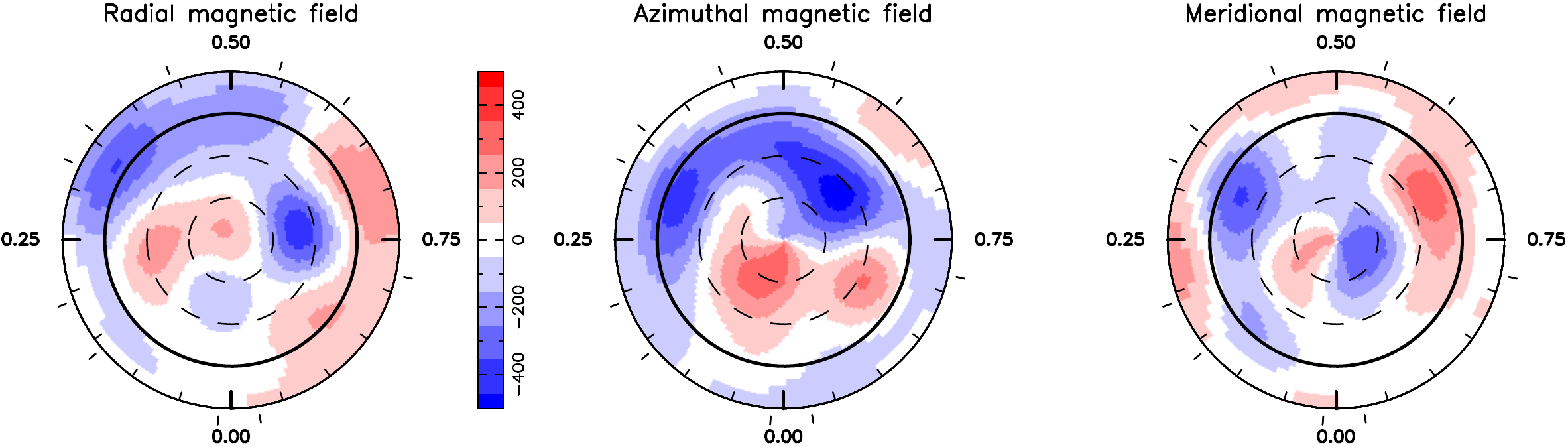}
\includegraphics[width=\textwidth]{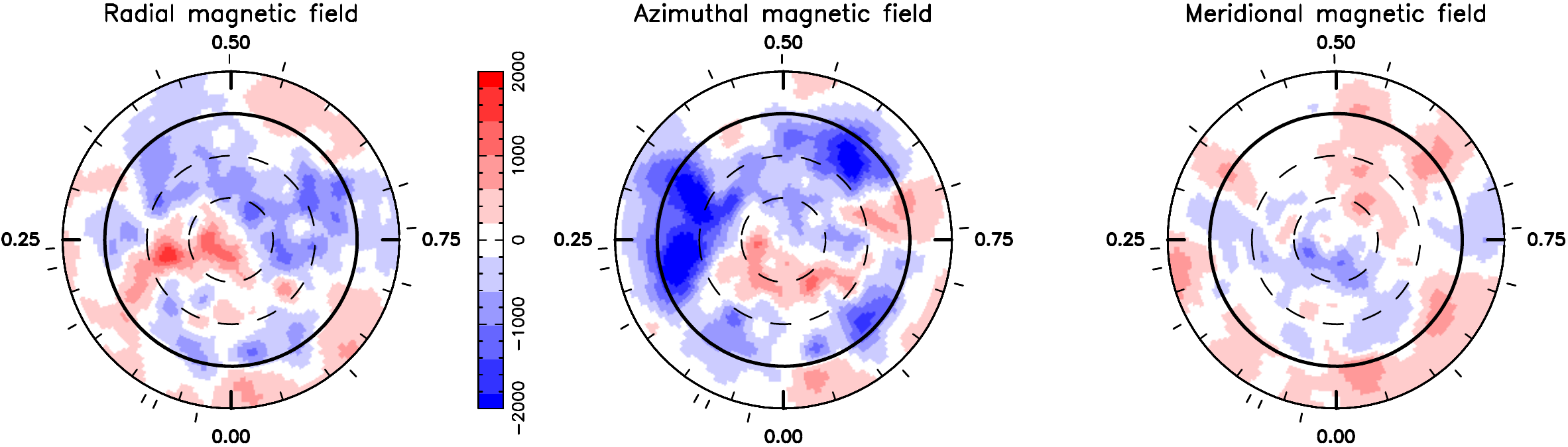}
\caption{Maps of the radial (left), azimuthal (middle) and meridional (right) components of the magnetic field \textbf{B} at the surfaces of \pa and (top) and \pb (bottom). Magnetic fluxes in the colourbar are expressed in G. The stars are shown in flattened polar projection as in Figure~\ref{fig:brightnessmaps}. This figure is best viewed in colour.}
\label{fig:magmaps}
\end{figure*}

The reconstructed magnetic fields of \pa and \pb (see Figure~\ref{fig:magmaps}) are quite different in both strength and topological properties. For \pa, the field mainly consists of a mostly non-axisymmetric poloidal component (at a level of 74~per~cent). The largest fraction of the reconstructed poloidal field energy (45~per~cent) is contained in the quadrupolar ($\ell = 2$) SH mode, with the remainder split evenly between the other modes (i.e., with $\ell = $1, 3 and 4). The large-scale topology of the poloidal component is tilted at $\simeq65\degr$ from the rotation axis (towards phase 0.41), and generates an intense radial field in excess of 400~G at mid latitudes around phase 0.7 (see top left panel of Figure~\ref{fig:magmaps}). We also find a mostly-axisymmetric toroidal component (containing 26~per~cent of the total energy), with 44~per~cent of the toroidal magnetic energy in the $\ell=1$ dipole, with a further 33~per~cent of the toroidal energy in the $\ell=4$ mode, combining to producing fields in excess of 400~G at low to mid latitudes (see centre top panel of Figure~\ref{fig:magmaps}). In total, we find an unsigned field strength of 250~G. %We note that the reconstructed magnetic features do not align significantly with those in the brightness maps for \pa. 

For \pb, the field mainly consists of a mostly non-axisymmetric poloidal component (58~per~cent), with 48~per~cent of the reconstructed energy contained in modes with $\ell \geq 4$ (with the remaining energy split fairly evenly between modes with $\ell<4$). The poloidal component can be approximated at large distances from the star by a 330~G dipole tilted at $\simeq45\degr$ from the rotation axis (towards phase 0.57), which when combined with higher order modes generates fields in excess of 1~kG at the surface (see bottom left panel of Figure~\ref{fig:magmaps}). We also find a significant, mostly axisymmetric toroidal component (42~per~cent of the total) with $\simeq31$~per~cent of the reconstructed energy contained in modes with $\ell =$~1 and 4. This complex topology generates fields in excess of 2~kG at low latitudes (see bottom-centre panel of Figure~\ref{fig:magmaps}). In total, we find an unsigned field strength of 0.86~kG. Comparing the brightness and magnetic maps of \pb (see Figures~\ref{fig:brightnessmaps} and \ref{fig:magmaps}, respectively), one can see there is some degree of spatial correlation between features. In particular, the moderately strong radial fields (around phases 0.4--0.5), and the strong azimuthal fields at low latitudes (at phases 0.15--0.35 and 0.55--0.65) loosely correlate with the high plage and spot coverage at these phases in the brightness map.

Lastly, we note that the SH expansions describing the reconstructed field presented in Figure~\ref{fig:magmaps} are limited to terms with $\ell \leq 5$ for \pa, as only marginal changes to the solution are observed when larger $\ell$ values are included, indicating that most of the detected \sv signal for \pa concentrates at larger spatial scales. In contrast, the SH expansion for \pb requires terms with $\ell \leq 18$ to fit the data to $\chi_{\text{r}}^{2} = 1$, indicating a more complex and compact magnetic field topology. Given that the amount of reconstructed structure depends on the star's \vsini, we may expect to have around 4 times the effective resolution for \pb as compared to \pa. Thus, to compare like-for-like, we reconstructed the brightness and magnetic maps for \pb while limiting the number of SH terms to $\ell \leq 5$ (matching that for \pa). The resulting maps cannot be fit to the same level of $\chi_{\text{r}}^{2}$, however, the resulting topology is very similar to that found when including significantly more terms: We find a mainly non-axisymmetric field split 60:40 between poloidal and toroidal components, with the largest fraction of poloidal energy in the $\ell \geq 4$ modes, and an unsigned field strength around 30~per~cent larger. Thus, as the data were fit in the same manner, we confirm that \pb has a significantly more complex field topology compared to \pa. This property is also reflected in the \sv profiles of \pb, as if one degraded them to the same velocity resolution as \pa, the Zeeman signatures would be weaker and more complex than those dominating the \sv profiles of \pa.

\begin{figure*}
\includegraphics[width=0.4\textwidth]{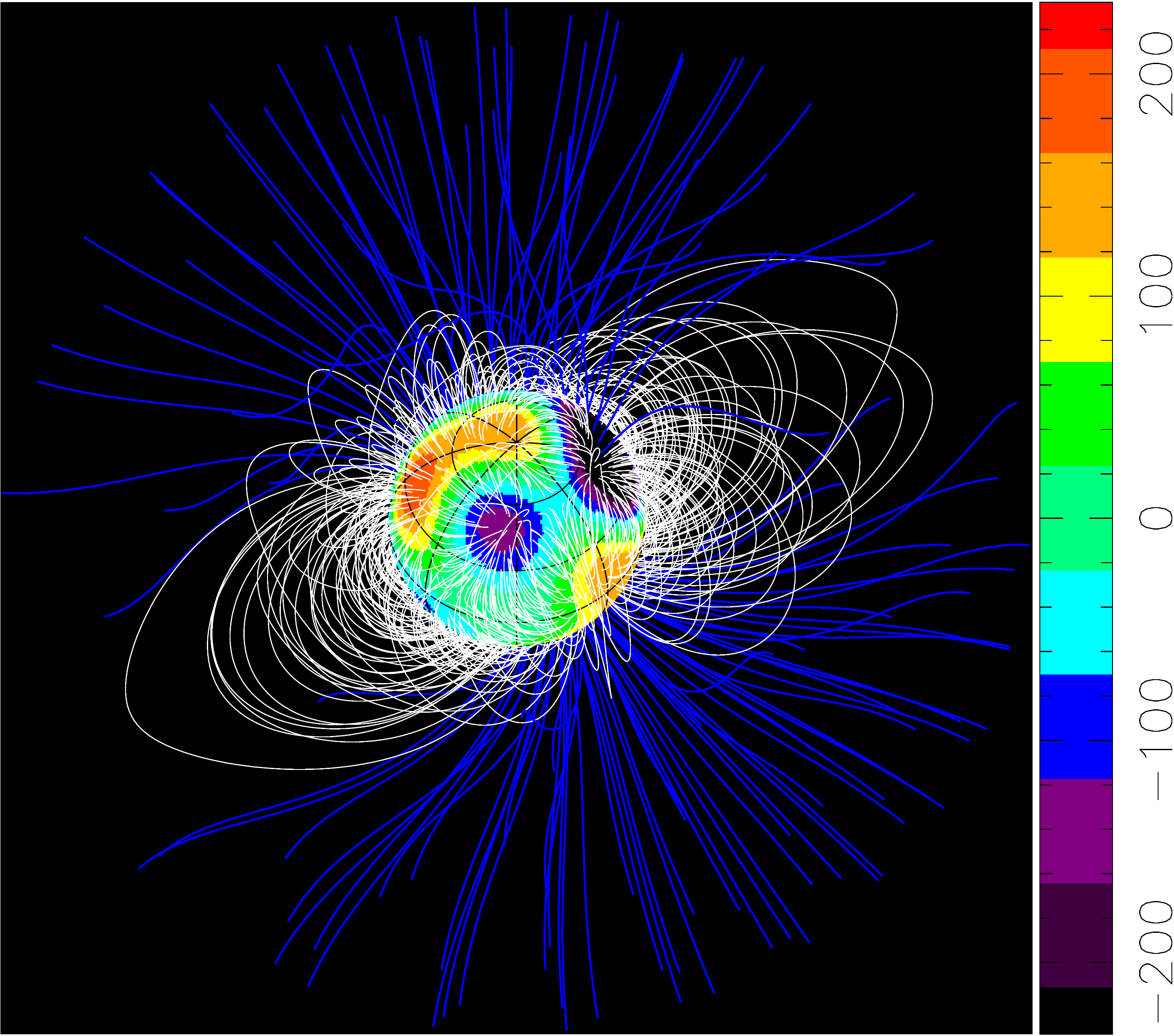}
\hspace{2cm}
\includegraphics[width=0.4\textwidth]{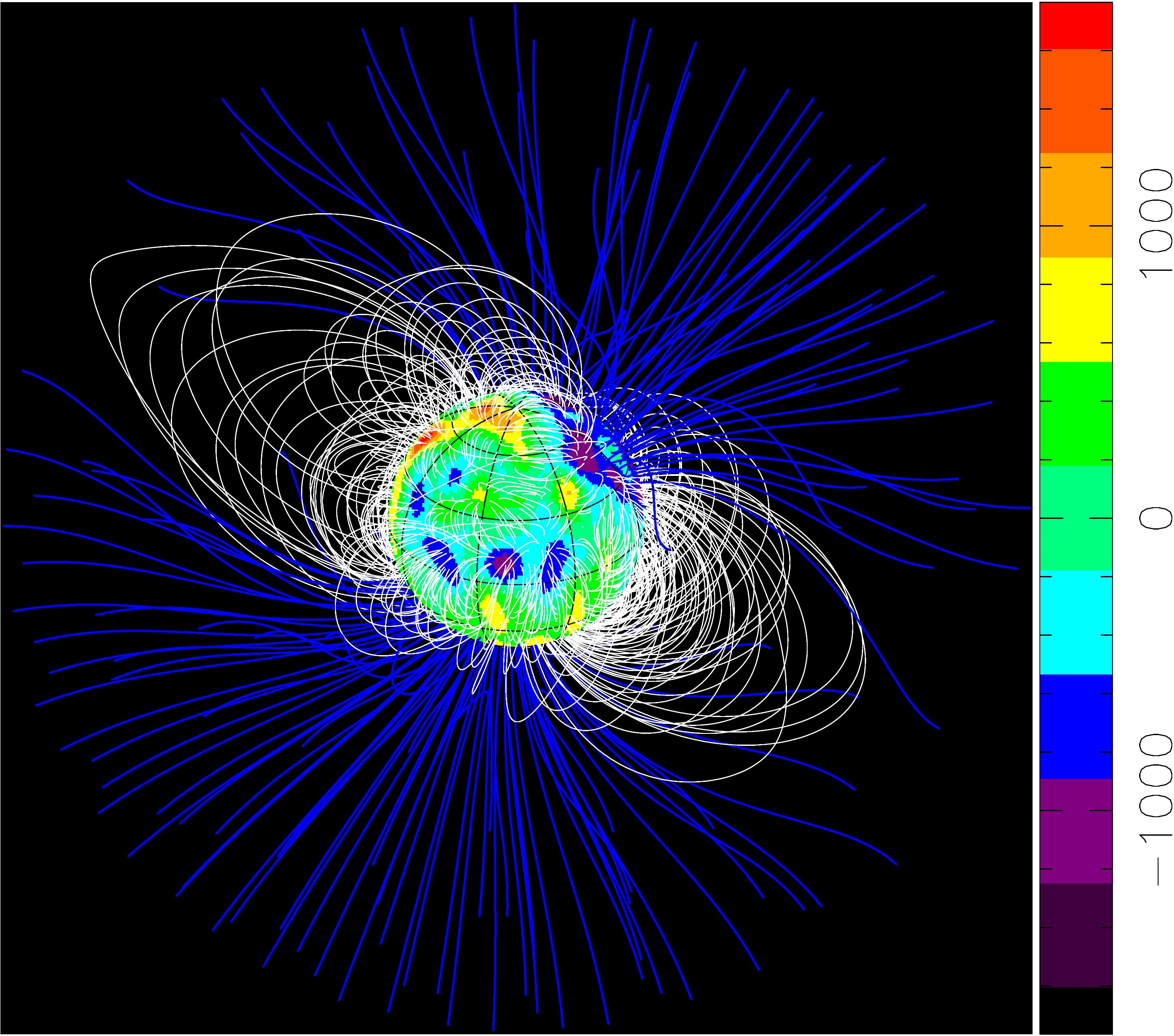}
\caption{Potential field extrapolations of the magnetic field reconstructed for \pa (left) and \pb (right), viewed at inclinations of $36\degr$ and $59\degr$, at phases 0.0 and 0.95, respectively. Open and closed field lines are shown in blue and white, respectively, whereas colours at the stellar surface depict the local values (in G) of the radial field (as shown in the left panels of Figure~\ref{fig:magmaps}). The source surfaces at which the field becomes radial are set at distances of 5.5~\rstar for \pa and 2.7~\rstar for \pb, as these are close to the co-rotation radii (where the Keplerian orbital period equals the stellar rotation period, and beyond which the field lines tend to open under the effect of centrifugal forces, \citealt{jardine2004}), but are smaller than the Alfv\'{e}n radii of $>10$~\rstar \citep{reville2016}. This figure is best viewed in colour. Full animations may be found for both \pa and \pb at http://imgur.com/a/qwAXg.}
\label{fig:extrap}
\end{figure*}

In Figure~\ref{fig:extrap} we present the extrapolated large-scale field topologies of \pa and \pb using the potential field approximation (e.g., \citealt{jardine2002}), and derived solely from the reconstructed radial field components. These potential topologies represent the lowest possible states of magnetic energy, and provide a reliable description of the magnetic field well within the Alfv\'{e}n radius \citep{jardine2013}. These plots show the largely quadrupolar field of \pa, and the complex field of \pb that has a large fraction of energy in higher order modes.

\subsection{Surface differential rotation}
\label{sec:diffrot}
Our observations of \pa and \pb were taken over reasonably long time-spans of 13~d (2.3 rotation cycles) and 26~d (9.3 rotation cycles), respectively, making them well suited to measuring differential rotation in the same way as has been carried out in several previous studies (e.g., \citealt{donati2003a,donati2010a,donati2014,donati2015}). We achieve this by assuming that the rotation rate at the surface of the star is varying with latitude $\theta$ as $\Omega_{\text{eq}} - d\Omega\sin^2{\theta}$, where $\Omega_{\text{eq}}$ is the rotation rate at the equator and $d\Omega$ is the difference in rotation rate between the equator and the pole. One can then reconstruct brightness and magnetic maps at a given information content for many pairs of $\Omega_{\text{eq}}$ and $d\Omega$, finding the corresponding reduced chi-squared $\chi_{\text{r}}^{2}$ at which the modelled spectra fit the observations. The topology of the resulting $\chi_{\text{r}}^{2}$ surface usually has a well defined minimum, and by fitting a parabola to this surface, we may estimate both $\Omega_{\text{eq}}$ and $d\Omega$, and their corresponding error bars. This process has proved reliable for estimating surface differential rotation on various kinds of active low-mass stars (e.g., \citealt{donati2003a,donati2010a}) including wTTSs \citep{skelly2008,skelly2010, donati2014,donati2015}, and we refer the reader to these papers for further details of this technique.

\begin{figure*}
\includegraphics[width=0.41\textwidth]{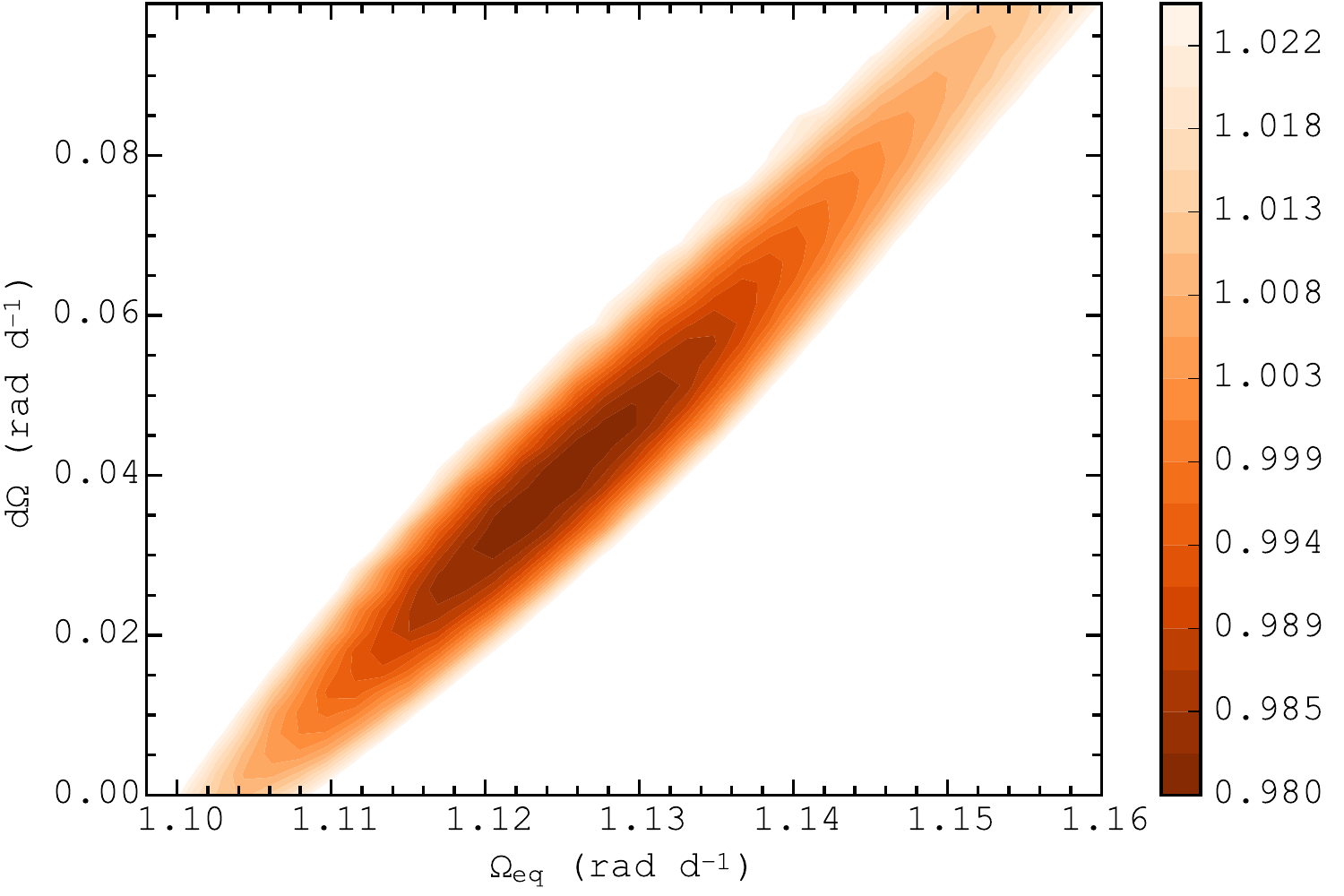}
\includegraphics[width=0.17\textwidth]{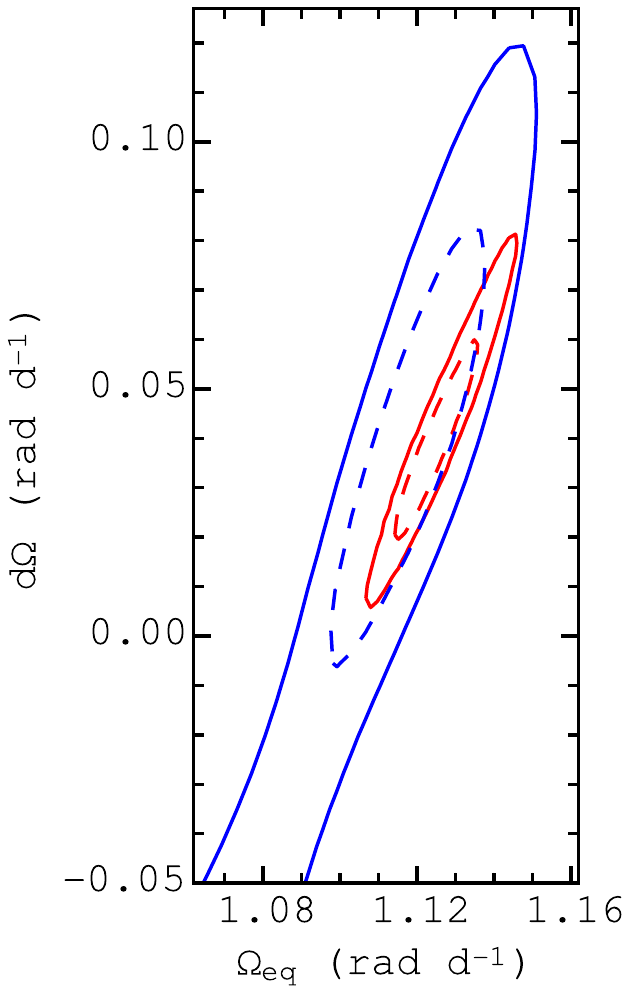}
\includegraphics[width=0.41\textwidth]{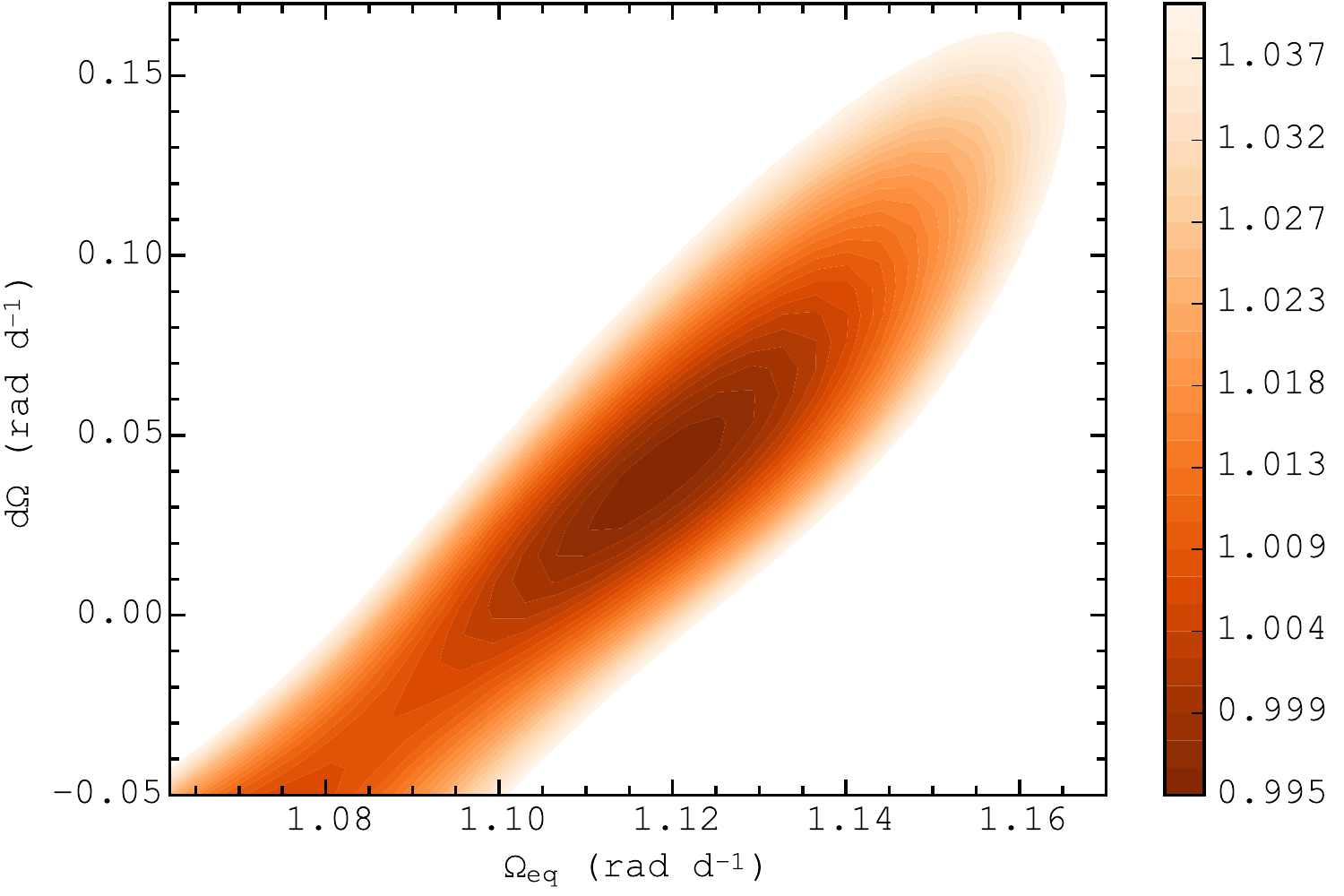}
\caption{Variations of $\chi_{\text{r}}^{2}$ as a function of $\Omega_{\text{eq}}$ and $d\Omega$, derived from modelling of our \si (left panel) and \sv (right panel) LSD profiles of \pa at a constant information content. In both cases, a clear and well defined parabola is observed, with the outer contour tracing the 4.5~per~cent increase in $\chi^{2}_{\text{r}}$ (or equivalently a $\chi^{2}$ increase of 11.8 for 260 fitted data points) that corresponds to a $3\sigma$ ellipse for both parameters as a pair. The centre panel shows how well the confidence ellipses from both measurements overlap, with $1\sigma$ and $2\sigma$ ellipses (respectively depicting the 68.3~per~cent and 95.5~per~cent confidence levels) shown in solid and dashed lines (in red and blue for Stokes $I$ and $V$ data respectively). This figure is best viewed in colour.}
\label{fig:par1379difrot}
\end{figure*}

The low \vsini of \pa means our maps have a coarser spatial resolution (as compared to \pb), reducing the accuracy to which we can measure differential rotation. However, the large, fairly narrow plage region extending across $\sim70\degr$ in latitude (see Figure~\ref{fig:brightnessmaps}) is well suited for measuring rotation periods and recurrence rates of profiles distortions across different latitudes. Figure~\ref{fig:par1379difrot} shows the $\chi_{\text{r}}^{2}$ surface we obtain (as a function of $\Omega_{\text{eq}}$ and $d\Omega$) for both Stokes $I$ and $V$, for \pa. We find a clear minimum at $\Omega_{\text{eq}} = 1.125 \pm 0.007$~\radd and $d\Omega = 0.039 \pm 0.014$ for \si data (corresponding to rotation periods of $5.585\pm0.035$~d at the equator and $5.786\pm0.07$~d at the poles; see left panel of Figure~\ref{fig:par1379difrot}), with the fits to the \sv data of $\Omega_{\text{eq}} = 1.118 \pm 0.011$~\radd and $d\Omega = 0.039 \pm 0.023$ showing consistent estimates, though with larger error bars (right panel of Figure~\ref{fig:par1379difrot}).

\begin{figure*}
\includegraphics[width=0.41\textwidth]{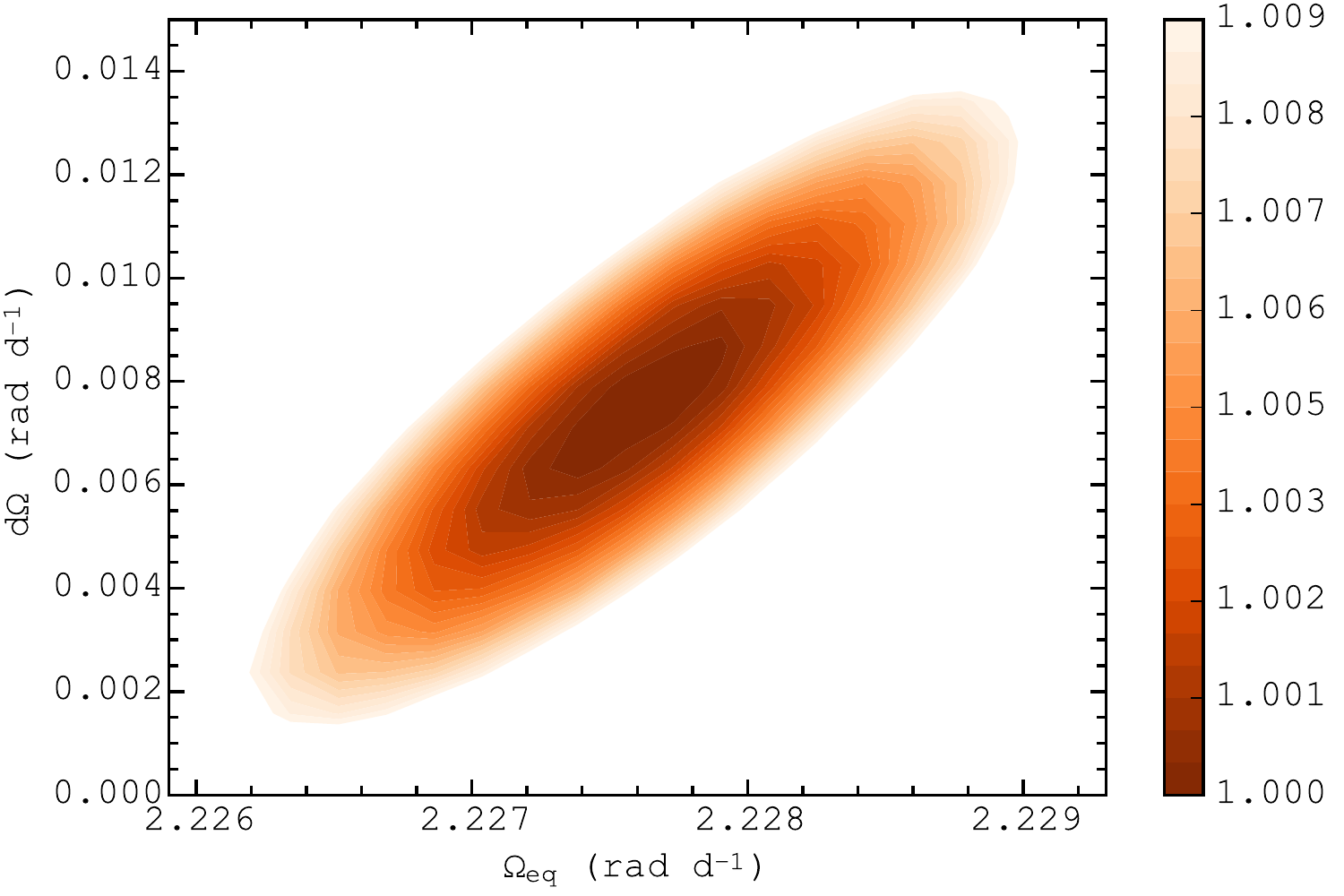}
\includegraphics[width=0.17\textwidth]{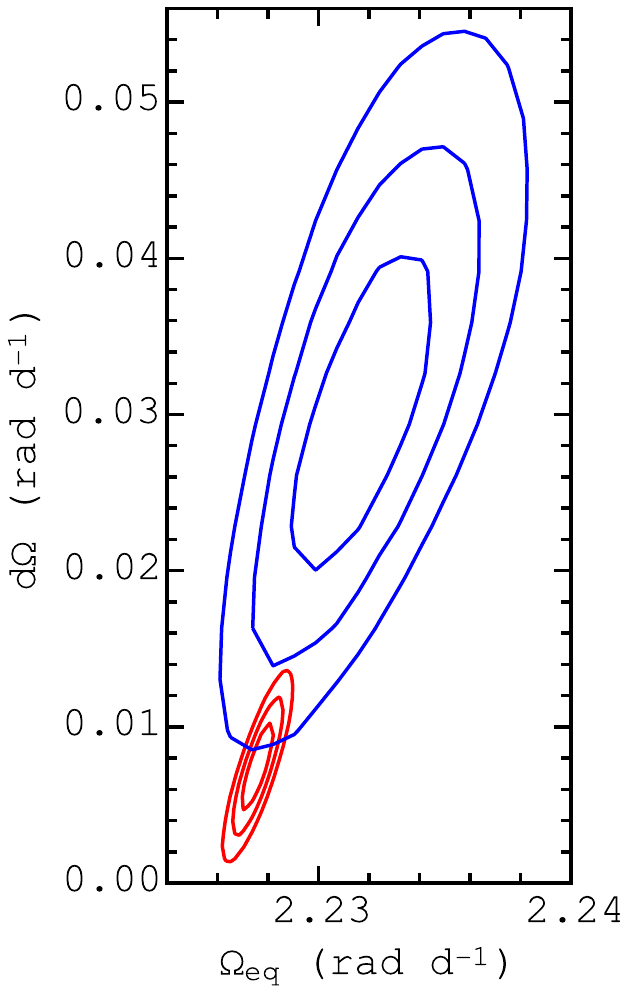}
\includegraphics[width=0.41\textwidth]{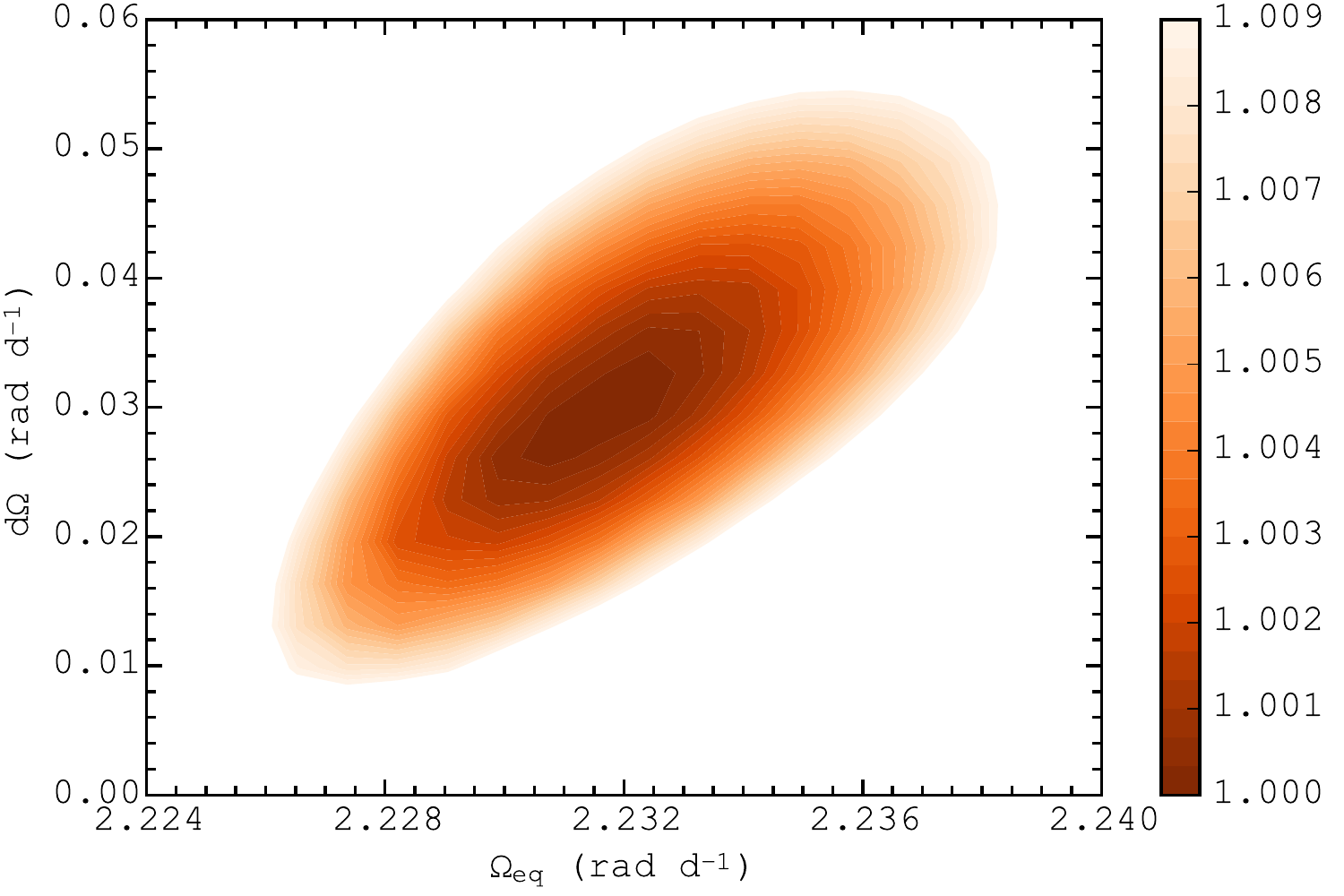}
\caption{Same as for Figure~\ref{fig:par1379difrot} but for \pb. For both Stokes $I$ and $V$, a well defined parabola is observed, with the outer contour tracing the 0.95~per~cent increase in $\chi^{2}_{\text{r}}$ (or equivalently a $\chi^{2}$ increase of 11.8 for 1246 fitted data points) that corresponds to a $3\sigma$ ellipse for both parameters as a pair. The centre panel shows the overlap of the confidence ellipses from both measurements, with $1\sigma$, $2\sigma$ and $3\sigma$ ellipses shown (respectively depicting the 68.3, 95.5 and 99.7~per~cent confidence levels) in red and blue for Stokes $I$ and $V$ data respectively. This figure is best viewed in colour.}
\label{fig:par2244difrot}
\end{figure*}

For \pb, we find a clear minimum in the $\chi^{2}$ surface for \si data at ${\Omega_{\text{eq}} = 2.2276 \pm 0.0004}$~\radd and ${d\Omega = 0.0075 \pm 0.0017}$~\radd (left panel of Figure~\ref{fig:par2244difrot}). Likewise, we also find a clear minimum for the fits to \sv data at ${\Omega_{\text{eq}} = 2.218 \pm 0.0018}$~\radd and ${d\Omega = 0.0306 \pm 0.0067}$ (right panel of Figure~\ref{fig:par2244difrot}). While these estimates do not agree within the formal error bars of the parabolic fit (to the region around the minimum), the minima found for each data set do however overlap at the $3\sigma$ level (see centre panel of Figure~\ref{fig:par2244difrot}). We attribute the discrepancy between minima to the temporal evolution of spots and plages over the observation window, confusing the measurements for \si data. This line of reasoning is supported by the fact that, for the same level of information content (i.e., unsigned magnetic field strength) we cannot fit the \sv data to as low a $\chi^{2}$ using the optimum ${\Omega_{\text{eq}}}$ and ${d\Omega}$ from the \si differential rotation measurements (we find a significant increase in $\chi^{2}$ of 10.7 for 1246 data points). Furthermore, we obtain better fits to the \si data when it is split into two groups that were obtained at similar times, and fit using $\Omega_{\text{eq}}$ and $d\Omega$ derived from \sv data (further discussed in Section~\ref{sec:rv}). Hence, we consider the values of ${\Omega_{\text{eq}}}$ and $d\Omega$ derived from \sv data to be more robust (giving rotation periods of $2.8153\pm0.0023$~d at the equator and $2.872\pm0.039$~d at the poles).

We note that the values of $\Omega_{\text{eq}}$ and $d\Omega$ determined above (for both brightness and magnetic maps, for both stars) do not change significantly when the LSD profiles are fit to different levels of information content (fitting to 30~per~cent less and 5~per~cent more information, compared to that for $\chi^{2}_{r} = 1$), demonstrating the robustness of this method against under or over-fitting the data. Furthermore, the rotation periods of \pa and \pb determined from our analysis agree (to within 1 and $2\sigma$) with the photometric periods found by \cite{rebull2001} of $5.62\pm0.009$~d and $2.82\pm0.002$~d, respectively.

\section{Filtering the activity jitter to search for hot Jupiters}
\label{sec:rv}
As well as studying the topology of magnetic fields, the MaTYSSE program also aims to detect potential hJs to quantitively assess the likelihood of the disc migration scenario (where giant planets form in the outer accretion disc and then migrate inward until they reach the central magnetospheric gaps of cTTSs, see e.g., \citealt{lin1996,romanova2006}). We achieve this in practice by filtering out the activity-related jitter from the RV curves of wTTS, by subtracting the first-order moments of our fits to the LSD profiles, from those of the observed data (following the methods outlined in \citealt{donati2014,donati2015}). Then, once the predicted activity jitter has been removed, one can look for periodic signals in the RV residuals that may probe the presence of hJs. This method has been used with great success for other stars in the MaTYSSE sample, leading to the detection of a hJ around both V830~Tau \citep{donati2015,donati2016a,donati2017} and TAP~26 \citep{yu2017}.

\begin{figure*}
\includegraphics[width=\textwidth]{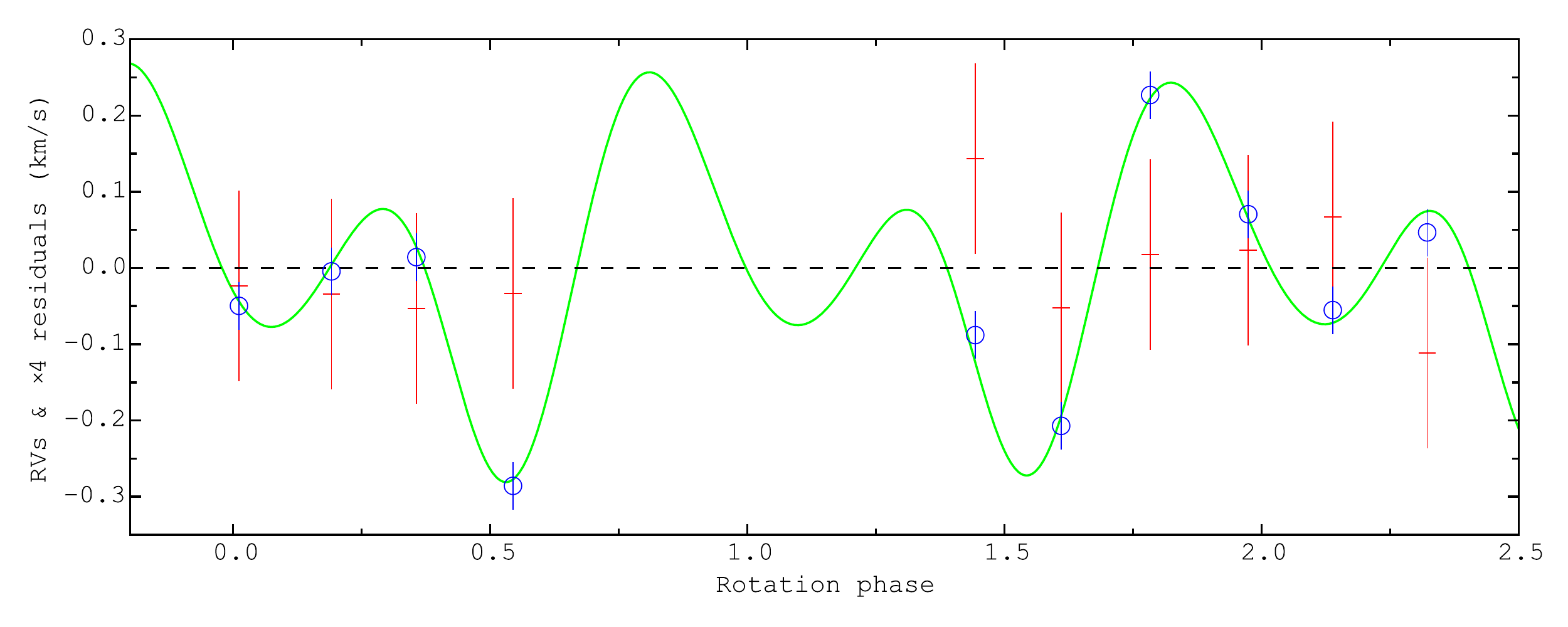} \\
\includegraphics[width=\textwidth]{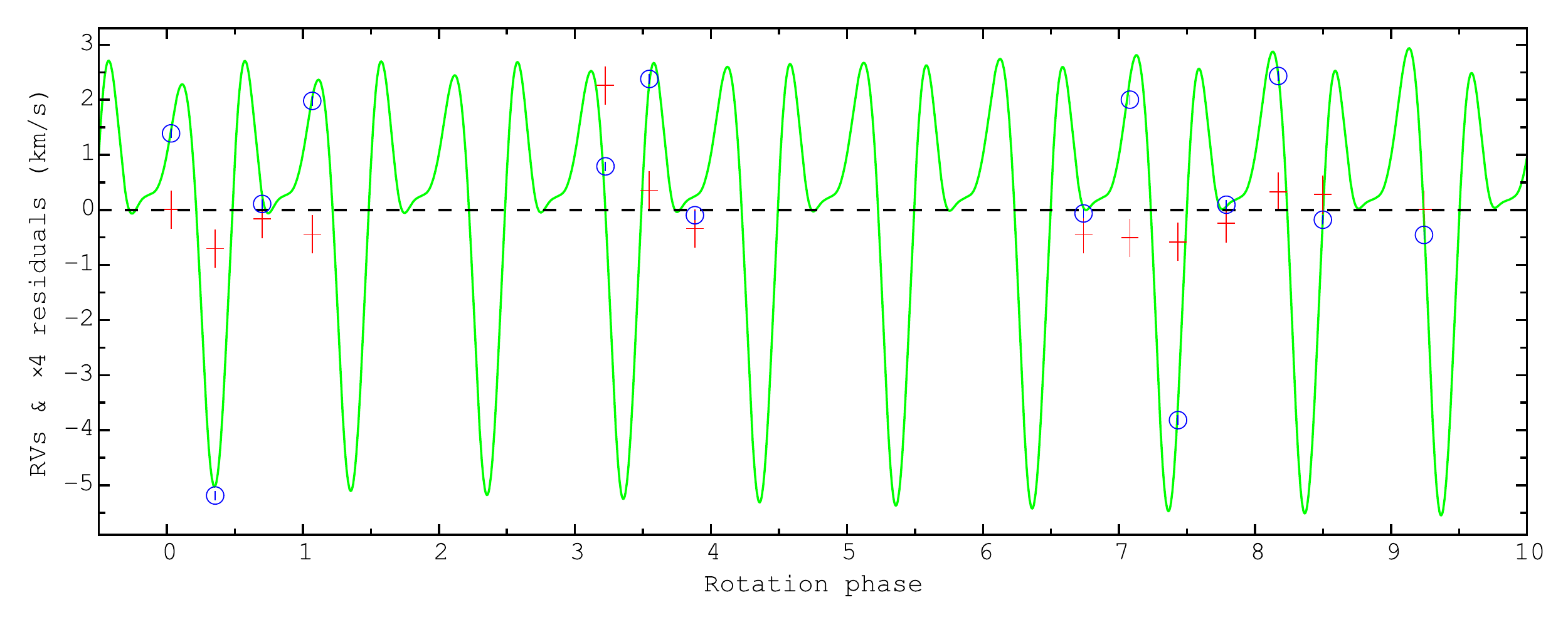}
\caption{RV variations (in the stellar rest frame) of \pa (top) and \pb (bottom) as a function of rotation phase, as measured from our observations (open blue circles) and predicted by the tomographic brightness maps of Figure~\ref{fig:brightnessmaps} (green line). RV residuals are also shown (red crosses, with values and error bars scaled by a factor of 4 for clarity), and exhibit a rms dispersion equal to 0.017~\kms for \pa and 0.086~\kms for \pb. Note that the RV residuals for \pb are those measured from the two separate maps. RVs are estimated as the first order moment of the \si LSD profiles rather than through Gaussian fits, due to their asymmetric and often irregular shape. This figure is best viewed in colour.}
\label{fig:rvcurves}
\end{figure*}

\begin{figure}
\includegraphics[width=\columnwidth]{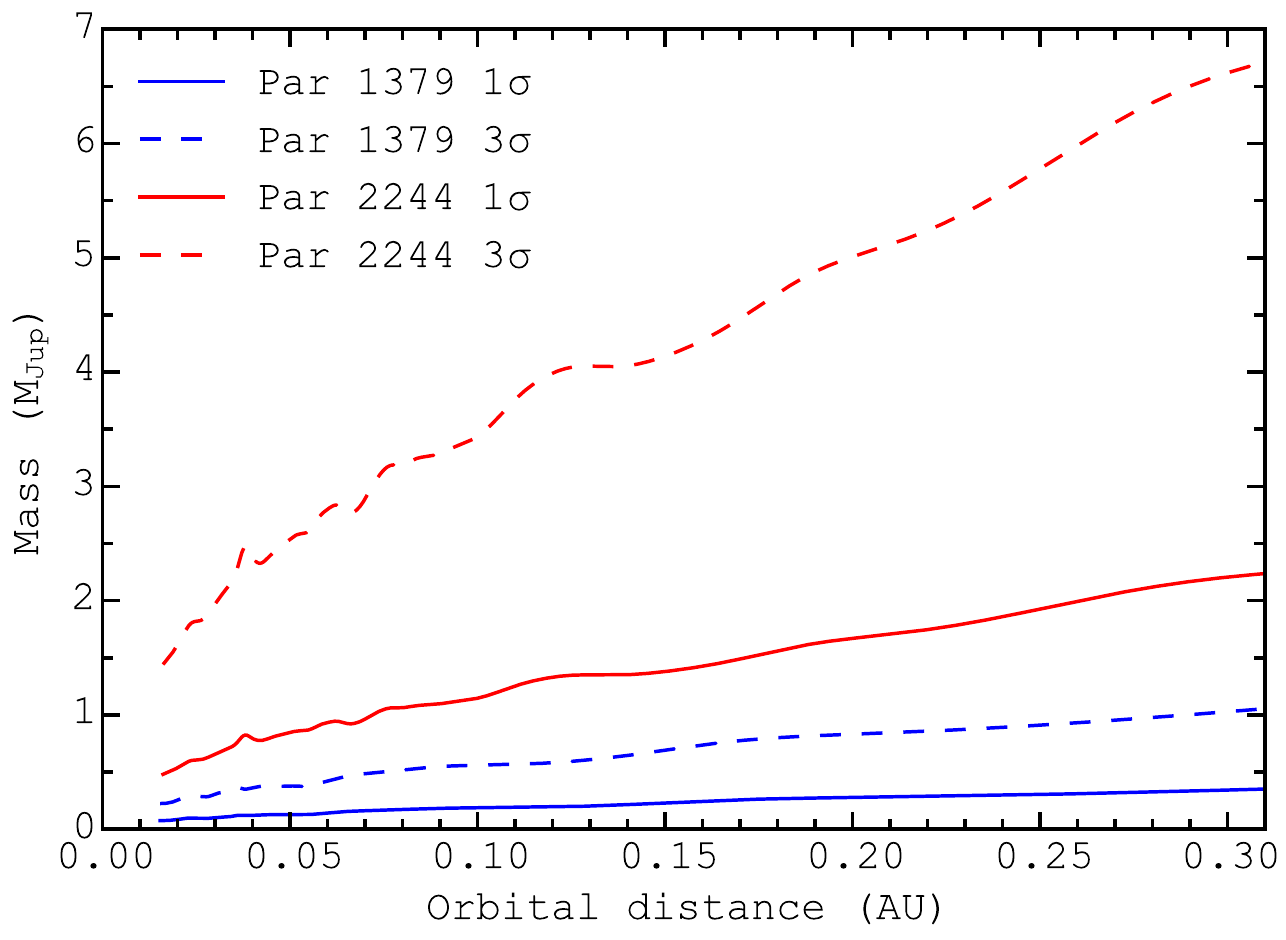} \\
\caption{The $1\sigma$ and $3\sigma$ upper limits (solid and dashed lines, respectively) on the recovered planet mass as a function of orbital distance, using the RVs shown in Figure~\ref{fig:rvcurves} for \pa (blue) and \pb (red). This figure is best viewed in colour.}
\label{fig:planetmass}
\end{figure}

Figure~\ref{fig:rvcurves} shows the predicted activity jitter and filtered RVs we derive for \pa and \pb. For \pa, the filtering process is very efficient, with the RV residuals exhibiting an rms dispersion of $\sim0.017$~\kms. These low residuals demonstrate that we are able to fit the simple surface features of \pa to a very high degree - well below the intrinsic RV precision of ESPaDOnS (around 0.03~\kms, e.g. \citealt{moutou2007,donati2008a}), and to a similar level as the intrinsic uncertainty of the filtering process itself (around 0.01~\kms in this case). Indeed, the filtered RVs are consistent with having zero amplitude (within their error bars). Hence, we find that \pa is unlikely to host a hJ with an orbital period in the range of what we can detect (i.e. not too close to the stellar rotation period or its first harmonics; see \citealt{donati2014}), with a $3\sigma$ error bar on the semi-amplitude of the RV residuals equal to 0.024~\kms, translating into a planet mass of $\simeq0.56$~\mjup orbiting at $\simeq0.1$~au (assuming a circular orbit in the equatorial plane of the star; see Figure~\ref{fig:planetmass}).

For \pb, the filtered RVs showed a significant rms dispersion of 0.15~\kms (using the optimum $\Omega_{\text{eq}}$ and $d\Omega$ from \si fitting; we find an rms dispersion of 0.14~\kms using the optimum fit to \sv data). Such a large dispersion is insignificant for assessing spot/plage coverage and magnetic field topology, but is very significant when searching for close-in giant planets. As mentioned previously, we attribute this large dispersion to the evolution of surface features over the observation gaps of 2 and 3 rotation cycles between data collection (around 6 and 8 nights, respectively; see Section~\ref{sec:observations}), as starspot lifetimes are expected to range from several weeks to months \citep[e.g.,][]{isik2007}. To take account of this evolution in our map reconstruction and RV filtering procedure, we split the \si LSD profiles into two groups of seven (with the first group spanning 18--29 Dec 2014, and the second group 06--13 Jan 2015, i.e., after the gap of 3 rotation cycles). Then, each set of 7 LSD profiles was fit to the same level of spot and plage coverage as that found for the best fit to the complete data set. Adopting the optimum parameters for \si data, this process yielded rms dispersions of the filtered RVs of 0.104~\kms and 0.131~\kms, for the first and second maps, respectively. However, when using the optimum parameters for \sv data, we find smaller rms dispersions of 0.084~\kms and 0.087~\kms for each map (an average of 0.086~\kms), respectively, suggesting that the differential rotation measurement from fitting \sv data is more robust against the intrinsic evolution of surface features. Indeed, the magnitude of these dispersions are similar to the 0.087~\kms uncertainties on the filtered RVs, showing that our tomographic modelling can account for the majority of the RV variability. These residuals are lower than our detection threshold of $\simeq0.1$~\kms (from preliminary simulations, see \citealt{donati2014}). Furthermore, we find no significant peaks in a Lomb-Scargle periodogram of the filtered RVs. Thus, we conclude that \pb is unlikely to host a hJ with an orbital period of what we can detect, with a $3\sigma$ error bar on the semi-amplitude of the RV residuals equal to 0.206~\kms, translating into a value of $\sim3.54$~\mjup for a planet orbiting at a distance of 0.1~au (again assuming a circular orbit in the equatorial plane of the star; see Figure~\ref{fig:planetmass}).

We note that we also fit both \si and \sv profiles simultaneously for each group of 7 profiles, with the resulting magnetic field topologies being largely similar to that found when fitting all 14 profiles, except that the unsigned field strength was around 20 per cent lower. However, we were unable to reliably constrain a measurement of differential rotation using only 7 profiles.

\section{Discussion}
\label{sec:discussion}
We have reported results from our spectropolarimetric observations collected with ESPaDOnS of the wTTSs \pa and \pb, in the framework of the international MaTYSSE Large Program.

We find that both stars have largely similar atmospheric properties, with photospheric temperatures of $4600 \pm50$~K and $4650\pm50$~K, and logarithmic gravities in cgs units of $3.9\pm0.2$ and $4.1\pm0.2$. These properties suggest that \pa and \pb are of a similar mass ($1.6\pm0.1$ and $1.8\pm0.1$~\msun), with respective radii of $2.7\pm0.2$ and $3.5\pm0.2$~\rsun, viewed at inclinations of 36\degr and 59\degr. We estimate their ages to be $1.8\pm0.6$ and $1.1\pm0.3$~Myr, with internal structures that are both around 70 per cent convective by radius (using \citealt{siess2000} evolutionary models).

The stars' rotation periods of 5.585~d and 2.8153~d for \pa and \pb, respectively, are similar to those of the (lower mass) wTTSs, V819~Tau and V830~Tau \cite{donati2015}, but are much slower than V410~Tau (a wTTSs with similar mass and position in the H-R diagram, rotating in 1.87~d). Furthermore, our targets' periods lie within the uniform period distribution between 1--8~d of the vast majority of stars in the ONC flanking fields \cite{rebull2001}.

We find that \pa harbours a dusty circumstellar disc with an inner radius of around 0.15~au, and with a dust mass ranging between $10^{-7}$--$10^{-5}$~\msun, with \pb surrounded by a either primordial dust, or a debris disc (starting around 20~au). This may explain why \pa rotates slower than \pb, despite being a similar age, suggesting that \pb has dissipated its (presumably less massive) accretion disc somewhat sooner than \pa, and has spent more time spinning up.

By using common tracers of accretion (see Section~\ref{sec:accretion}), we find that \pa and \pb may be accreting at levels of ${\log{\dot{M}_{\mathrm{acc}}} \simeq -10}$ to $-11$~\msunyr. Given that this accretion rate is so low, our measurements are likely to be strongly affected by chromospheric emission, and so they must be taken as upper limits. These very low accretion rates suggest that these targets may be classed as wTTSs.

As the accretion rates for \pa are low, there may be other explanations for the absorption features seen in H$\alpha$ that are unrelated to inflow along an accretion funnel. Using our map of surface magnetic fields (see Section~\ref{sec:magfield}), we have compared the observed variations in H$\alpha$ in Figure~\ref{fig:halpha} with our potential field extrapolation (see Figure~\ref{fig:par1379extrap}). At cycle 0.191, we see the (observed) maximum of redshifted absorption in H$\alpha$, with the potential field extrapolation showing several large active regions, with many closed magnetic loops, some of which are aligned along the line of sight. At cycle 0.544, we see no redshifted absorption in H$\alpha$, with few regions of closed loops in the field extrapolation, but rather mainly open field lines. Finally, cycle 1.783 also shows no redshifted absorption, but rather blueshifted emission, with the extrapolation showing a region of many closed loops, aligned with the line of sight.

Given that the absorption appears when we see closed magnetic loops, it may be caused by an infall of material along these loops from flare or prominence material (as seen in the late-type rapid-rotator LO Peg, see \citealt{eibe1999}). Furthermore, the strongest absorption occurs around 140~\kms - the same as the free-fall velocity of the star, and so infalling material from the top of the larger loops could create such absorption features. Moreover, the blueshifted emission (seen prominently at cycle 1.783) may trace the same prominence material that falls down towards the star in front of the disc around phase 0.2; the material may have be seen off the stellar limb (as a result of the low inclination), and is thus in emission and moving towards the observer (while still falling onto the star) around half a rotation cycle later. Alternatively, the blue-shifted emission may be caused by erupting prominence material along sight-aligned loops.

If the redshifted absorption were due to inflow along an accretion funnel, we would expect to see maximum absorption when many open field lines (that are able to connect to the inner disc) are aligned along the line of sight. This is not the case, as there is no absorption at, for example, cycle 0.544 (see Figure~\ref{fig:par1379extrap}), where many open field lines are seen. Additionally, we note that while \cite{megeath2012} place \pa near the limit between wTTSs and cTTSs, \cite{rebull2001} report a very regular periodic light curve for \pa that does not appear like that of a cTTS, with \cite{rebull2000} reporting a lack of UV excess, further supporting the non-accreting hypothesis.

Using our tomographic imaging code (adapted for wTTSs, see \citealt{donati2014}), we derived surface brightness maps and magnetic topologies of both stars. Cool spots and warm plages are found on both \pa and \pb, with the latter exhibiting more complex features with higher contrast, that appear to show some intrinsic evolution over our $\sim1$~month observing window.

The reconstructed magnetic fields for \pa and \pb are significantly different in both strength and topology. \pa harbours a predominately non-axisymmetric poloidal field (3/4 of the total field), with the largest fraction of energy in the quadrupolar mode, and where the large-scale magnetosphere of the poloidal field is inclined at $\simeq65\degr$ to the rotation axis. In contrast, the field of \pb is split 3:2 into a mostly non-axisymmetric poloidal component (with half of the reconstructed energy in modes with $\ell \geq 4$),  tilted at $\simeq45\degr$ from the rotation axis, and a mostly axisymmetric toroidal component.

The magnetic field of \pb is fairly similar in strength and topology to that of V410~Tau, the wTTS lying closest in the H-R diagram that has been mapped with ZDI ($M_{\star} = 1.4\pm0.2$~\msun, age $\simeq1.2$~Myr, average unsigned field strength 0.49~kG, \citealt{skelly2010}). The fields of both of these stars are split fairly evenly between a mostly non-axisymmetric poloidal component, and a toroidal component, with a similarly high number of modes required to fit the data ($\ell \leq18 $ for \pb, $\ell = 15$ for V410~Tau).

Compared to the lower mass wTTSs, V819~Tau and V830~Tau, \pb has a similar field strength, but is much more complex \citep{donati2015}. The field strength of \pa is also weaker than that of Tap~35 (1500~G), but is similar to the 700~G field of Tap~10 \citep{basri1992}. In contrast, the much less complex field of \pa is also much weaker than any other wTTSs that has been mapped to date, suggesting it is likely to be more structurally evolved than \pb (assuming the magnetic topology is related to the development of a radiative core, based on \citealt{siess2000} models), becoming largely radiative already, despite model predictions.

\begin{figure*}
\begin{center}
\includegraphics[width=\textwidth]{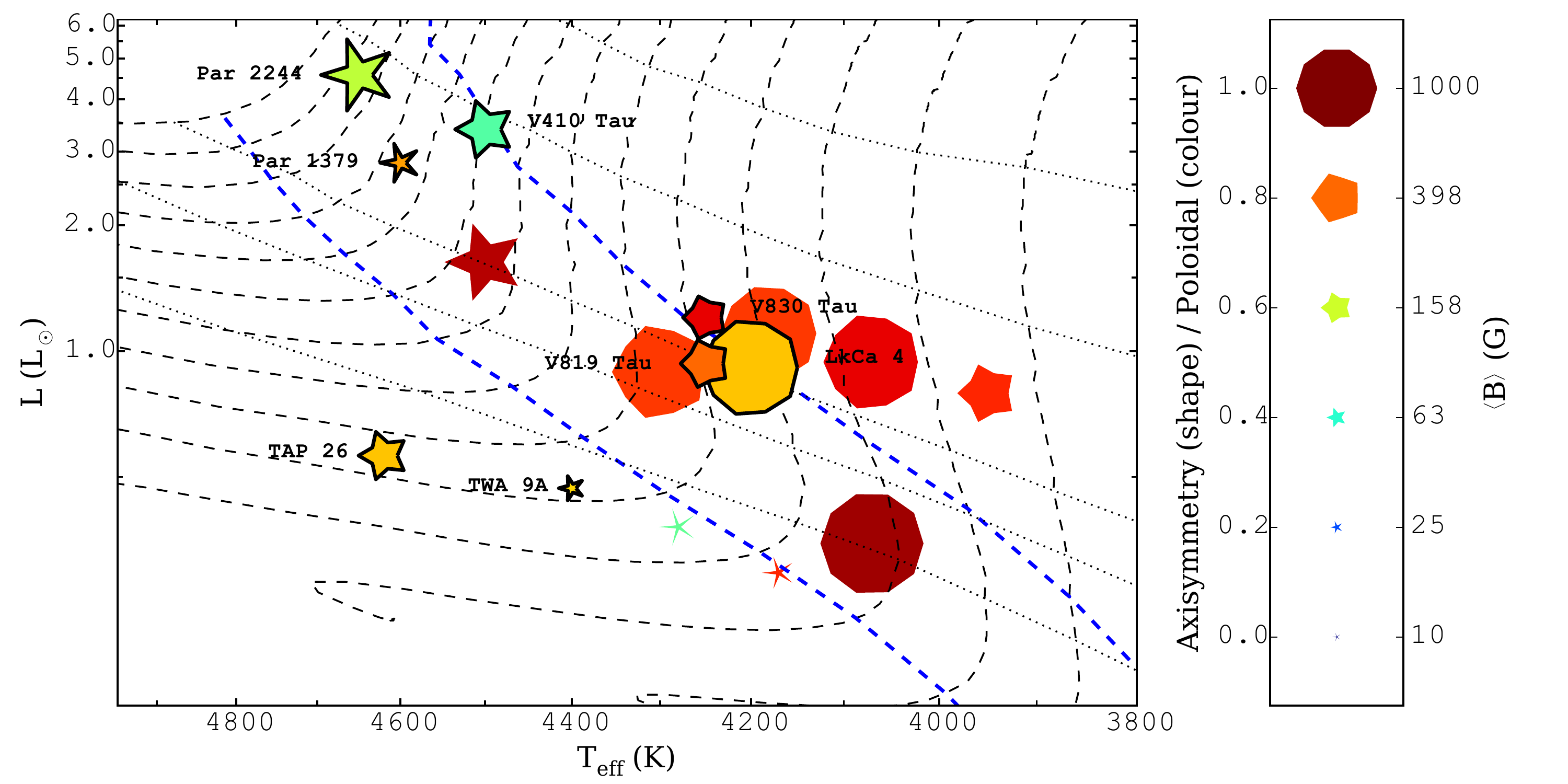} 
\caption{H-R diagram showing the MaTYSSE wTTSs (black line border and labelled) and the MaPP cTTSs (no border). The size of the symbols represents the surface-averaged magnetic field strength (with a larger symbol meaning a stronger field), the colour of the symbol represents the fraction of the field that is poloidal (with red being completely poloidal), and the shape of the symbols represents the axisymmetry of the poloidal field component (with higher axisymmetry shown as a more circular symbol). Also shown are evolutionary tracks from \protect \cite{siess2000} (black dashed lines, ranging from 0.5--1.9\msun), with corresponding isochrones (black dotted lines, for ages of 0.5, 1, 3, 5 \& 10 Myr), and lines showing 100\% and 50\% convective interoir by radius (blue dashed).}
\label{fig:confuse}
\end{center}
\end{figure*}

A comprehensive analysis of the similarity between magnetic fields of wTTSs and cTTSs is still premature at this stage. For completeness, however, we show in Figure~\ref{fig:confuse} an H-R diagram of the cTTSs from the MaPP programme, as well as the (analysed) MaTYSSE wTTSs. Figure~\ref{fig:confuse} also indicates the fraction of the field that is poloidal, the axisymmetry of the poloidal component, and shows PMS evolutionary tracks from \cite{siess2000}. One can see that the wTTSs studied thus far generally show a wider range of field topologies compared to cTTSs, with large scale fields that can be more toroidal and non-axisymmetric (also see discussion in \citealt{donati2015}). Clearly, given our limited sample of wTTSs with which to compare field strength and topology, further studies of MaTYSSE stars are required before we can carry out a full analysis.

Our data indicate that significant latitudinal shear has occurred for both brightness and magnetic maps over the observation timescales for each star (14 and 25 nights, respectively). For \pa, we estimated the amount of differential rotation to be non-zero at a confidence level of 99.5~per~cent, with a shear rate $1.4\times$ smaller than that of the Sun (with an equator-pole lap time of $\simeq160$~d, as opposed to $\simeq110$~d for the Sun). Furthermore, the estimates derived from brightness and magnetic maps agree to within their error bars. For \pb, we estimate the amount of differential rotation to be non-zero at a confidence level of over 99.99~per~cent (for both brightness and magnetic maps). We find that the intrinsic evolution of surface features over our observation window resulted in a much lower shear rate for the brightness maps, as compared to the magnetic maps (with shear rates $7.3\times$ and $1.8\times$ smaller than the Sun, for \si and \sv data, respectively). However, by accounting for this evolution, we show that the estimate from the magnetic maps is more robust, and so determine the equator-pole lap time to be $\simeq205$~d. Our results show higher shear rates than those for other wTTSs that have had a similar measurement, namely TWA~6, V410~Tau, V819~Tau, V830~Tau, LkCa~4 and TAP~26 \citep{skelly2008,skelly2010,donati2014,donati2015,yu2017}. However, all these stars (apart from V410~Tau) have significantly different masses, and occupy a different part of the H-R diagram. Furthermore, our measured shear rates are similar to that found for a cTTS with similar properties, namely V2129~Oph (where the shear is $1.5\times$ smaller than the Sun, \citealt{donati2011}).

%v830tau = 0.0124±0.0029
%v819tau = 0.008±0.008
%LkCa4 = 0.010±0.006

Using our tomographic maps to predict the activity related RV jitter, we were able to filter the RV curves of both stars in the search for potential hJs (in the same manner as \citealt{donati2014,donati2015}). For \pa, we find that the activity jitter is filtered down to a rms RV precision of 0.017~\kms, a value similar to the intrinsic uncertainty of the filtering process itself, and lower than the RV stability of ESPaDOnS. With no significant peaks in a Lomb-Scargle periodogram of the filtered RVs, we find that \pa is unlikely to host a hJ with an orbital period of what we can detect, with a $3\sigma$ error bar on the semi-amplitude of the RV residuals equal to 0.024~\kms, translating into a planet mass of $\simeq0.56$~\mjup orbiting at $\simeq0.1$~au. For \pb, we find a significantly larger rms dispersion of 0.086~\kms, however, these residuals are lower than our estimated detection threshold for hJs ($\simeq0.1$~\kms, see \citealt{donati2014}). Furthermore, we find no significant peaks in a Lomb-Scargle periodogram of the filtered RVs, and thus conclude that \pb is unlikely to host a hJ with an orbital period of what we can detect, with a $3\sigma$ error bar on the semi-amplitude of the RV residuals equal to 0.206~\kms, translating into a value of $\sim3.54$~\mjup for a planet orbiting at a distance of 0.1~au.

\section{Summary}
\label{sec:summary}
We report the results of our spectropolarimetric monitoring of the wTTSs \pa and \pb, within the MaTYSSE (Magnetic Topologies of Young Stars and the Survival of close-in giant Exoplanets) programme. We have determined out target stars to be of a similar mass (1.6 and 1.8~\msun) and age (1.8 and 1.1~Myr), with \pa hosting an evolved low-mass dusty circumstellar disc, and with \pb showing evidence of a young debris disc. Using several accretion diagnostics, we find that the stars may be accreting at a very low level, however, our derived accretion rates are strongly influenced by chromospheric emission (due to stellar activity), and so are likely unreliable. Using tomographic imaging, we have modelled the rotational modulation of line profile distortions and Zeeman signatures, yielding brightness and magnetic maps of the surface. We find that \pa harbours a weak (250~G), mostly poloidal field tilted $65\degr$ from the rotation axis. In contrast, \pb hosts a stronger field (860~G) split 3:2 between poloidal and toroidal components, with most of the energy in higher order modes, and with the poloidal component tilted $45\degr$ from the rotation axis. Compared to the lower mass wTTSs, V819 Tau and V830 Tau, \pb has a similar field strength, but is much more complex, whereas the much less complex field of \pa is also much weaker than any other mapped wTTS. We find moderate surface differential rotation of $1.4\times$ and $1.8\times$ smaller than Solar. Using our tomographic maps to predict the activity related radial velocity (RV) jitter, and filter it from the RV curves, we find RV residuals with dispersions of 0.017~\kms and 0.086~\kms for \pa and \pb, respectively. We find no evidence for close-in giant planets around either star, with $3\sigma$ upper limits of 0.56 and 3.54~\mjup (at an orbital distance of 0.1~au).

\section*{Acknowledgements}
This paper is based on observations obtained at the CFHT, operated by the National Research Council of Canada, the Institut National des Sciences de l'Univers of the Centre National de la Recherche Scientifique (INSU/CNRS) of France and the University of Hawaii. We thank the CFHT QSO team for its great work and effort at collecting the high-quality MaTYSSE data presented in this paper. MaTYSSE is an international collaborative research programme involving experts from more than 10 different countries (France, Canada, Brazil, Taiwan, UK, Russia, Chile, USA, Switzerland, Portugal, China and Italy). We also warmly thank the IDEX initiative at Universit\'{e} F\'{e}d\'{e}rale Toulouse Midi-Pyr\'{e}n\'{e}es (UFTMiP) for funding the STEPS collaboration program between IRAP/OMP and ESO and for allocating a `Chaire d'Attractivit\'{e}' to GAJH ,allowing her to regularly visit Toulouse to work on MaTYSSE data. SGG acknowledges support from the Science \& Technology Facilities Council (STFC) via an Ernest Rutherford Fellowship [ST/J003255/1]. SHPA acknowledges financial support from CNPq, CAPES and Fapemig. This work has made use of the VALD database, operated at Uppsala University, the Institute of Astronomy RAS in Moscow, and the University of Vienna, and the SVO Filter Profile Service supported from the Spanish MINECO through grant AyA2014-55216. We thank Christophe Pinte for the use of his \textsc{mcfost} radiative transfer code.

%%%%%%%%%%%%%%%%%%%%%%%%%%%%%%%%%%%%%%%%%%%%%%%%%%

\bibliographystyle{mnras}
\bibliography{matysse_par1379_par2244_v2.bbl}

%%%%%%%%%%%%%%%%% APPENDICES %%%%%%%%%%%%%%%%%%%%%
\appendix
\section{Par 1379 H$\alpha$ and H$\beta$ line profiles, and potential field extrapolations, Par 2244 H$\alpha$ line profiles}
\label{app:halpha}

\begin{figure*}
\begin{center}
\includegraphics[height=0.65\textheight]{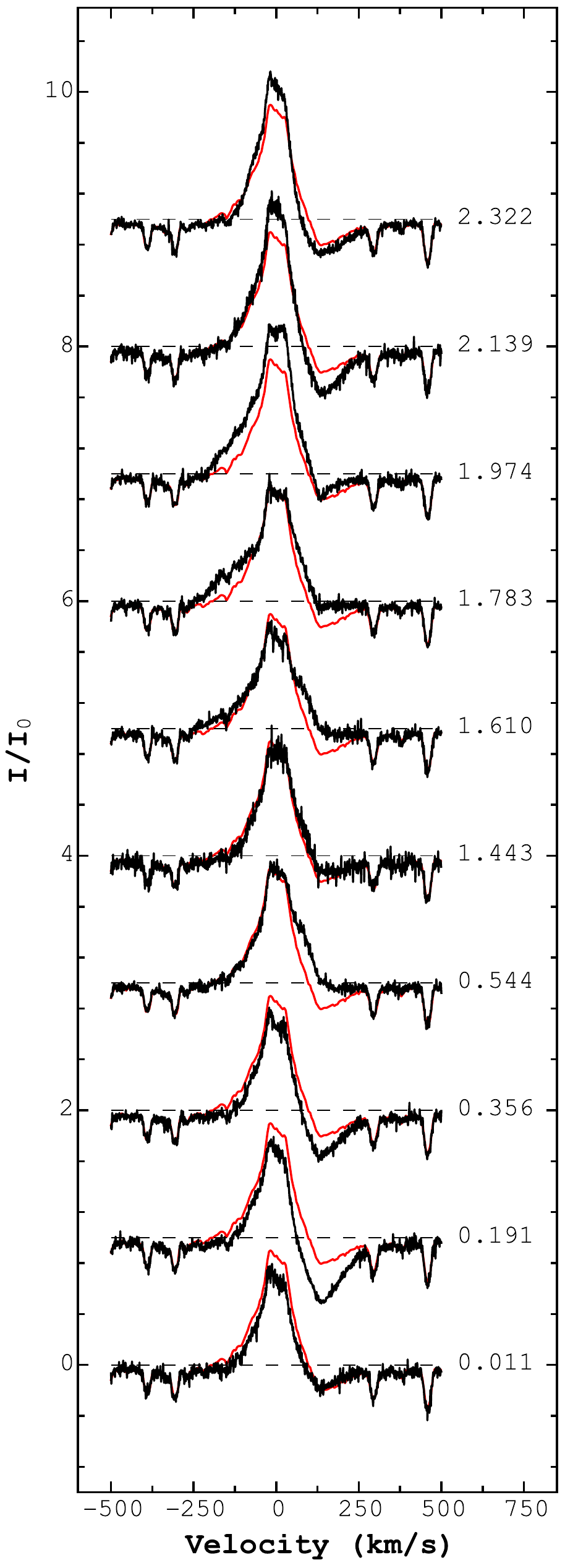}
\includegraphics[height=0.65\textheight]{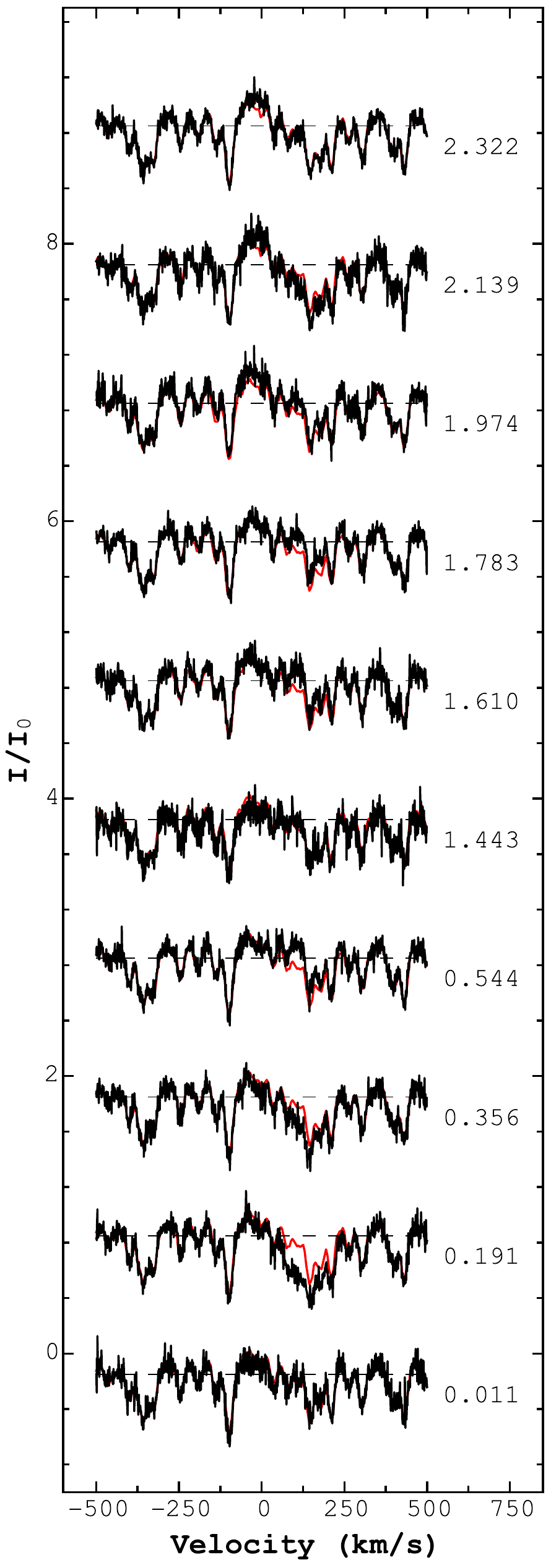}
\hspace{0.5cm}
\includegraphics[height=0.65\textheight]{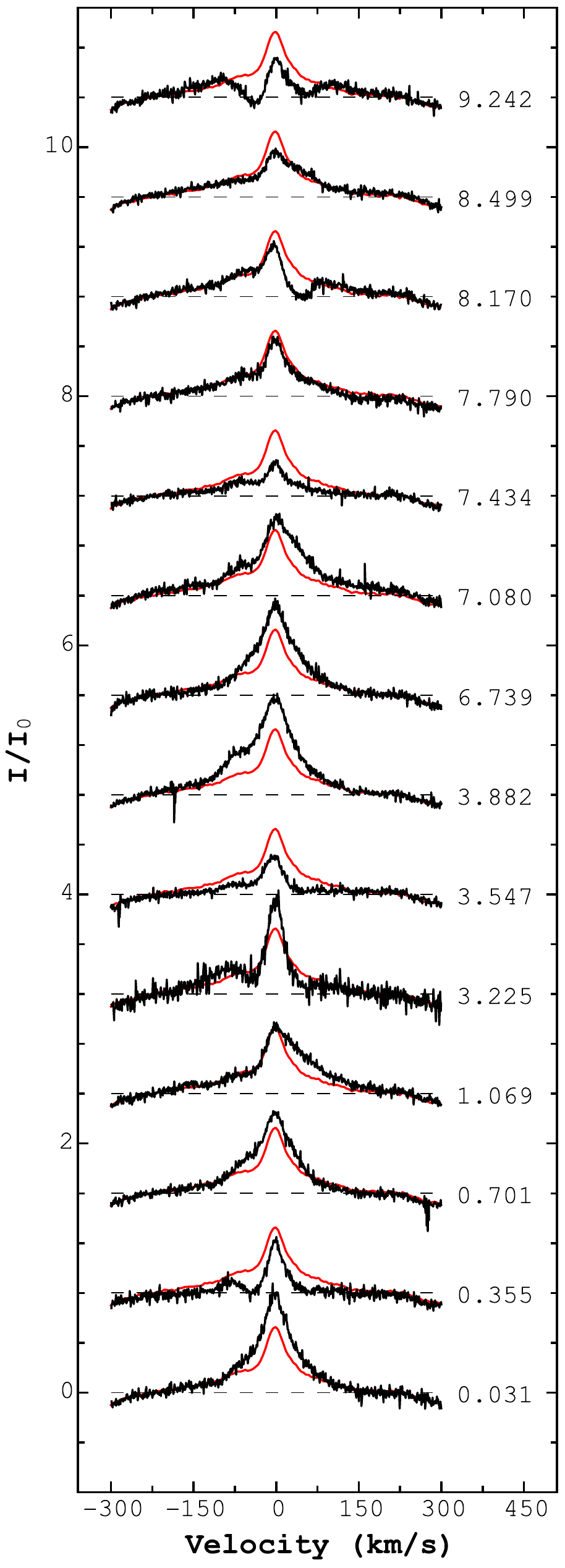} 
\caption{H$\alpha$ and H$\beta$ line profiles for \pa (left, centre) and H$\alpha$ line profiles for \pb (right), all shown as black solid lines, where the mean line profile is shown in red, and the continuum as a black dashed line, with the cycle number displayed on the right of each profile.}
\label{fig:halpha}
\end{center}
\end{figure*}

\begin{figure*}
\begin{center}
\includegraphics[width=0.45\textwidth]{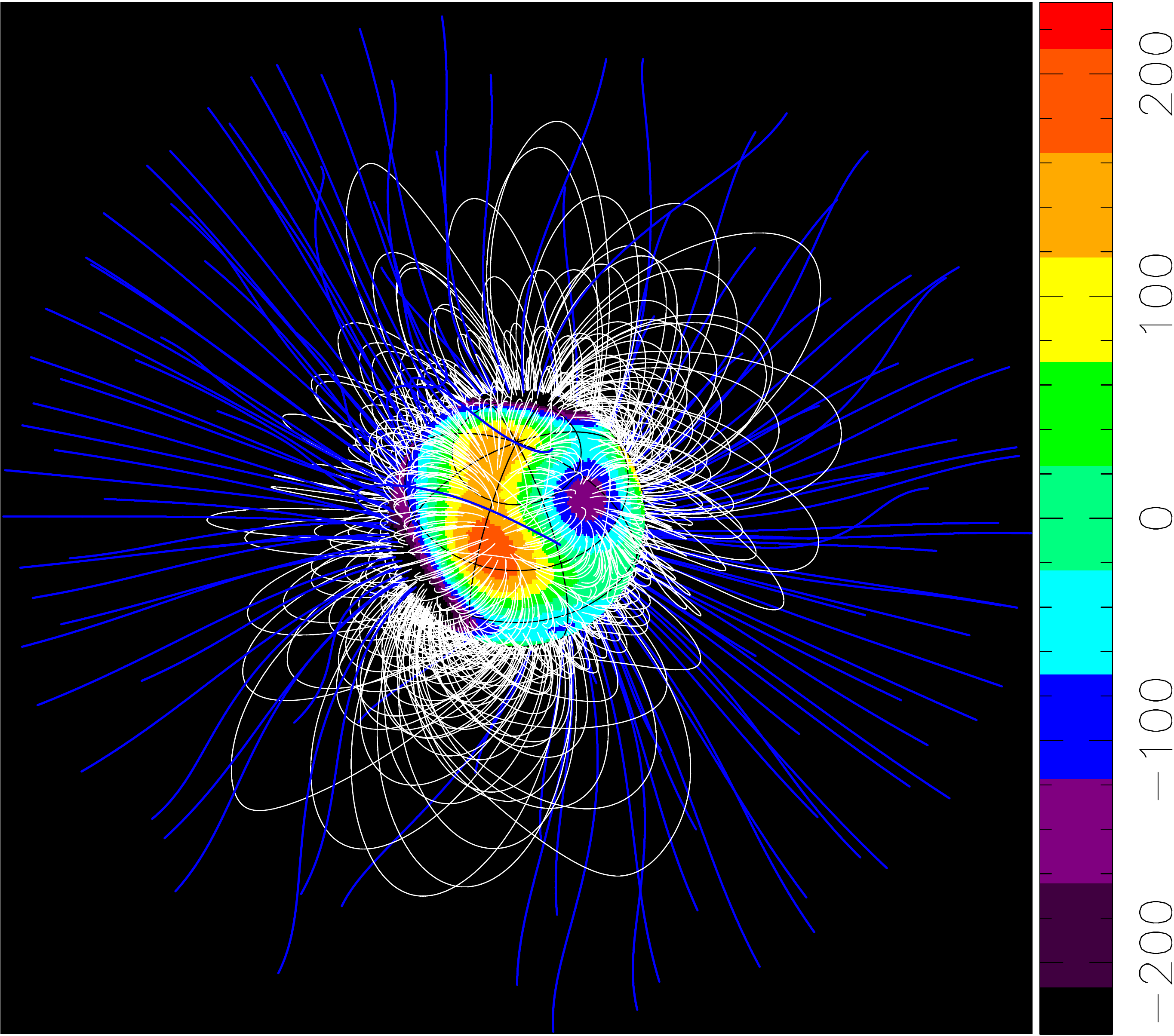} \\
\includegraphics[width=0.45\textwidth]{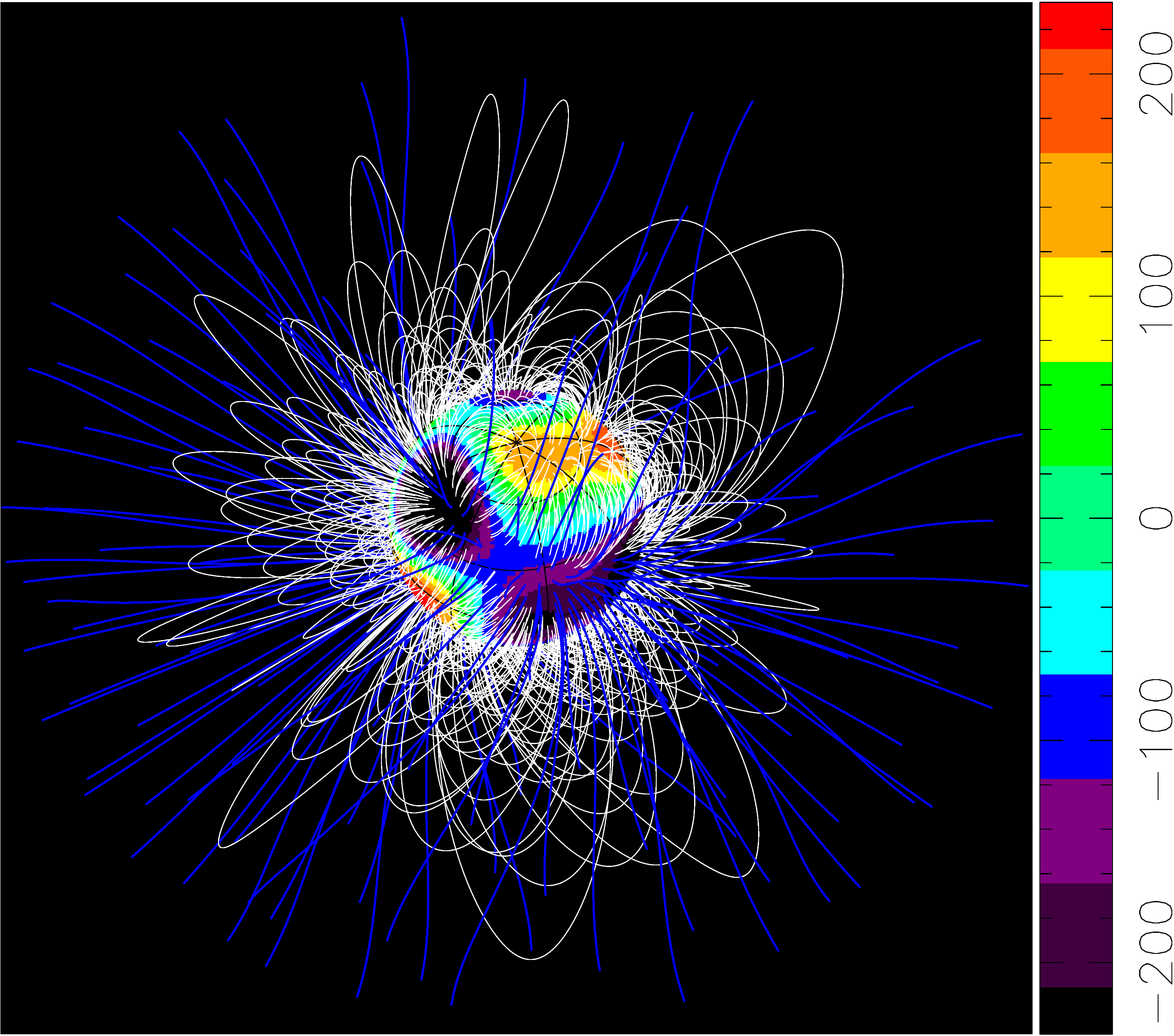} \\
\includegraphics[width=0.45\textwidth]{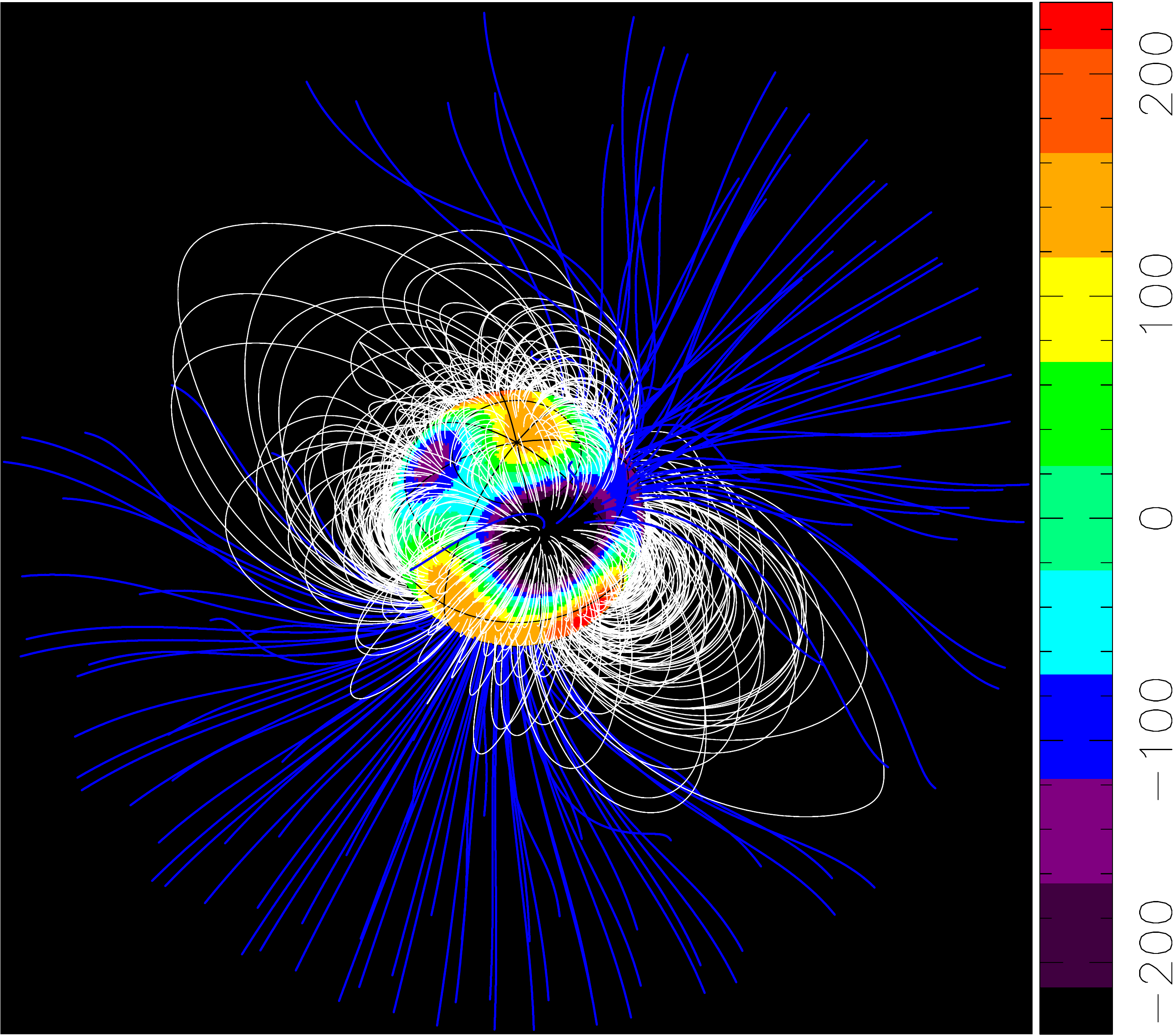}
\caption{Potential extrapolations of the magnetic field reconstructed for \pa (as for Figure~\ref{fig:extrap}), shown at phases 0.191 (top), 0.544 (middle) and 1.783 (bottom). A full animation may be found online at http://imgur.com/a/qwAXg.}
\label{fig:par1379extrap}
\end{center}
\end{figure*}

\section{SED fitting probability and histogram plots}
\label{app:sedprob}

\begin{figure*}
\includegraphics[width=0.9\textwidth]{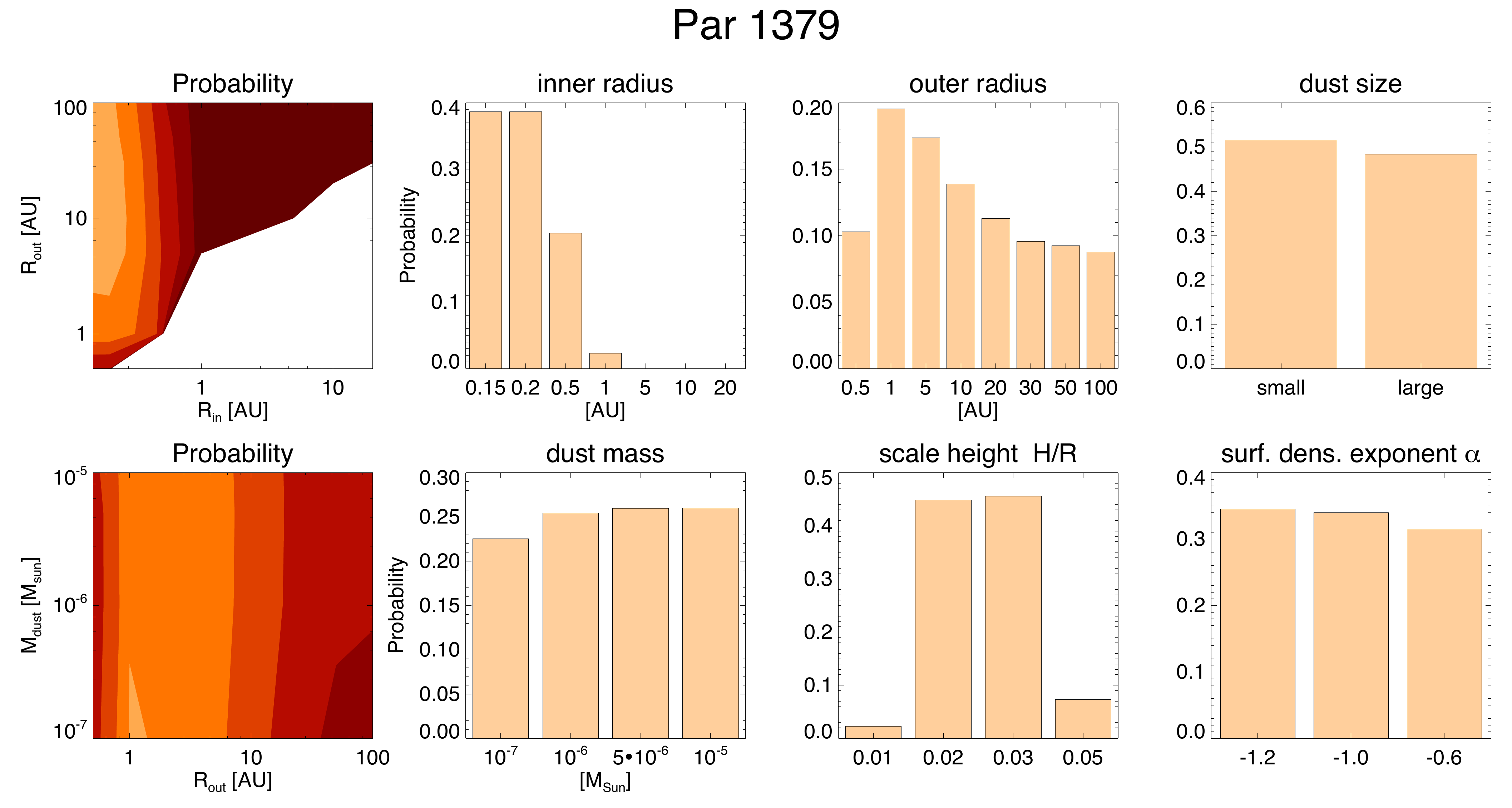}\\
\includegraphics[width=0.9\textwidth]{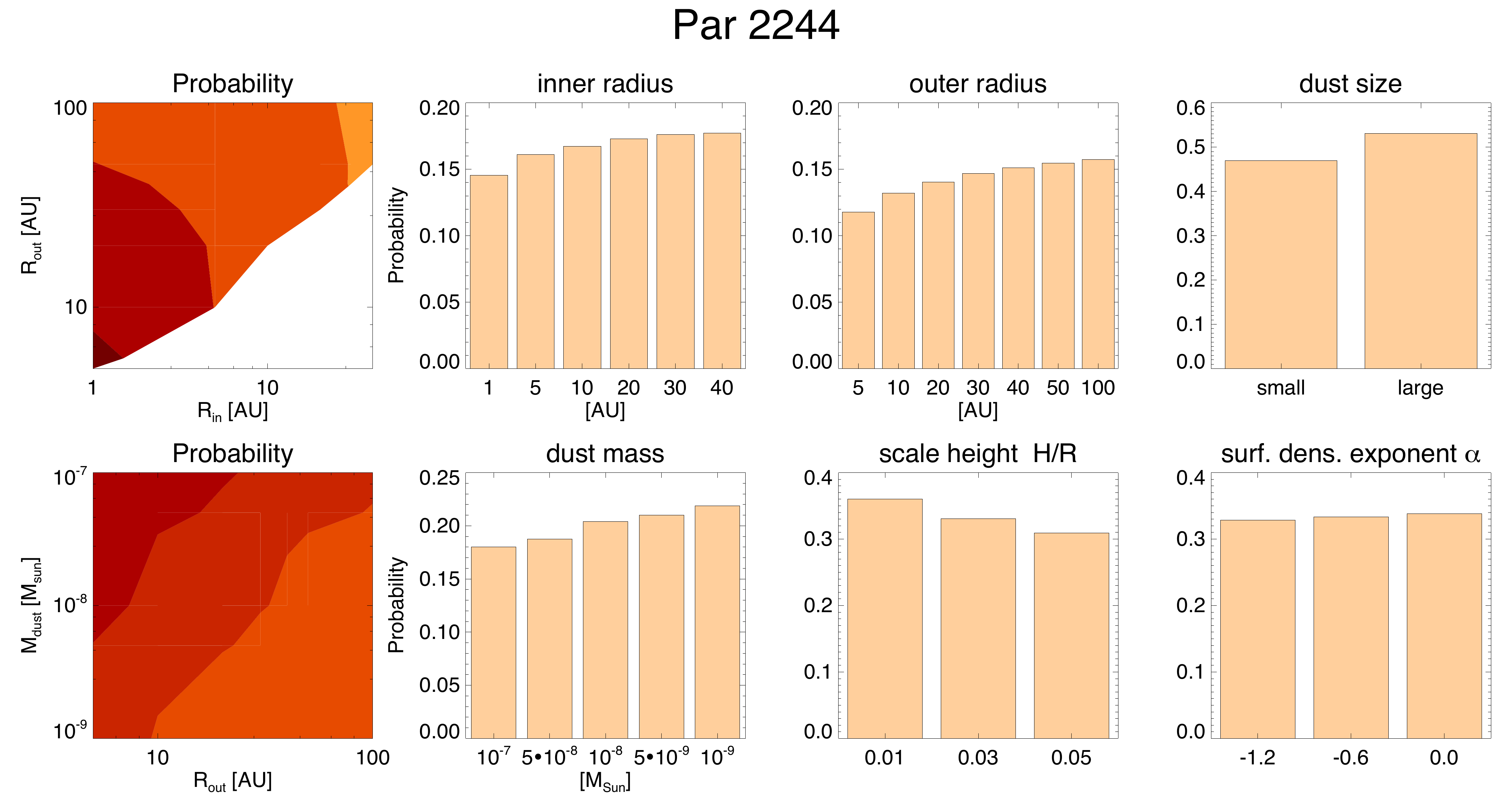} \\
\caption{2-D probability and histograms of the grid of SED models calculated for \pa and \pb. The colour code in the $R_{\rm out}$ vs $R_{\rm in}$ and $M_{\rm dust}$ vs $R_{\rm out}$ contour plots is $P = 0.01, 0.02, 0.03, 0.04$ and 0.05, from dark red to orange, with lighter colours being a higher probability.}
\label{fig:Histograms}
\end{figure*}

%%%%%%%%%%%%%%%%%%%%%%%%%%%%%%%%%%%%%%%%%%%%%%%%%%

% Don't change these lines
\bsp	% typesetting comment
\label{lastpage}
\end{document}